\documentclass[12pt]{article}
\usepackage{stan-papers}

\title{The Stan Math Library: \\ Reverse-Mode Automatic
  Differentiation \\ in C++}

\author{Bob Carpenter \\ {\small Columbia University}
        \and Matthew D.\ Hoffman \\ {\small Adobe Research}
        \and Marcus Brubaker 
        \\ {\small \begin{tabular}{c}University of Toronto, \\ Scarborough\end{tabular}}
        \and Daniel Lee \\ {\small Columbia University}
        \and Peter Li \\ {\small Columbia University}
        \and Michael Betancourt \\ {\small University of Warwick}
}
\date{\vspace*{8pt}\normalsize \today}

\begin{document}

\maketitle
\thispagestyle{empty}

\begin{abstract} 
  \noindent
  As computational challenges in optimization and statistical
  inference grow ever harder, algorithms that utilize derivatives are
  becoming increasingly more important.  The implementation of the
  derivatives that make these algorithms so powerful, however, is a
  substantial user burden and the practicality of these algorithms
  depends critically on tools like automatic differentiation that
  remove the implementation burden entirely.  The Stan Math Library is
  a C++, reverse-mode automatic differentiation library designed to be
  usable, extensive and extensible, efficient, scalable, stable,
  portable, and redistributable in order to facilitate the
  construction and utilization of such algorithms.

  Usability is achieved through a simple direct interface and a
  cleanly abstracted functional interface.  The extensive built-in
  library includes functions for matrix operations, linear algebra,
  differential equation solving, and most common probability
  functions.  Extensibility derives from a straightforward
  object-oriented framework for expressions, allowing users to easily
  create custom functions. Efficiency is achieved through a
  combination of custom memory management, subexpression caching,
  traits-based metaprogramming, and expression templates.  Partial
  derivatives for compound functions are evaluated lazily for improved
  scalability.  Stability is achieved by taking care with arithmetic
  precision in algebraic expressions and providing stable, compound
  functions where possible. For portability, the library is
  standards-compliant C++ (03) and has been tested for all major
  compilers for Windows, Mac OS X, and Linux.  It is distributed under
  the new BSD license.

  This paper provides an overview of the Stan Math Library's
  application programming interface (API), examples of its use, and a
  thorough explanation of how it is implemented.  It also demonstrates
  the efficiency and scalability of the Stan Math Library by comparing
  its speed and memory usage of gradient calculations to that of
  several popular open-source C++ automatic differentiation systems
  (Adept, Adol-C, CppAD, and Sacado), with results varying
  dramatically according to the type of expression being
  differentiated.
\end{abstract}



\section{Reverse-Mode Automatic Differentiation}

Many contemporary algorithms require the evaluation of a derivative
of a given differentiable function, $f$, at a given input value, 
$\left( x_{1}, \ldots, x_{N} \right)$, for example a gradient,
\begin{equation*}
  \left(\frac{ \partial f }{ \partial x_{1} } \left( x_{1}, \ldots, x_{N} \right),
  \cdots,
  \frac{ \partial f }{ \partial x_{N} } \left( x_{1}, \ldots, x_{N} \right) \right),
\end{equation*}
or a directional derivative,%
\footnote{A special case of a directional derivative computes derivatives with
respect to a single variable by setting $\vec{v}$ to a vector with a
value of 1 for the single distinguished variable and 0 for all other
variables.}
\begin{equation*}
\vec{v} ( f ) \left( x_{1}, \ldots, x_{N} \right) = 
\sum_{n = 1}^{N} v_{n} \frac{ \partial f }{ \partial x_{n} } \left( x_{1}, \ldots, x_{N} \right).
\end{equation*}
Automatic differentiation computes these values automatically,
using only a representation of $f$ as a computer program.  For example, 
automatic differentiation can take a simple C++ expression such as 
\code{x~*~y~/~2} with inputs $\code{x}=6$ and $\code{y}=4$ and 
produce both the output value, 12, and the gradient, $(2,3)$.

Automatic differentiation is implemented in practice by transforming
the subexpressions in the given computer program into nodes of an
\textit{expression graph} (see \reffigure{expression-graph}, below,
for an example), and then propagating chain rule evaluations along
these nodes \citep{GriewankWalther:2008,Giles:2008}. In
\textit{forward-mode automatic differentiation}, each node $k$ in the
graph contains both a value $x_k$ and a \textit{tangent}, $t_k$, which
represents the directional derivative of $x_k$ with respect to the
input variables.  The tangent values for the input values are
initialized with values $\vec{v}$, because that represents the
appropriate directional derivative of each input variable.  The
complete set of tangent values is calculated by propagating tangents
forward from the inputs to the outputs with the rule
\begin{equation*}
t_i
=
\sum_{j \in \mathrm{children}[i]}
\frac{ \partial x_i }{ \partial x_j } \, t_j.
\end{equation*}
In one pass over the expression graph, forward-mode computes a
directional derivative of any number of output (dependent) variables
with respect to a vector of input (independent) variables and
direction $\vec{v}$; the special case of a derivative with respect to
a single independent variable computes a column of the Jacobian of a
multivariate function.  The proof follows a simple inductive argument
starting from the inputs and working forward to the output(s).

In \textit{reverse-mode automatic differentiation}, each node $k$ in
the expression graph contains a value $x_k$ and an \textit{adjoint}
$a_k$, representing the derivative of a single output node with
respect to $x_k$.  The distinguished output node's adjoint is
initialized to 1, because its derivative with respect to itself is 1.
The full set of adjoint values is calculated by propagating backwards
from the outputs to the inputs via
\begin{equation*}
a_j
=
\sum_{i \in \mathrm{parents}[j]}
\frac{ \partial x_i }{ \partial x_j } a_i.
\end{equation*}
This enables the derivative of a single output variable to be done
with respect to multiple input variables in a single pass over the
expression graph.  A proof of correctness follows by induction
starting from the base case of the single output variable and working
back to the input variables. This computes a gradient with respect to
a single output function or one row of the Jacobian of a multi-output
function.

Both forward- and reverse-mode require partial derivatives of each
node $x_i$ in the expression graph with respect to its daughters $x_j$.

Because gradients are more prevalent in contemporary algorithms,
reverse-mode automatic differentiation tends to be the most efficient
approach in practice.  In this section we take a more detailed look
at reverse-mode automatic differentiation and compare it to other
differential algorithms.

\subsection{Mechanics of Reverse-Mode Automatic Differentiation}

As an example, consider the log of the normal probability density function
for a variable $y$ with a normal distribution with mean $\mu$ and
standard deviation $\sigma$,
\renewcommand{\theequation}{\arabic{equation}}
\begin{equation}\label{normal-log-density.equation}
f(y, \mu, \sigma) = \log \left( \distro{Normal}(y|\mu,\sigma) \right)
= -\frac{1}{2} \left( \frac{y - \mu}{\sigma} \right)^2
- \log \sigma
- \frac{1}{2} \log (2 \pi)
\end{equation}
and its gradient,
\begin{align}
\frac{ \partial f }{ \partial y} (y, \mu, \sigma)
&= -(y - \mu) \sigma^{-2} \nonumber
\\[6pt]
\frac{ \partial f }{ \partial \mu} (y, \mu, \sigma)
&= (y - \mu) \sigma^{-2} \label{gradient-normal-log-density.equation}
\\[6pt]
\frac{ \partial f }{ \partial \sigma} (y, \mu, \sigma)
&= (y - \mu)^2 \sigma^{-3} - \sigma^{-1}. \nonumber
\end{align}

\newcommand{\gmnode}[3]{\put(#1,#2){\circle{20}}\put(#1,#2){\makebox(0,0){$#3$}}}
\newcommand{\gmnodeobs}[3]{\put(#1,#2){\color{yellow}\circle*{20}}\put(#1,#2){\color{black}\circle{20}}\put(#1,#2){\makebox(0,0){$#3$}}}
\newcommand{\gmnoderoot}[3]{\put(#1,#2){\color{red}\circle*{20}}\put(#1,#2){\color{black}\circle{20}}\put(#1,#2){\makebox(0,0){$#3$}}}
\newcommand{\gmdata}[3]{\put(#1,#2){\makebox(16,16){\footnotesize $#3$}}}
\begin{figure}
\begin{center}
\begin{picture}(200,200)
\gmnoderoot{130}{200}{-}
\put(142,200){\mbox{\color{blue}{\footnotesize $v_{10}$}}}
\gmnode{100}{170}{-}
\put(112,170){\mbox{\color{blue}{\footnotesize $v_{9}$}}}
\gmdata{150}{160}{\mbox{\color{gray}{$0.5\log 2 \pi$}}}
\gmnode{70}{140}{*}
\put(82,140){\mbox{\color{blue}{\footnotesize $v_{7}$}}}
\gmdata{30}{100}{\mbox{\color{gray}{$-.5$}}}
\gmnode{100}{110}{\mbox{\footnotesize pow}}
\put(112,110){\mbox{\color{blue}{\footnotesize $v_{6}$}}}
\gmdata{120}{70}{\mbox{\color{gray}{$2$}}}
\gmnode{190}{80}{\mbox{\footnotesize $\log$}}
\put(202,80){\mbox{\color{blue}{\footnotesize $v_{8}$}}}
\gmnode{70}{80}{/}
\put(82,80){\mbox{\color{blue}{\footnotesize $v_{5}$}}}
\gmnode{40}{50}{-}
\put(52,50){\mbox{\color{blue}{\footnotesize $v_{4}$}}}
\gmnodeobs{10}{20}{\mbox{\footnotesize $y$}}
\put(22,20){\mbox{\color{blue}{\footnotesize $v_{1}$}}}
\gmnodeobs{70}{20}{\mbox{\footnotesize $\mu$}}
\put(82,20){\mbox{\color{blue}{\footnotesize $v_{2}$}}}
\gmnodeobs{130}{20}{\mbox{\footnotesize $\sigma$}}
\put(142,20){\mbox{\color{blue}{\footnotesize $v_{3}$}}}
\put(123,193){\vector(-1,-1){16}}
\put(137,193){\color{gray}{\vector(1,-1){16}}}
\put(93,163){\vector(-1,-1){16}}
\put(107,163){\vector(1,-1){76}}
\put(63,133){\color{gray}{\vector(-1,-1){16}}}
\put(77,133){\vector(1,-1){16}}
\put(93,103){{\vector(-1,-1){16}}}
\put(107,103){\color{gray}{\vector(1,-1){16}}}
\put(63,73){\vector(-1,-1){16}}
\put(77,73){\vector(1,-1){46}}
\put(183,73){\vector(-1,-1){46}}
\put(33,43){\vector(-1,-1){16}}
\put(47,43){\vector(1,-1){16}}
\end{picture}
\end{center}
\mycaption{expression-graph}{Expression graph for the normal log
  density function given in \refeq{normal-log-density}.  Each circle
  corresponds to an automatic differentiation variable, with the
  variable name given to the right in blue.  The independent variables
  are highlighted in yellow on the bottom row, with the dependent
  variable highlighted in red on the top of the graph.  The function
  producing each node is displayed inside the circle, with operands
  denoted by arrows.  Constants are shown in gray with gray arrows
  leading to them because derivatives need not be propagated to
  constant operands.}
\end{figure}
The mathematical formula for the normal log density corresponds to the
expression graph in \reffigure{expression-graph}.  Each subexpression
corresponds to a node in the graph, and each edge connects the node
representing a function evaluation to its operands.  Becuase $\sigma$
is used twice in the formula, it has two parents in the graph.
\begin{figure}
\begin{center}
\begin{tabular}{c||c|cc}
{\it var} & {\it value} & \multicolumn{2}{|c}{\it partials}
\\ \hline \hline
$v_1$ & $y$ 
\\[2pt]
$v_2$ & $\mu$
\\[2pt]
$v_3$ & $\sigma$
\\[2pt]
$v_4$ & $v_1 - v_2$ & $\partial v_4 / \partial v_1 = 1$ 
                   & $\partial v_4 / \partial v_2 = -1$
\\[4pt]
$v_5$ & $v_4 / v_3$ & $\partial v_5 / \partial v_4 = 1/v_3$
                    & $\partial v_5 / \partial v_3 = -v_4 v_3^{-2}$
\\[4pt]
$v_6$ & $\left(v_5\right)^2$
      & \multicolumn{2}{c}{$\partial v_6 / \partial v_5 = 2 v_5$}
\\[4pt]
$v_7$ & $(-0.5) v_6$ & \multicolumn{2}{c}{$\partial v_7 / \partial v_6
                                          = -0.5$}
\\[4pt]
$v_8$ & $\log v_3$ & \multicolumn{2}{c}{$\partial v_8 / \partial v_3 = 1/v_3$}
\\[4pt]
$v_9$ & $v_7 - v_8$ & $\partial v_9 / \partial v_7 = 1$
                    & $\partial v_9 / \partial v_8 = -1$
\\[4pt]
$v_{10}$ & $v_9 - (0.5 \log 2\pi)$ 
         & \multicolumn{2}{c}{$\partial v_{10} / \partial v_9 = 1$}
\end{tabular}
\end{center}
\mycaption{forward-pass}{Example of gradient calculation for the the
  log density function of a normally distributed variable, as defined
  in \refeq{normal-log-density}.  In the forward pass to construct the
  expression graph, a stack is constructed from the first input
  $v_1$ up to the final output $v_{10}$, and the values of the variables
  and partials are computed numerically according to the formulas
  given in the value and partials columns of the table.}
\end{figure}

\reffigure{forward-pass} illustrates the forward pass used by
reverse-mode automatic differentiation to construct the expression
graph for a program.  The expression graph is constructed in the
ordinary evaluation order, with each subexpression being numbered and
placed on a stack.  The stack is initialized here with the dependent
variables, but this is not required.  Each operand to an expression is
evaluated before the expression node is created and placed on the
stack.  As a result, the stack provides a topological sort of the
nodes in the graph (i.e., a sort in which each node representing an
expression occurs above its subexpression nodes in the stack---see
\citep[Section~2.2.3]{knuth:97}).  \reffigure{forward-pass} lists in
the right column for each node, the partial derivative of the function
represented by the node with respect to its operands.  In the Stan
Math Library, most of these partials are evaluated lazily during the
reverse pass based on function's value and its operands' values.

\begin{figure}
\[
\begin{array}{rcl|l}
{\it var} & {\it operation} & {\it adjoint} & {\it result}
\\ \hline \hline
a_{1:9} & = & 0 & a_{1:9} = 0
\\
a_{10} & = & 1 & a_{10} = 1
\\ \hline
a_{9} & {+}{=} & a_{10} \times (1) & a_9 = 1
\\
a_{7} & {+}{=} & a_9 \times (1) & a_7 = 1
\\
a_{8} & {+}{=} & a_9 \times (-1) & a_8 = -1
\\
a_{3} & {+}{=} & a_8 \times (1 / v_3) & a_3 = -1 / v_3
\\
a_{6} & {+}{=} & a_7 \times (-0.5) & a_6 = -0.5
\\
a_{5} & {+}{=} & a_6 \times (2 v_5) & a_5 = -v_5
\\
a_{4} & {+}{=} & a_5 \times (1 / v_3) & a_4 = -v_5 / v_3
\\
a_{3} & {+}{=} & a_5 \times (-v_4 v_3^{-2}) & a_3 = -1 / v_3 + v_5 v_4 v_3^{-2}
\\
a_{1} & {+}{=} & a_4 \times (1) & a_1 = -v_5 / v_3
\\
a_{2} & {+}{=} & a_4 \times (-1) & a_2 = v_5 / v_3
\end{array}
\]
\mycaption{autodiff-stack}{The variables $v_i$ represent values and
  $a_i$ represent corresponding adjoints.  In the reverse pass, the
  stack is traversed from the final output down to the inputs, and as
  each variable is visited, each of its operands is updated with the
  variable's adjoint times the partial with respect to the operand.
  After the reverse pass finishes, $(a_1,a_2,a_3)$ is the gradient of
  the density function evaluated at $(y,\mu,\sigma)$, which matches
  the correct result given in \refeq{gradient-normal-log-density}
  after substitution for $v_4$ and $v_5$.}
\end{figure}
\reffigure{autodiff-stack} shows the processing for reverse mode,
which involves an adjoint value for each node.  The adjoints for all
nodes other than the root are initialized to 0; the root's adjoint is
initialized to 1, because $\partial x / \partial x = 1$.  The backward
sweep walks down the expression stack, and for each node, propagates
derivatives from it down to its operands using the chain rule.
Because the nodes are in topological order, by the time a node is
visited, its adjoint will represent the partial derivative of the root
of the overall expression with respect to the expression represented
by the node.  Each node then propagates its derivatives to its
operands by incrementing its operands' adjoints by the product of the
expression node's adjoint times the partial with respect to the
operand.  Thus when the reverse pass is completed, the adjoints of the
independent variables hold the gradient of the dependent variable
(function value) with respect to the independent variables (inputs).

\subsection{Comparison to Alternative Methods of Computing Gradients}

\subsubsection{Finite Differencing}

Finite differencing a numerical approximation to gradient evaluations. 
Given a positive difference $\epsilon > 0$, an approximate derivative 
can be calculated as
\[
\frac{\partial f}{\partial x_n} (x) 
\approx 
\frac{f(x_1,\ldots,x_n + \epsilon, \ldots, x_N) - f(x_1, \ldots, x_N)}
     {\epsilon}
\]
or a bit more accurately with a centered interval as
\[
\frac{\partial f}{\partial x_n} (x) 
\approx 
\frac{f(x_1,\ldots,x_n + \epsilon/2, \ldots, x_N) 
      - f(x_1,\ldots,x_n - \epsilon/2, \ldots, x_N)}
     {\epsilon}.
\]
Although straightforward and general, calculating gradients using
finite differences is slow and imprecise.  Finite differencing is slow
because it requires $N + 1$ function evaluations ($2N + 1$ in the
centered case) to compute the gradient of an $N$-ary function.  The
numerical issues with finite differencing arise because a small
$\epsilon$ is required for a good approximation of the derivative, but
a large $\epsilon$ is required for floating-point precision.  Small
$\epsilon$ values are problematic for accuracy because because
subtraction of differently scaled numbers loses a degree of precision
equal to their difference in scales; for example, subtracting a number
on the scale of $10^{-6}$ from a number on the scale of $10^0$ loses 6
orders of precision in the $10^{-6}$-scaled term \citep{higham:2002}.
Thus in practice, the arithmetic precision of finite differencing is
usually at best $10^{-7}$ rather than the maximum of roughly
$10^{-14}$ possible with double-precision floating point calculations.
There are more accurate ways to calculate derivatives with multiple
differences on either side of the point being evaluated and
appropriate weighting; see \cite{Fornberg:1988} for a general
construction and error analysis.

\subsubsection{Symbolic Differentiation}

Whereas automatic differentiation simply evaluates a derivative at a
given input, symbolic differentiation, such as Mathematica
\citep{mathematica:2014} or SymPy \citep{sympy:2014}, calculate the
entire derivative function analytically by applying the chain rule
directly to the original function's subexpressions.   

An advantage of symbolic differentation is that algebraic
manipulations may be easily performed on either the input formulas or
output formulas to generate more efficient or numerically stable code.
For example, $\log (1 + x)$ can be rendered using the more stable
\code{log1p} function than literally adding 1 to $x$ (which either
loses precision or underflows if $x$ is close to 0) and taking the
logarithm.  While this can be done with reverse-mode automatic
differentiation, it is more challenging and rarely undertaken (though
an expression template approach similar to that of Adept
\citep{Hogan:2014} could be improved upon to reduce local expressions
statically);  \cite{VandevoordeJosuttis:2002} provide an excellent
overview of expression templates.

Symbolic differentiation is less expressive than reverse-mode
automatic differentiation. Reverse-mode automatic differentiation can
apply to any computer program involving differentiable functions and
operators.  This includes programs with conditionals, loops for
iterative algorithms, or (recursive) function calls.  A second
difficulty for symbolic differentiation is that most symbolic math
libraries (e.g., SymPy) are univariate and do not neatly reduce
multivariate expressions involving matrices.  A third difficulty for
symbolic differentiation is in dealing with repeated subexpressions.
The adjoint method on which reverse-mode automatic differentiation is
based automatically applies an efficient dynamic programming approach
that removes the kind of repeated calculations that would be involved
in an unfolded expression tree.  Yet another difficulty is that
symbolic differentiation requires multiple passes which either require
interpretation of the derivatives (what most implementations of
reverse-mode automatic differentiation do, including Stan's), or a
pass to compile the output of the symbolic differentiation.

The disadvantages of limited expressiveness and requiring a further
compilation stage can be combined into an advantage; symbolic
differentiation can generate very efficient code in cases where the
same expression must be evaluated and differentiated multiple times.

A hybrid method is to use reverse-mode automatic differentiation to
generate code for derivatives. This approach has the generality of
reverse-mode automatic differentiation and can be advantageous in the
circumstance where the same expression is evaluated multiple times.
The difficulty is in parsing and dealing with all of the constructions
in a complex language such as C++; existing code-generating tools such
as Tapenade \citep{Hascoet:2013} have only been developed for
relatively simple languages like C and Fortran.

\section{Calculating Gradients and Jacobians}\label{calc-gradients.section}

Reverse-mode automatic differentiation in the Stan Math Library can be
used to evaluate gradients of functions from $\reals^N$ to $\reals$ or
Jacobians of differentiable functions from $\reals^N$ to $\reals^M$,
returning values for a specified input point $x \in \reals^N$.


\subsection{Direct Calculation of Gradients of Programs}

The following complete C++ program calculates derivatives of the
normal log density function \refeq{normal-log-density} with respect to
its mean ($\mu$) and standard deviation ($\sigma$) parameters for a
constant outcome ($y$).  The first block assigns the constant and
independent variables, the second block computes the dependent
variable and prints the value, and the final block computes and prints
the gradients.
\begin{smallcode}
#include <cmath>
#include <stan/math.hpp>

int main() {
  using std::pow;
  double y = 1.3;
  stan::math::var mu = 0.5, sigma = 1.2;
    
  stan::math::var lp = 0;
  lp -= 0.5 * log(2 * stan::math::pi());
  lp -= log(sigma);
  lp -= 0.5 * pow((y - mu) / sigma, 2);
  std::cout << "f(mu, sigma) = " << lp.val() << std::endl;

  lp.grad();
  std::cout << " d.f / d.mu = " << mu.adj()
            << " d.f / d.sigma = " << sigma.adj() << std::endl;
  return 0;
}
\end{smallcode}
Constants like \code{y} are assigned to type \code{double} variables
and independent variables like \code{mu} and \code{sigma} are assigned
to type \code{stan::math::var}.  The result \code{lp} is assigned type
\code{stan::math::var} and calculated using ordinary C++ operations
involving operators (e.g., \code{*}, \code{/}), compound assignments
(e.g., \code{-=}), and library functions (e.g., \code{pow},
\code{log}, \code{pi}).  The value is available through the method
(member function) \code{val()} as soon as operations have been
applied.  The call to the method \code{grad()} propagates derivatives
from the dependent variable \code{lp} down through the expression
graph to the independent variables.  The derivatives of the dependent
variable with respect to the independent variables may then be
extracted from the independent variables with the method \code{adj()}.

The gradient evaluated at the input can also be extracted as a
standard vector using the method \code{grad()} of \code{stan::math::var}.
Included libraries are not duplicated from the previous code.
\begin{smallcode}
#include <vector>

int main() {
  double y = 1.3;
  stan::math::var mu = 0.5, sigma = 1.2;
  
  stan::math::var lp = 0;
  lp -= 0.5 * log(2 * stan::math::pi());
  lp -= log(sigma);
  lp -= 0.5 * pow((y - mu) / sigma, 2);

  std::vector<stan::math::var> theta;
  theta.push_back(mu);   theta.push_back(sigma);
  std::vector<double> g;
  lp.grad(theta, g);
  std::cout << " d.f / d.mu = " << g[0]
            << " d.f / d.sigma = " << g[1] << std::endl;
  return 0;
}
\end{smallcode}
The standard vector \code{theta} holds the dependent variables and the
standard vector \code{g} is used to hold the result.  The function
\code{grad()} function is called on the dependent variable to
propagate the derivatives and fill \code{g} with the gradient.  The
method \code{grad()} will resize the vector \code{g} if it is not the
right size.

\subsection{Coding Template Functions for Automatic Differentiation}

The previous example was implemented directly in the main function
block using primitive operations.  The Stan Math Library
implements all of the built-in C++ boolean and arithmetic operators
as well as the assignment and compound assignment operators.  
It also implements all of the library functions.  Examples of
operator, assignment, and function implementations are given later.

From a user perspective, a function can be automatically
differentiated with respect to some input parameters if the arguments
to be differentiated can all be instantiated to \code{stan::math::var}
types.  For the most flexibility, functions should be separately
templated in all of their arguments so as to support any combination
of primitive (e.g., \code{double}, \code{int}) and autodiff
(\code{stan::math::var}) instantiations.

For example, the following templated C++ function computes the log
normal density as defined in \refeq{normal-log-density}.
\begin{smallcode}
#include <boost/math/tools/promotion.hpp>

template <typename T1, typename T2, typename T3>
inline 
typename boost::math::tools::promote_args<T1, T2, T3>::type
normal_log(const T1& y, const T2& mu, const T3& sigma) {
  using std::pow;  using std::log;  
  return -0.5 * pow((y - mu) / sigma, 2.0)
    - log(sigma)
    - 0.5 * log(2 * stan::math::pi());
}
\end{smallcode}

\subsubsection{Argument-Dependent Lookup for Function Resolution}

In order to allow built-in functions such as \code{log()} and
\code{pow()} to be instantiated with arguments of type
\code{stan::math::var} and primitive C++ types, the primitive version
of the function is brought in with a \code{using} statement.  For
example, \code{using~std::pow} brings the version of \code{pow()} that
applies to \code{double} arguments into scope.  The definition of
\code{pow()} for autodiff variables \code{stan::math} is brought in
through argument-dependent lookup
\cite[Section~3.4]{cpp-standard:2003}, which brings the namespace of
any argument variables into scope for the purposes of resolving
function applications.

\subsubsection{Traits Metaprogram for Computing Return Types}

Boost's traits-based metaprogram \code{promote\_args}
\citep{Boost:2011} is designed to calculate return types for highly
templated functions like \code{normal\_log()};
\cite{VandevoordeJosuttis:2002} provide an excellent overview of
trait-based template metaprogramming.

In general, for an automatically differentiated function, the return
type should be \code{double} if all the input arguments are primitive
integers or double-precision floating point values, and
\code{stan::math::var} if any of the arguments is of type
\code{stan::math::var}.

Given a sequence of types $\code{T1},~...,~\code{TN}$, the template
structure \code{promote\_args<T1,~...,~TN> } defines a typedef named
\code{type}, which is defined to be \code{double} if all the inputs
are \code{int} or \code{double}, and \code{stan::math::var} if any of the
input types is \code{stan::math::var}.  The keyword \code{typename} is
required to let the C++ parser know that the member variable is a
typedef rather than a regular value.  The Boost promotion mechanism is
used throughout the Stan Math Library to define both return types and
types of variables for intermediate results.  For instance,
\code{y~-~mu} would be assigned to an intermediate variable of type
\code{promote\_args<T1,~T2>::type}.

The fully templated definition allows the template parameters
\code{T1}, \code{T2}, or \code{T3} to be instantiated as independent
variables\, \mbox{\rm (\code{stan::math::var})} or constants\, \mbox{\rm
  (\code{double}, \code{int})}.  For example, it can be used in place
of the direct definitions in the previous programs as
\begin{smallcode}
...
double y = 1.3;
stan::math::var mu = 0.5, sigma = 1.2;

stan::math::var lp = normal_log(y, mu, sigma);
...
\end{smallcode}
No explicit template specifications are necessary on the
function---C++ infers the template type parameters from the types of
the arguments.

\subsection{Calculating the Derivatives of Functors with Functionals}\label{stan-gradient-functional.section}

The Stan Math Library provides a fully abstracted approach to
automatic differentiation that uses a C++ functor to represent a
function to be differentiated and a functional for the gradient
operator.

\subsubsection{Eigen C++ Library for Matrices and Linear Algebra} 

The Eigen C++ library for matrix operations and linear algebra
\citep{Eigen:2013} is used throughout the Stan Math Library. The type
\code{Matrix<T,~R,~C>} is the Eigen type for matrices containing
elements of type \code{T}, with row type \code{R} and column type
\code{C}; the automatic differentiation type \code{stan::math::var}
can be used for the element type \code{T}. The Stan Math
Library uses three possible instantiations for the row and column
type, \code{Matrix<T,~Dynamic,~1>} for (column) vectors,
\code{Matrix<T,~1,~Dynamic>} for row vectors, and
\code{Matrix<T,~Dynamic,~Dynamic>} for matrices. These three instances
are all specialized with their own operators in the Eigen library.
Like the standard template library's \code{std::vector} class, these
all allocate memory for elements dynamically on the C++ heap following
the resource allocation is instantiation (RAII) pattern
\citep[p.~389]{stroustrup:94}.  The RAII pattern is used to
encapsulate memory allocation and freeing by allocating memory when an
object is instantiated and freeing it when an object goes out of
scope.

\subsubsection{Definining Functors in C++}

A functor in C++ is a function that defines \code{operator()} and can
hence behaves syntactically like a function.  For example, the normal
log likelihood function for a sequence of observations may be defined
directly as a functor as follows.
\begin{smallcode}
#include <Eigen/Dense>

using Eigen::Matrix;  
using Eigen::Dynamic;

struct normal_ll {
  const Matrix<double, Dynamic, 1> y_;

  normal_ll(const Matrix<double, Dynamic, 1>& y) : y_(y) { }

  template <typename T>
  T operator()(const Matrix<T, Dynamic, 1>& theta) const {
    T mu = theta[0];   
    T sigma = theta[1];
    T lp = 0;
    for (int n = 0; n < y_.size(); ++n)
      lp += normal_log(y_[n], mu, sigma);
    return lp;
  }
};
\end{smallcode}
The variable \code{y\_} is used to store the data vector in the
structure.  The \code{operator()} defines a function whose argument
argument type is the Eigen type for vectors with elements of type
\code{T}; the result is also defined to be of type \code{T}.

\subsubsection{Calculating Gradients with Functionals}

A functional in C++ is a function that applies to a functor.  The Stan
Math Library provides a functional \code{stan::math::gradient()} that 
differentiates functors implementing a constant operator over Eigen
vectors with automatic differentiation scalars, i.e.,
\begin{smallcode}
var operator()(const Matrix<var, Dynamic, 1>& x) const;
\end{smallcode}

The example functor \code{normal\_ll} defined in the prvious section
provides a templated operator, the template parameter \code{T} of
which can be instantiated to \code{stan::math::var} for
differentiation and to \code{double} for testing. The following code
calculates the gradient of the functor at a specified input.
\begin{smallcode}
Matrix<double, Dynamic, 1> y(3);
y << 1.3, 2.7, -1.9;
normal_ll f(y);

Matrix<double, Dynamic, 1> x(2);
x << 1.3, 2.9;

double fx;
Matrix<double, Dynamic, 1> grad_fx;
stan::math::gradient(f, x, fx, grad_fx);
\end{smallcode}
The argument \code{f} is the functor, which is the log likelihood
function instantiated with a vector of data, \code{x}. The argument
vector \code{x} is filled with the parameter values. The scalar
\code{fx} is defined to hold the function value and an Eigen vector
\code{grad\_fx} is defined to hold the gradient.  The functional
\code{stan::math::gradient} is then called to calculate the value and
the gradient from the function \code{f} and input \code{x}, setting
\code{fx} to the calculated function value and setting \code{grad\_fx}
as the gradient. See \refsection{functionals} for information on how
the \code{stan::math::gradient} functional is implemented.

\subsection{Calculating Jacobians}

With reverse-mode automatic differentiation, a single forward pass can
be used to construct the complete expression graph, then a reverse
pass can be carried out for each output dimension.

\subsubsection{Direct Jacobian Calculation}

Suppose that \code{f} is a functor that accepts an dimensional Eigen
$N$-vector of autodiff variables as input (typically through
templating) and produces an Eigen $M$-vector as output.
The following code calculates the Jacobian of \code{f} evaluated 
at the input \code{x}.
\begin{smallcode}
Matrix<double, Dynamic, 1> x = ...;   // inputs
  
Matrix<var, Dynamic, 1> x_var(x.size());
for (int i = 0; i < x.size(); ++i) x_var(i) = x(i);

Matrix<var, Dynamic, 1> f_x_var = f(x_var);

Matrix<double, Dynamic, 1> f_x(f_x_var.size());
for (int i = 0; i < f_x.size(); ++i) f_x(i) = f_x_var(i).val();

Matrix<double, Dynamic, Dynamic> J(f_x_var.size(), x_var.size());
for (int i = 0; i < f_x_var.size(); ++i) {
  if (i > 0) stan::math::set_zero_all_adjoints();
  f_x_var(i).grad();
  for (int j = 0; j < x_var.size(); ++j)
    J(i,j) = x_var(j).adj();
}
\end{smallcode}
First, the arguments are used to create autodiff variables.  Next, the
function is applied and the result is converted back to a vector of
doubles.  Then for each output dimension, automatic differentiation
calculates the gradient of that dimension and uses it to populate a
row of the Jacobian.  After the first gradient is calculated, all
subsequent gradient calculations begin by setting the adjoints to
zero; this is not required for the first gradient because the adjoints
are initialized to zero.

\subsubsection{Functional Jacobian Calculation}

Alternatively, the Jacobian functional can be applied directly to the
function \code{f}, as follows.
\begin{smallcode}
...
Matrix<double, Dynamic, Dynamic> J;
Matrix<double, Dynamic, 1> f_x;
stan::math::jacobian(f, x, f_x, J);
\end{smallcode}
Like the \code{grad()} functional, the \code{jacobian()} functional
will automatically resize \code{J} and \code{f\_x} if necessary.

\section{Automatic Differentiation Variable Base Classes}\label{autodiff-base-classes.section}

As demonstrated in the previous section, the client-facing data type
of the Stan Math Library is the type \code{stan::math::var}, which is
used in place of primitive \code{double} values in functions that are
to be automatically differentiated.  

\subsection{Pointers to Implementations}

The \code{var} class is implemented following the pointer to
implementation (Pimpl) pattern \citep{sutter:98,sutter:01}.  The
implementation class is type \code{vari}, and is covered in
the next subsection.  Like the factory pattern, the Pimpl pattern
encapsulates the details of the implementation class(es), which the
application programmer interface (API) need not expose to clients.
For example, none of the example programs in the last section involved
the \code{vari} type explicitly.  As described in the next section,
Pimpl classes often encapsulate the messy and error-prone pointer
manipulations required for allocating and freeing memory.

\subsection{Resource Allocation is Initialization}

Like many Pimpl classes, \code{var} manages memory using the resource
allocation is initialization (RAII) pattern \citep{stroustrup:94}.
With RAII, class instances are managed on the stack and passed to
functions via constant references.  Memory management is handled
behind the scenes of the client-facing API.  Memory is allocated when
an object is constructed or resized.  Memory is freed when the object
goes out of scope.  

Other classes implemented using Pimpl and RAII include
\code{std::vector} and \code{Eigen::Matrix}, both of which allocate
memory directly on the C++ heap and deallocate it when their variables
go out of scope and their destructors are called.  As explained below,
memory is managed in a custom arena for \code{stan::math::var} instances
rather than being allocated on the general-purpose C++ heap.

\subsection{The \code{var} Class}

The core of the \code{stan::math::var} class is a pointer to a
\code{stan::math::vari} implementation.
\begin{smallcode}
class var {
public:
  var() : vi_(static_cast<vari*>(0U)) { }
  var(double v) : vi_(new vari(v)) { }

  double val() const { return vi_->val_; }
  double adj() const { return vi_->adj_; }  

private:
  vari* vi_;
};
\end{smallcode}
The default constructor \code{var()} uses a null value for its
\code{vari} to avoid memory-allocation that would just be overwritten.
When the constructor is given a \code{double} value, a new \code{vari}
instance is constructed for the value.  The memory management for
\code{vari} is handled through a specialization of
\code{operator}~\code{new} as described below.

The \code{var} class additionally defines a copy constructor,
constructors for the other primitive types, and the full set of
assignment and compound assignment operators.  The destructor is
implicit because there is nothing to do to clean up instances; all of
the work is done with the memory management for \code{vari}. The class
is defined in the \code{stan::math} namespace, but the namespace
declarations are not shown to save
space.

\subsection{The \code{chainable} Base Class}

The design of Stan's variable implementation classes is
object-oriented, in that subclasses extend \code{chainable} and
implement the virtual method \code{chain()} in a function-specific way
to propagate derivatives with the chain rule.  The class \code{vari}
extends the base class \code{chainable} and holds a value and an
adjoint.  Extensisons of \code{vari} will hold pointers to operands
for propagating derivatives and sometimes store partial derivatives
directly.

\begin{smallcode}
struct chainable {
  chainable() { }
  virtual ~chainable() { }

  virtual void chain() { }
  virtual void init_dependent() { }
  virtual void set_zero_adjoint() { }

  static inline void* operator new(size_t nbytes) {
    return ChainableStack::memalloc_.alloc(nbytes);
  }
};
\end{smallcode}
The virtual method \code{chain()} has no body, but allows extensions
of \code{chainable} to implement derivative propagation.  The static
specialization of \code{operator}~\code{new} provides arena-based
memory management for extensions of \code{chainable} (see
\refsection{memory}).  The initialization of dependents and set-zero
methods are also virtual; they are used to initialize the adjoints
during the derivative propagation (reverse) pass.

\subsection{The \code{vari} Class}

The \code{stan::math::vari} class extends the base class
\code{stan::math::chainable} and holds a value and adjoint for a variable
instantiation.
\begin{smallcode}
class vari : public chainable {
public:
  const double val_;
  double adj_;

  vari(double v) : val_(v), adj_(0) { 
    ChainableStack::var_stack_.push_back(this);
  }

  virtual ~vari() { }

  virtual void init_dependent() { adj_ = 1; }
  virtual void set_zero_adjoint() { adj_ = 0; }
};
\end{smallcode}
The initialization method for dependent (output) variables sets the
adjoint to 1; it is called on the output variable whose derivative is
being calculated before propagating derivatives down to input
(independent) variables.  The set-zero method sets the adjoint to 0,
which is the value required for all other variables in the expression
graph before derivative propagation; because adjoints are initialized
to zero, it is only called in situations where multiple derivatives
are required, as in Jacobian calculations.

\section{Calculating Gradients}

There is a static method in the \code{stan::math::chainable} class which
performs the derivative propagation.
\begin{smallcode}
static void grad(chainable* vi) {
  typedef std::vector<chainable*>::reverse_iterator it_t;
  vi->init_dependent(); 
  it_t begin = ChainableStack::var_stack_.rbegin();
  it_t end = ChainableStack::var_stack_.rend();
  for (it_t it = begin; it < end; ++it)
    (*it)->chain();
}
\end{smallcode}
The function initializes the dependent variable to 1 using the
\code{init\_dependent} method on \code{chainable}.  Then a reverse
iterator is used to iterate over the variable stack from the top
(\code{rbegin}) down to the bottom (\code{rend}), executing each
\code{chainable} instance's \code{chain()} method.  Because the stack
of \code{chainable} instances is sorted in topological order, each
node's parents (superexpressions in which it directly appears) will
have all been visited by the time the node's chain-rule propagation
method is called.

\subsection{Jacobians and Zeroing Adjoints}\label{jacobians.section}

To calculate Jacobians efficiently, the expression graph is
constructed once and then reused to calculate the gradient required
for each row of the Jacobian.  Before each gradient calculation after
the first, the static method \code{stan::math::set\_zero\_all\_adjoints()}
is called.  This function walks over the variable stack
\code{stan::math::ChainableStack::var\_stack\_}, resetting each value to zero
using the virtual method \code{set\_zero\_adjoint()}.
\begin{smallcode}
static void set_zero_all_adjoints() {
  for (size_t i = 0; i < ChainableStack::var_stack_.size(); ++i)
    ChainableStack::var_stack_[i]->set_zero_adjoint();
}
\end{smallcode}
As in all of the client-facing API methods, pointer and global
variable manipulation in encapsulated.


\section{Memory Management} \label{memory.section}

Instances of \code{vari} and its extensions are allocated using the
specialized static \code{operator}~\code{new} definition in
\code{chainable}.  The specialization references the
\code{alloc(size\_t)} method of the global variable
\code{ChainableStack::memalloc\_}.  This global variable holds the
custom arena-based memory used for \code{vari} instances.  Using an
arena allows the memory for a gradient calculation to be efficiently
allocated elemtentwise and then freed all at once after derivative
calculations have been performed (see, e.g., \cite{GayAiken:2001,Gay:2005}).

The core of the \code{stan::math::ChainableStack} template class
definition, which holds the memory arena for reverse-mode autodiff, is
as follows.
\begin{smallcode}
template<typename T, typename AllocT>
struct AutodiffStackStorage {
  static std::vector<T*> var_stack_;
  static std::vector<AllocT*> var_alloc_stack_;
  static stack_alloc memalloc_;
};

struct chainable_alloc {
  chainable_alloc() {
    ChainableStack::var_alloc_stack_.push_back(this);
  }
  virtual ~chainable_alloc() { };
};

typedef AutodiffStackStorage<chainable,chainable_alloc>
  ChainableStack;
\end{smallcode}
The variables are all declared static, so they serve as global
variables in the program.  The variable \code{var\_stack\_} holds the
stack of nodes in the expression graph, each represented as a pointer
to a \code{chainable}.  This stack is traversed in reverse order of
construction to propagate derivatives from expressions to their
operands.  The variable \code{memalloc\_} holds the byte-level memory
allocator, which is described in the next section.

The typedef \code{ChainableStack} specifies the template parameters
for the template class \code{AutodiffStackStorage}.   Rather than
creating instances,  the global variables are accessed through the
typedef instantiation (e.g., \code{ChainableStack::var\_stack\_}).

The \code{var\_alloc\_stack} variable holds objects allocated for
autodiff that need to have their \code{delete()} methods called.  Such
variables are required to specialize \code{chainable\_alloc}, which
pushes them onto the \code{var\_alloc\_stack\_} during construction.
The destructor is virtual for \code{chainable\_alloc} because
extensions with customized destructors will be placed on the
\code{var\_alloc\_stack\_}.

\subsection{Byte-Level Memory Manager:  \code{stack\_alloc}}

The byte-level memory manager uses an arena-based strategy where
blocks of memory are allocated and used to hold the \code{vari}
instances created during the forward expression-graph building pass of
automatic differentiation.  After the partials are propagated and final
gradients calculated in the reverse pass, the memory for the
expression graph is reclaimed all at once.

The memory used for \code{chainable} instances is managed as a
standard vector of \code{char*} blocks.  New variables are filled into
the blocks until they will no longer fit, at which point a new block
is allocated.  The data structures for memory management are defined
as member variables in the \code{stan::math::stack\_alloc} class.
\begin{smallcode}
class stack_alloc {
private: 
  std::vector<char*> blocks_;
  std::vector<size_t> sizes_;
  size_t cur_block_;
  char* cur_block_end_;
  char* next_loc_;
  ...
\end{smallcode}
The variable \code{blocks\_} holds blocks of memory allocated on the
heap.  The parallel vector \code{sizes\_} stores the size of each
block.  The remaining variables indicate the index of the current
block, the end of the current block's available memory, and the next
location within the current block to place a new variable
implementation.

Given a request for a given block of memory, if it fits, it will be
allocated starting at the next location in the current block.
Otherwise, a new block is allocated that is twice the size of the
current block and the request is attempted again.  

Because memory allocation is such a low-level operation in the
algorithm and because CPU branch misprediction is so expensive, the
ability of the GNU compiler family to accept guidance in terms of how
to sequence generated assembly code is used through the following
macros.
\begin{smallcode}
#ifdef __GNUC__
#define likely(x)      __builtin_expect(!!(x), 1)
#define unlikely(x)    __builtin_expect(!!(x), 0)
#else
#define likely(x)     (x)
#define unlikely(x)   (x)
#endif
\end{smallcode}
Then a condition is wrapped in \code{likely(...)} or \code{unlikely(...)} to
give the compiler a hint as to which branch to predict by default.
This improved throughput by a factor of nearly 10\% compared to the
unadorned code.    The allocation method in \code{stack\_alloc} is
defined as
\begin{smallcode}
inline void* alloc(size_t len) {
  char* result = next_loc_;
  next_loc_ += len;
  if (unlikely(next_loc_ >= cur_block_end_))
    result = move_to_next_block(len);
  return (void*)result;
}
\end{smallcode}
Thus running out of memory in a block is marked as unlikely.  

The \code{move\_to\_next\_block} method, not shown here, allocates a
new block of memory and updates the underlying memory stacks.  No
movement of existing data is required and thus no additional memory
overhead is required to store the original and copy.  All of the
memory allocation is also done to ensure 8-byte alignment for maximum
access speed of pointers in 64-bit architectures.

\subsection{Memory Lifecycle}

An instance of \code{vari} is constructed for each subexpression
evaluated in the program that involves at least one \code{var}
argument.  For each \code{vari} constructed, memory is allocated in
the arena and a pointer to the memory allocated is pushed onto the
variable stack.  Both the bytes and the stack of pointers must persist
until the call to \code{grad} is used to compute derivatives, at which
point it may be recovered.

The \code{AutodiffStackStorage} class holds the variable stacks and
an instance of \code{stack\_alloc} for the bytes used to
create \code{chainable} instances.  The \code{AutodiffStackStorage}
class also implements a no-argument \code{recover\_memory()} method
which frees the stack of pointers making up the variable stack, and
resets the byte-level \code{stack\_alloc} instance.  The underlying
blocks of memory allocated within the \code{stack\_alloc} instance are
intentionally \emph{not} freed by the \code{recover\_memory()} method.
Rather, the underlying memory pool is saved and reused when the next
call to automatic differentiation is made.  A method is available to
clients in the \code{stack\_alloc} class to completely free the
underlying memory pools.

The functionals for automatic differentiation all provide
exception-safe memory recovery that ensures whenever a derivative
functional is called memory cannot leak even if an operation within
the functor throws an exception.  The functionals do not themselves
free the underlying memory pool, but clients can do that manually
after functionals are called if desired.

\section{Gradient Functional}\label{functionals.section}

The functional \code{gradient()} encapsulates the gradient
calculations shown in the direct implementation as follows (the
namespace qualifications of functions and classes are omitted for
readability).
\begin{smallcode}
template <typename F> 
void gradient(const F& f,
              const Matrix<double, Dynamic, 1>& x,
              double& fx,
              Matrix<double, Dynamic, 1>& grad_fx) {
  try {
    Matrix<var, Dynamic, 1> x_var(x.size());
    for (int i = 0; i < x.size(); ++i)
      x_var(i) = x(i);
    var fx_var = f(x_var);
    fx = fx_var.val();
    grad(fx_var.vi_);
    grad_fx.resize(x.size());
    for (int i = 0; i < x.size(); ++i)
      grad_fx(i) = x_var(i).adj();
  } catch (const std::exception& /*e*/) {
    recover_memory();
    throw;
  }
  recover_memory();
}
\end{smallcode}
Client code, as our earlier example shows, does not need to specify
the template parameter \code{F} explicitly, because it can be
automatically inferred from the static type of the argument functor
\code{f}.  Because the functor argument reference is declared
\code{const} here, its \code{operator()} implementation must also be
marked \code{const}.  The final two arguments, \code{fx} and
\code{grad\_fx}, will be set to the value of the function and the value
of the gradient respectively.  

All of the code is wrapped in a \code{try} block with a corresponding
\code{catch} that recovers memory and rethrows whatever standard
exception was caught.  If no exception is thrown in the block of the
\code{try}, then the function terminates with a call to recover
memory (which only resets memory---it does not free any of the
underlying blocks that have been allocated, saving them for reuse).

The main body of the algorithm first copies the input vector of
\code{double} values, \code{x}, into a vector of \code{var} values,
\code{x\_var}.  Then the functor \code{f} is applied to this newly
created vector \code{x\_var}, resulting in a type \code{var} result.
This usage requires \code{f} to implement a constant \code{operator()}
which takes a vector of \code{var} as its single argument and returns
a \code{var}.  Any other information required to compute the function,
such as data vectors (\code{double} or \code{int} values) or external
callbacks such as a logging or error reporting mechanism, must be
accessible to the \code{operator()} definition for the functor
\code{f}.

The evaluation of \code{f} may throw an exception, which will result
in the automatic differentiation memory being recovered.  If no
exception is thrown, the function returns a \code{var}, the value of
which is then assigned to the argument reference \code{fx}.

Next, the static \code{grad} function is called directly on the
variable implementation of the result.  The derivatives are then
extracted from the dependent variables in the argument vector and
assigned to the argument reference vector \code{grad\_fx}, first
resizing \code{grad\_fx} if necessary.

\section{Constants}

There are two equivalent ways to create constant autodiff variables
from constant \code{double} values.  The first approach uses the unary
constructor for \code{var} instances, which creates a new \code{var}
instance \code{sigma} on the C++ function-call stack.
\begin{smallcode}
var sigma(0.0);
\end{smallcode}
As shown in \refsection{autodiff-base-classes}, an underlying
\code{vari} instance will be constructed in the memory arena.

The second approach to constructing a constant autodiff variable uses
assignment.  Because the unary constructor of \code{var} is implicit
(i.e., not declared \code{explicit}), assignment of a \code{double} to
a \code{var} will construct a new \code{var} instance.  
\begin{smallcode}
var tau = 0.0;
\end{smallcode}
The behavior of \code{sigma} and \code{tau} are the same because they
have exactly the same effect on memory and the autodiff stack.

It is also possible to construct instances using integers or other
base types such as \code{unsigned long} or \code{float}.  The
additional constructors are not shown.  Because the copy constructor
is declared explicit, there is no ambiguity in the constructors.

A \code{vari} inherits the no-op \code{chain()} method implementation
from its superclass \code{chainable}.  Therefore, a constant will not
propagate any derivative information.  Where possible, it is best to
use \code{double} values directly as arguments to functions and
operators, because it will cut down on the size of the variable stack
and thus speed up gradient calculations.

\section{Functions and Operators}

This section describes how functions are implemented in Stan for
reverse-mode automatic differentiation.  

\subsection{Unary Operand Storage Class}

Most of the unary functions in the Stan Math Library are implemented
as extensions of a simple helper class, \code{op\_v\_vari}, which
stores the value of a unary function and its operand.
\begin{smallcode}
struct op_v_vari : public vari {
  vari* avi_;

  op_v_vari(double f, vari* avi) : vari(f), avi_(avi) { }
};
\end{smallcode}
Constructing a \code{op\_v\_vari} instance passes the function value
\code{f} to the superclass constructor, which stores it in member
variable \code{val\_}.  The second argument to the constructor is a
\code{vari} pointer, which is stored here in the member variable
\code{avi\_}.  The \code{vari} will be set to The \code{op\_v\_vari}
inherits the no-op \code{chain()} method from \code{chainable}.

\subsection{Unary Function Implementation}

As a running example, consider the natural logarithm function, the
derivative of which is
\[
\log'(x) = \frac{1}{x}.
\]
The implementation of the \code{log} function for \code{var}
constructs a specialized \code{vari} instance as follows.
\begin{smallcode}
inline var log(const var& a) {
  return var(new log_vari(a.vi_));
}
\end{smallcode}
The argument \code{a.vi\_} supplied to the constructor is a pointer to
the \code{vari} implementing \code{a}; this pointer is stored as the
operand for the \code{log} function.

The class \code{log\_vari} is defined as follows.
\begin{smallcode}
struct log_vari : public op_v_vari {
  log_vari(vari* avi) :
    op_v_vari(std::log(avi->val_), avi) { }

  void chain() {
    avi_->adj_ += adj_ / avi_->val_;
  }
};
\end{smallcode}
The constructor takes a pointer, \code{avi}, to the operand's
implementation.  The value of the operand is given by
\code{avi->val\_}.  The log of the operand's value is computed using
\code{std::log()} and passed to the \code{op\_v\_vari} constructor,
which stores it, along with the operand implementation itself,
\code{avi}.  

The virtual \code{chain()} method of the superclass \code{chainable}
is overriden to divide the variable's adjoint, \code{adj\_}, by the
operand's value, \code{avi\_->val\_}, and use the result to increment
the operand's adjoint, \code{avi\_->adj\_}.  For comparison, the
operation shown in \reffigure{autodiff-stack} for a log is $a_3 \
{+}{=} \ a_8 \times (1 / v_3)$, in which $a_3$ and $v_3$ are the adjoint and value
of the operand and $a_8$ the adjoint of the result.

The \code{chain()} method is not called until the reverse pass of
automatic differentiation is executed, so the derivative $1 / x$ times
the adjoint is evaluated lazily as \Verb|adj_ / avi_->val_|, which
requires only a single division.  The eager approach (as used by other
packages such as Sacado and Adept) store $1/x$ as the partial in the
forward pass and then multiply that stored value by the adjoint in the
reverse pass.  The eager approach requires additional storage for the
partial $1/x$, which must be set and later accessed.  The eager
approach also carries out an extra operation in multiplying by the
inverse rather than just dividing as in the lazy approach, effectively
evaluating \Verb|adj_ * (1 / avi_->val_)| rather than 
\Verb|adj_ / avi_->val_|.

Most other Stan functions are implemented in the same way as
\code{log}, with their own specialized \code{chain()} implementation
and storage.  The savings for operations like addition and subtraction
are even larger because there is no need to do any multiplication at
all.

\subsection{Binary Functions}

As an example, consider the \code{pow()} function, defined by
\[
\mbox{pow}(x,y) = x^y,
\]
with derivatives
\[
\frac{\partial}{\partial x} x^y = y \, x^{y-1}
\]
and
\[
\frac{\partial}{\partial y} x^y = x^y \log x.
\]

\subsubsection{Binary Operands Storage Class}

Binary functions of two autodiff variables are implemented using the
following specialized \code{vari} implementation.
\begin{smallcode}
struct op_vv_vari : public vari {
  vari* avi_;   vari* bvi_;

  op_vv_vari(double f, vari* avi, vari* bvi)
  : vari(f), avi_(avi), bvi_(bvi) { }
};
\end{smallcode}
Like its unary counterpart, this class stores the value (\code{f}) and
pointers to the operands (\code{avi\_}, \code{bvi\_}).  The number of
operands need not be stored explicitly because it is implicit in the
identity of the class. 

For mixed operations of \code{var} and \code{double} variables, the
following base \code{vari} is provided.
\begin{smallcode}
struct op_vd_vari : public vari {
  vari* avi_;
  double bd_;

  op_vd_vari(double f, vari* avi, double b)
  : vari(f), avi_(avi), bd_(b) {  }
};
\end{smallcode}
Although not strictly necessary given the above, the following
symmetric class is provided for naming convenience.
\begin{smallcode}
struct op_dv_vari : public vari {
  double ad_;
  vari* bvi_;

  op_dv_vari(double f, double a, vari* bvi)
  : vari(f), ad_(a), bvi_(bvi) { }
};
\end{smallcode}

\subsubsection{Binary Function Implementation}

There are three function implementations for \code{pow()} based on the
type of the arguments, and each has its own specialized
implementation, the first of of which follows the same pattern as
\code{log}, only with two arguments.
\begin{smallcode}
inline var pow(const var& base, const var& exponent) {
  return var(new pow_vv_vari(base.vi_,exponent.vi_));
}
\end{smallcode}
The second specialization is similar, only the \code{double} argument
is passed by value to the specialized \code{vari} constructor.
\begin{smallcode}
inline var pow(double base, const var& exponent) {
  return var(new pow_dv_vari(base,exponent.vi_));
}
\end{smallcode}
The last implementation with a \code{double} exponent first checks if
there is a built-in special case that can be used, and if not,
constructs a specialized \code{vari} for power.
\begin{smallcode}
inline var pow(const var& base, double exponent) {
  if (exponent == 0.5) return sqrt(base);
  if (exponent == 1.0) return base;
  if (exponent == 2.0) return square(base);
  return var(new pow_vd_vari(base.vi_,exponent));
}
\end{smallcode}
By using \code{sqrt}, \code{square}, or a no-op, less memory is
allocated and fewer arithmetic operations are needed during the
evaluation of \code{chain()} for the remaining expressions, thus
saving both time and space.

\subsubsection{Variable Implementations}

The \code{pow\_vv\_vari} class for \code{pow(var,var)} is defined as follows.
\begin{smallcode}
struct pow_vv_vari : public op_vv_vari {
  pow_vv_vari(vari* avi, vari* bvi)
  : op_vv_vari(std::pow(avi->val_,bvi->val_),avi,bvi) { }

  void chain() {
    if (avi_->val_ == 0.0) return; 
    avi_->adj_ += adj_ * bvi_->val_ * val_ / avi_->val_;
    bvi_->adj_ += adj_ * std::log(avi_->val_) * val_;
  }
};
\end{smallcode}
The constructor calls the \code{op\_vv\_vari} constructor to store the
value and the two operand pointers.  The \code{chain()} implementation
first tests if the base is 0, and returns because the derivatives are
both zero, and thus there is nothing to do for the chain rule.  This
test is not for efficiency, but rather because it avoids evaluating
the logarithm of zero, which evaluates to special not-a-number value
in IEEE floating-point arithmetic.%
\footnote{An even more efficient alternative
would be to evaluate this condition ahead of time and just return a
constant zero \code{var} at the top level.}
In the normal case of execution when the base is not zero, the
\code{chain()} method increments the operands' adjoints based on the
derivatives given above.  The expression for the derivative for the
power operand (\code{bvi\_}) conveniently involves the value of the
function itself, just as \code{exp} did.

The implementations for mixed inputs only need to compute a single
derivative propagation.
\begin{smallcode}
struct pow_vd_vari : public op_vd_vari {
  pow_vd_vari(vari* avi, double b) :
    op_vd_vari(std::pow(avi->val_,b),avi,b) {
  }

  void chain() {
    avi_->adj_ += adj_ * bd_ * val_ / avi_->val_;
  }
};
\end{smallcode}
The base argument (\code{avi}) is a variable implementation pointer
that is stored and accessed in \code{chain()} as \code{avi\_}.  The
exponent argument (\code{b}) is stored and accessed as \code{bd\_}.

\subsection{Arithmetic Operators}

Basic arithmetic operators are implemented in exactly the same way as
functions.  For example, addition of two variables is implemented
using a support class \code{add\_vv\_vari}.
\begin{smallcode}
inline var operator+(const var& a, const var& b) {    
  return var(new add_vv_vari(a.vi_,b.vi_));
}
\end{smallcode}
The implementation of \code{add\_vv\_vari} follows that of \code{exp},
with the same naming conventions.  The derivatives are
\[
\frac{\partial}{\partial x} \left( x + y \right) = 1
\]
and
\[
\frac{\partial}{\partial y} \left( x + y \right) = 1.
\]
As a result, the \code{chain()} method can be reduced to just adding
the expression's adjoints to that of its operands.
\begin{smallcode}
void chain() {
  avi_->adj_ += adj_;
  bvi_->adj_ += adj_;
}
\end{smallcode}
This is particularly efficient because the derivative values of 1 need
not be stored or mutliplied.

The mixed input operators are specialized in the same way as the mixed
input functions.  The other built-in operators for unary subtraction,
multiplication, and division are implemented similarly.

\subsection{Boolean Operators}

The boolean operators are implemented directly without creating new
variable implementation instances.  For example, equality of two
variables is implemented as follows.
\begin{smallcode}
inline bool operator==(const var& a, const var& b) {
  return a.val() == b.val();
}
\end{smallcode}
Mixed input types are handled similarly.
\begin{smallcode}
inline bool operator==(const var& a, double b) {
  return a.val() == b;
}
\end{smallcode}
The other boolean operators are defined the same way.

\subsection{Memory Usage}

The amount of memory required for automatic differentiation is the sum
of the memory required for each expression in the expression graph.
These expressions are represented with data structures like
\code{log\_vari}.  The storage cost in the arena is as follows for a
function with $K$ operands.
\begin{center}
\begin{tabular}{r|cc}
{\it Description} & {\it Type} & {\it Size (bytes)}
\\ \hline
value & \code{double} & 8
\\
operand pointers & \code{size\_t} & $K \times 8$
\\
vtable pointer  & \code{size\_t} & 8
\\
\code{var\_stack\_} ptr & \code{size\_t} & 8
\\ \hline \hline
{\sc total} & & 24 + $K \times 8$
\end{tabular}
\end{center}
Each expression stores its value using double-precision floating
point, requiring 8 bytes.  Pointers are stored to the operand(s),
requiring 8 bytes per pointer.  The number of operands is not stored,
but is rather implicit in the identity of the class used for the
\code{vari} and how it computes \code{chain()}.

Wrapping the \code{vari*} in a \code{var} does not increase memory
usage.  The \code{var} itself is allocated on the stack and because it
has no virtual functions, occupies the same amount of memory as a
\code{vari*} pointer.

In addition to the memory required for the \code{vari} instances, each
expression gets a pointer on the variable stack, \code{var\_stack\_}.
The price for object orientation in the virtual \code{chain()} method
is 8 bytes for the vtable pointer, which most implementations of C++ 
includes in order to dynamically resolve the implementation of virtual 
methods.  The cost for storing a heterogeneous collection of expressions 
on the variable stack is either a virtual function lookup using the vtable 
or pointer chasing to implementations with function pointers.

\section{Assignment and Compound Assignment}

\subsection{Assignment Operator}

The assignment operator, \code{operator=}, is defined implicitly for
\code{var} types.  Default assignment of \code{var} instances to
\code{var} instances is handled through the C++ defaults, which
perform a shallow copy of all member variables, in this case copying
the \code{vari*} pointer.

For assigning \code{double} values to \code{var} instances, the
implicit constructor \code{var(double)} is used to construct a new
variable instance and its \code{vari*} pointer is copied into the
left-hand side \code{var}.  

Assigning \code{var} typed expressions to \code{double} variables is
forbidden.  

\subsection{Compound Assignment Operators}

The compound assignment operators such as \code{+=} are used as
follows.
\begin{smallcode}
var a = 3;
var b = 4;
a += b;
\end{smallcode}
The above is intended to have the same effect as the following.
\begin{smallcode}
a = a + b;
\end{smallcode}
The compound operators are declared as member functions in \code{var}.
\begin{smallcode}
struct var {
...
  inline var& operator*=(const var& b);
  inline var& operator*=(double b);

};
\end{smallcode}
The left-hand side is implicitly the variable in which the operators
are declared.

The implementations are defined outside the class declaration.  For variable
arguments, the definition is as follows; the implementation for
\code{double} arguments is similar.
\begin{smallcode}
inline var& var::operator*=(const var& b) {
  vi_ = new multiply_vv_vari(vi_,b.vi_);
  return *this;
}
\end{smallcode}
A new \code{multiply\_vv\_vari} instance is created and the pointer to
it assigned to the member variable \code{vi\_} of the left-hand side
variable in the compound assignment.  The expression \code{*this} is
the value of the left-hand side variable in the assignment, which will
be returned by reference according to the declared return type
\code{var\&}.

The effect of \code{a~+=~b;} is to modify \code{a}, assigning it to
\code{a~+~b}.  This is a destructive operation on \code{var} instances
such as \code{a}, but it is not a destructive operation on variable
implementations.  A sequence such as 
\begin{smallcode}
var a = 0;
a + = b * b;
a + = c * c;
\end{smallcode}
winds up creating the same expression graph as
\begin{smallcode}
var a = b * b + c * c;
\end{smallcode}
with the same value pointed to by \code{a.vi\_} in the end.

\section{Variadic Functions}\label{variadic-functions.section}

Some functions, such as sum, can take vectors of arbitrary size of
arguments.  For such variadic functions, \code{op\_vector\_vari}
provides a base class with member variables for the array of operands
and the array's size.
\begin{smallcode}
struct op_vector_vari : public vari {
  const size_t size_;
  vari** vis_;

  op_vector_vari(double f, 
             const std::vector<stan::math::var>& vs) 
  : vari(f), size_(vs.size()), 
    vis_(static_cast<vari**>(operator new(sizeof(vari*) 
                                          * vs.size()))) {
    for (size_t i = 0; i < vs.size(); ++i)
      vis_[i] = vs[i].vi_;
  }
};
\end{smallcode}
Because it is used in an extension of \code{vari}, the specialized
memory arena \code{operator}~\code{new} is used to allocate the
memory.  Rather than use a \code{std::vector}, which allocates memory
on the standard C++ heap, using the memory arena avoids fragmentation
and also avoids the overhead of calling destructors.

\subsection{Log Sum of Exponentials}

One example that is useful in statistical models to prevent overflow
and underflow is the log sum of exponentials operation, defined for an
$N$-vector $x$ by
\[
\mbox{log\_sum\_exp}(x) 
= \log \sum_{n=1}^N \exp(x_n).
\]
The function is symmetric in its arguments and the derivative with
respect to $x_n$ is
\[
\frac{\partial}{\partial x_n} \mbox{log\_sum\_exp}(x)
= \frac{\exp(x_n)}{\sum_{n=1}^N \exp(x_n)}.
\]

The log sum of exponential operator has the folloiwng implementation
class. 
\begin{smallcode}
struct log_sum_exp_vector_vari : public op_vector_vari {
  log_sum_exp_vector_vari(const std::vector<var>& x)
  : op_vector_vari(log_sum_exp_as_double(x), x) { }

  void chain() {
    for (size_t i = 0; i < size_; ++i)
      vis_[i]->adj_ += adj_ * std::exp(vis_[i]->val_ - val_);
  }
};
\end{smallcode}
The \code{chain()} method here loops over the operand pointers stored
in the array \code{vis\_}, and for each one, increments its
adjoint with the adjoint of the result times the derivative with
respect to the operand, as given by
\[
\exp(x_n - \mbox{log\_sum\_exp}(x))
\ = \
\frac{\exp(x_n)}
     {\exp(\mbox{log\_sum\_exp}(x))}
\ = \
\frac{\exp(x_n)}
     {\sum_{n=1}^N \exp(x_n)}.
\]
The function to compute the value, \code{log\_sum\_exp\_as\_double} is
not shown here;  it uses the usual algebraic trick to preserve the
most significant part of the result and avoid overflow on the
remaining calculations, by
\[
\log \sum_{n=1}^N \exp(x_n)
= \max(x) + \log \sum_{n=1}^N \exp(x_n - \max(x)).
\]
The data structure only requires storage for the operands and an
additional size pointer.  There is also only a single virtual function
call to \code{chain} that propagates derivatives for each argument.

\subsection{Sum Accumulation}

The Stan Math Library was developed to calculate gradients of log
density functions, which are typically composed of sums of simpler log
probability functions.  A typical structure used in a simple linear
regression would be the following (though see
\refsection{vectorization} for the vectorized implementation of
\code{normal\_log}).
\begin{smallcode}
var lp = 0;
for (int n = 0; n < N; ++n)
  lp += normal_log(y[n], x[n] * beta, sigma);
\end{smallcode}
This results in an expression tree structured as
\[
(( \cdots ((p_1 + p_2) + p_3) + p_4) \cdots + p_{N-1}) + p_N).
\]
This tree has $N-1$ nodes representing addition, each with two
operands, for a total of $N-1$ virtual function calls to
\code{chain()} requiring two addition operations each, for a total of
$2(N-1)$ addition operations.

It is a much more efficient use of resources (time and memory) to
produce a single node with $N$ operands, requiring only a single
virtual function call involving a total of $N$ addition operations.
This can be achieved with the following variable implementation class.
\begin{smallcode}
struct sum_v_vari : public vari {
  vari** v_;
  size_t length_;

  sum_v_vari(const vector<var>& v) 
  : vari(var_sum(v)), 
    v_(ChainableStack::memalloc_.alloc_array<vari*>(v.size())),
    length_(v.size())
  {
    for (size_t i = 0; i < length_; i++)
      v_[i] = v(i).val();
  }

  void chain() {
    for (size_t i = 0; i < N_; i++)
      v_[i]->adj_ += adj_;
  }
};
\end{smallcode}
For simplicitly, only the constructor for a standard vector of
autodiff variables is shown.  This class provides more general
constructors and is used throughout the Stan Math Library.

The \code{var\_sum} function sums the values of the variables in the
vector.  
\begin{smallcode}
inline static double 
var_sum(const vector<var>& v) {
  double result = 0;
  for (size_t i = 0; i < v.size(); ++i)
    result += v[i].val();
  return result;
} 
\end{smallcode}
This sum is then passed to the superclass constructor, which stores
the value (and the adjoint, which it initializes to zero).

The following member template function of \code{stack\_alloc} allows
arrays of any type to be allocated within the memory arena for
automatic differentiation.
\begin{smallcode}
template <typename T>
inline T* alloc_array(size_t n) {
  return static_cast<T*>(alloc(n * sizeof(T)));
}
\end{smallcode}
The \code{alloc()} method does the allocation of the bytes in the
arena, taking care to align them on an 8-byte boundary.  

With the implementation in place, the sum function for vectors is
trivial.
\begin{smallcode}
inline var sum(const vector<var>& x) {
  return var(new sum_v_vari(x));
}
\end{smallcode}
More efficient special cases can be added for inputs of size zero, one
and two, if these are common.
\begin{smallcode}
  if (x.size() == 0) return 0.0;
  if (x.size() == 1) return x[0];
  if (x.size() == 2) return x[0] + x[1];
\end{smallcode}

To deal with running sums in the context of log density definitions, 
the intermediate values are pushed onto a \code{vector<var>} and then
the \code{sum()} function is used to return their value.  The linear
regression density would then be implemented as follows.
\begin{smallcode}
vector<var> acc;
for (int n = 0; n < y.size(); ++n)
  acc.push_back(normal_log(y[n], x[n] * beta, sigma));
var lp = sum(acc);
\end{smallcode}

\section{Matrix Data Structures and Arithmetic}

The Stan Math Library uses the Eigen C++ Library for storing
basic matrix data structures, performing matrix arithmetic, and
carrying out linear algebra operations such as factoring and
calculating determinants.

\subsection{Matrix Data Structures}

Matrix and vector types are containers of univariate autodiff
variables, i.e., instances of \code{var}.  That is, the design is
intrinsically organized around univariate automatic differentiation
variables.  This is not required by the \code{chainable} base class,
but only \code{vari} instances are used.

The Stan Math Library uses the following three Eigen types.
\begin{center}
\begin{tabular}{l|l}
{\it Type} & {\it Description}
\\ \hline
\code{Matrix<T, Dynamic, Dynamic>} & matrix
\\
\code{Matrix<T, Dynamic, 1>} & column vector
\\
\code{Matrix<T, 1, Dynamic>} & row vector
\end{tabular}
\end{center}
The template parameter \code{T} indicates the type of elements in the
matrix or vector.  Here, the template parameter will be instantiated
to \code{double} for constants or \code{stan::math::var} for automatic
differentiation through the matrix or vector.

\subsection{Type Inference}

The arithmetic operations are typed, which allows inference of the
result type.  For example, multiplying two matrices produces a matrix,
multiplying a matrix by a column vector produces a column vector, and
multiplying a row matrix by a matrix produces a row matrix.  As
another example, multiplying a row vector by a column vector produces
a scalar.  Two matrices can be added, as can two column vectors or two
row vectors.  Adding a scalar to a matrix (vector) is defined to add
the scalar to each element of the matrix (vector).

\subsection{Dot Products}

For example, multiplying a row vector by a column vector of the same
size produces a scalar; if they are not the same size, an exception is
raised.  If both arguments are \code{double}, Eigen matrix
multiplication is used to produce the result.  

The naive approach to dot products is to just evaluate the product and
sums.  With two $N$-vectors $x$ and $y$, the resulting expression
introduces $2N - 1$ nodes ($N$ multiplication, $N-1$ addition), one
for each operator in
\[
\left( x_1 \times y_1 \right)
+ \left( x_2 \times y_2 \right) 
+ \cdots +
\left( x_N \times y_N \right).
\]

Matrix arithmetic functions such as multiplication essentially follow
the design of the variadic functions as discussed in
\refsection{variadic-functions}.   A specialized \code{vari} extension
is used to cut the number of nodes allocated from $2N$ to 1 and cut the
number of edges traversed during derivative propagation from $4N$ to $2N$.  

The following function definition applies to multiplying a row vector
by a column vector;  there is also a \code{dot\_product} operation
that applies to row or column vectors in either position.
\begin{smallcode}
var multiply(const Matrix<var, 1, Dynamic>& v1,
             const Matrix<var, Dynamic, 1>& v2) {
  return var(new dot_product_vari(v1, v2));
}
\end{smallcode}
The following is a simplified version of the actual
\code{dot\_product\_vari} class; the library implementation includes
extra methods for other uses and is highly templated for flexibility).
\begin{smallcode}
struct dot_product_vari : public vari {
  vari** v1_;    vari** v2_;
  size_t size_;
  
  dot_product_vari(vari** v1, vari** v2, size_t size) 
  : vari(dot(v1,v2,size)), v1_(v1), v2_(v2), size_(size) { }

  dot_product_vari(const Matrix<var, 1, Dynamic>& v1,
                   const Matrix<var, Dynamic, 1>& v2)
  : vari(dot(v1,v2)),
    v1_(static_cast<vari**>(operator new(sizeof(vari*) * size))),
    v2_(static_cast<vari**>(operator new(sizeof(vari*) * size))),
    size_(v1.size()) {
      for (size_t n = 0; n < size_; ++n) {
        v1_[n] = v1[n].vi_;  
        v2_[n] = v2[n].vi_;
      }
  }

  void chain() {
    for (size_t n = 0; n < size_; ++n) {
      v1_[n]->adj_ += adj_ * v2_[n]->val_;
      v2_[n]->adj_ += adj_ * v1_[n]->val_;
    }
  }
};
\end{smallcode}
The first constructor, for arrays of \code{vari*}, will be used later
for matrix multiplication.  The second constructor, for vectors of
\code{var}, is what is used in the function definition above.  Both
the constructors compute the result using a call to an overloaded
function, neither definition of which is shown.  The arena-based
\code{operator}~\code{new} is used to allocate the member arrays
\code{v1\_} and \code{v2\_}, then the body of the constructor
populates them with the implementation pointers of the input operands.
The chain rule is also straightforward because the partial with
respect to each operand component is the corresponding component of
the other operand vector.

\subsubsection{Mixed Type Operations}

Taking the dot product of a vector with \code{double} entries and one
with \code{var} entries cannot be done directly using the \code{Eigen}
multiplication operator because it only applies to inputs with
identical entry types.  

A very naive approach would be to promote the \code{double} entries to
\code{var} and then multiply two \code{var} matrices.  This introduces
$N$ unnecessary nodes into the expression graphs and involves $N$
unnecessary gradient propagation steps to constants.  A slightly less
naive approach would be to write the loop directly, but that is
problematic for the same reason as using a loop directly for two
vectors of \code{var} entries.  Instead, the Stan Math Library
introduces a custom \code{var} vector times \code{double} vector dot
product function that introduces a single node into the expression
graph and only requires $N$ propagations for the dot product of
$N$-vectors.

\subsection{Matrix Multiplication}

To multiply two matrices, a matrix is created for the result and each
entry is populated with the appropriate product of a row vector and
column vector.  A naive implementation for two matrices both composed
of \code{var} entries could just use Eigen's built-in matrix
multiplication, but this has the same problem as dot products, only
$N^2$-fold.

To avoid introducing unnecessary nodes into the expression graph, an
array of \code{vari*} is allocated in the autodiff memory arena and
populated with the corresponding implementation pointers from the
operand matrix rows or columns.  For the result matrix, an instance of
\code{dot\_product\_vari} is constructed using the corresponding row
and column variable implementation pointer arrays.  The resulting
storage is much more economical than if a separate dot product were
created for each entry; for instance, in multiplying a pair of $N
\times N$ matrices, separate operand storage for each result entry
would require a total of $\mathcal{O}(N^3)$ memory, whereas the
current scheme requires only $\mathcal{O}(N^2)$ memory.

\subsection{Specialized Matrix Multiplication}

The Stan Math Library provides a custom
\code{multiply\_self\_transpose} function because the memory required
can be cut in half and the speed improved compared to applying the
transposition and then the general matrix multiplication function. The
same vectors make up the original matrix's rows and the transposed
version's columns, so the same array of \code{vari**} can be reused.

Reducing memory usage often has the pleasant side effect of increasing
speed.  In the case of multiplying a matrix by its own transpose,
fewer copy operations need to be performed.  In general, memory
locality of algorithms will also be improved because of less need to
bring copies into memory.  

If the matrix is lower triangular, only the non-zero portions of the
matrix need to be stored and traversed and saved in the expression
graph.  Multiplying a lower-triangular matrix by its own transposition
is common in multivariate statistics, where it is used to reconstruct
a positive definite matrix $\Omega$, such as a correlation,
covariance, or precision matrix, from its Cholesky factor, $\Omega =
L\,L^{\top}$.  For efficiency in both time and memory, the Stan Math
Library provides a built-in \code{multiply\_self\_transpose\_lower\_tri} function.

\section{Linear Algebra}

Matrix arithmetic is just repeated standard arithmetic.  Linear
algebra operations such as calculating determinants or matrix
divisions are much more complicated in their internal structure.  In
both cases, specialized derivatives can be much more efficient than
just automatically differentiating the library functions.

\cite{Giles:2008} provides a rigorous definition of matrix automatic
differentiation and a translation of some of the key matrix derivative
results \citep{PetersenPedersen:2008,MagnusNeudecker:2007} to
automatic differentiation terms.

\subsection{Log Determinants}

Many probability formulas require the logarithm of the absolute value
of the determinant of a matrix; it shows up in change of variables
problems and as the normalizing term of the multivariate normal
density.  If $x$ is an $N \times N$ matrix, then the partials of the
log absolute determinant are given by
\[
\frac{\partial}{\partial x} \log | \mbox{det}(x) |
= \left( x^{-1} \right)^{\top}.
\]
On an element by element basis, this reduces to
\[
\frac{\partial}{\partial x_{m,n}} \log | \mbox{det}(x) |
= \left( \left( x^{-1} \right)^{\top}\right)_{m,n}.
\]

\subsubsection{Precomputed Gradients Implementation}

In some cases, such as log determinants, it is easier to compute
partial eagerly at the same time as the function value.  The following
utility variable implementation will be used to store vector and
matrix derivatives with precomputed gradients.
\begin{smallcode}
struct precomputed_gradients_vari : public vari {
  const size_t size_;
  vari** varis_;
  double* gradients_;

  precomputed_gradients_vari(double val, size_t size, 
                             vari** varis, double* gradients)
        : vari(val), size_(size),
          varis_(varis), gradients_(gradients) { }

  void chain() {
    for (size_t i = 0; i < size_; ++i) 
      varis_[i]->adj_ += adj_ * gradients_[i];
  }
};
\end{smallcode}
In addition to the value and adjoint stored by the parent class
(\code{vari}), this class adds a parallel array of operands and
gradients along with their size.  The \code{chain()} method adds the
adjoint times the stored partial to the operands' adjoint.

\subsubsection{Log Determinant Implementation}

The implementation of derivatives for the log determinant involves
performing a Householder QR decomposition of the double values of the
matrix, at which point the log absolute determinant and inverse can
easily be extracted.

\begin{smallcode}
template <int R, int C>
inline var 
log_determinant(const Eigen::Matrix<var,R,C>& m) {
  Matrix<double,R,C> m_d(m.rows(),m.cols());
  for (int i = 0; i < m.size(); ++i)
    m_d(i) = m(i).val();

  Eigen::FullPivHouseholderQR<Matrix<double,R,C> > hh
    = m_d.fullPivHouseholderQr();

  double val = hh.logAbsDeterminant();

  vari** operands = ChainableStack::memalloc_
                  .alloc_array<vari*>(m.size());
  for (int i = 0; i < m.size(); ++i)
    operands[i] = m(i).vi_;

  Matrix<double,R,C> m_inv_transpose 
    = hh.inverse().transpose();
  double* gradients = ChainableStack::memalloc_
                        .alloc_array<double>(m.size());
  for (int i = 0; i < m.size(); ++i)
    gradients[i] = m_inv_transpose(i);

  return var(new precomputed_gradients_vari(
                   val,m.size(),operands,gradients));
}
\end{smallcode}
First, the values of the \code{var} matrix are extracted and used to
set the values in the \code{double} matrix \code{m\_d}.  Then the
decomposition is performed using Eigen's \code{FullPivHouseHholderQR}
class, which according to its documentation provides very good
numerical stability \citep[Section~5.1]{GolubVanLoan:96}. Next, the
value of the log absolute determinant is extracted from the
decomposition class.  Then the operands are set to the argument
matrix's variable implementations to be used in the reverse pass of
automatic differentiation.  Next, the inverse is transposed and used
to populate the double array \code{gradients}.  The final step is to
allocate (on the arena) a new precomputed gradients variable
implementation (see the previous section for definitions); the
precomputed-gradients implementation is then wrapped in a \code{var}
for return.
 
The memory for the gradients and the operands is allocated using the
arena-based allocator \code{memalloc\_}.  The precomputed gradients
structure cannot store standard vectors or Eigen vectors directly
because they manage their own memory and would not be properly
deleted.%
\footnote{The Stan Math Library memory manager provides an additional
  stack to store pointers to \code{chainable} implementations that
  need to have their destructors called.}

The forward pass requires the value of the log determinant.  It would
be possible to be lazy and reconstruct the input matrix from the
operands and then calculate its inverse in the reverse pass.  This
would save a matrix of double values in storage, but would require an
additional QR decomposition.

Critically for speed, all matrix calculations are carried out directly
on \code{double} values rather than on \code{var} values.  The primary
motivation is to reduce the size of the expression graph and
subsequent time spent propagating derivatives in the reverse pass.  A
pleasant side effect is that the matrix operations are carried out
with all of Eigen's optimizations in effect; many of these are only
possible with the memory locality provided by \code{double} values.%
\footnote{\cite{Fog:2014} provides an excellent overview of memory locality,
parallelization, and branch prediction, which is very useful for those
wishing to optimize C++ code.}

\subsubsection{Specialized Log Determinant Implementation}

The Stan Math Library provides a specialized log determinant
implementation for symmetric, positive-definite matrices, which are
commonly used as metrics for geometrical applications or as
correlation, covariance or precision matrices for statistical
applications.  Symmetric, positive-definite matrices can be Cholesky factored
using the $\mbox{LDL}^{\top}$ algorithm
\citep[Chapter~4]{GolubVanLoan:96}, which Eigen provides through its
\code{LDLT} class.  Like the QR decomposition, after the decomposition
is performed, extracting the log determinant and inverse is efficient.

\section{Differential Equation Solver}

The Stan Math Library provides a differential equation solver and
allows derivatives of solutions to be calculated with respect to
parameters and/or initial values.  The system state is coded as a
\code{std::vector} and the system of equations coded as a functor.

\subsection{Systems of Ordinary Differential Equations}

Systems of differential equations describe the evolution of a
multidimensional system over time from a given starting point.  

The state of an $N$-dimensional system at time $t$ will be represented
by a vector $y(t) \in \reals^N$. The change in the system is
determined by a function $f(y,\theta,t)$ that returns the change in
state given current position $y$, parameter vector $\theta \in
\reals^K$, and time $t \in \reals$,
\[
\frac{\totald}{\totald t} y = f(y,\theta,t).
\]
The notation $f_n(y,\theta,t)$ will be used for the $n$-th component
of $f(y,\theta,t)$, so that the change in $y_n$ can be given by
\[
\frac{\totald}{\totald t} y_n = f_n(y,\theta,t).
\]

Given an initial position 
\[
\xi = y(0) \in \reals^N
\]
at initial time $t = 0$, parameter values $\theta$, and system
function $f$, it is possible to solve for state position $y(t)$ at
other times $t$ using numerical methods.%
\footnote{It is technically possible to solve systems backward in time
  for $t < t_0$, but the Stan Math Library is currently restricted to
  forward solutions; signs can be reversed to code backward time
  evolution.}
The Stan Math Library also supports initial positions given
at times other than 0, which can be reduced to systems with initial
time 0 with offsets.  \cite[Chapter~17]{PressEtAl:2007} provides an
overview of ODE solvers.

\subsubsection{Example: Simple Harmonic Oscillator}

The simple harmonic osillator in two dimensions ($N = 2$) with a
single parameter $\theta$ is given by the state equations
\[
\frac{\totald}{\totald t} y_1 = y_2
\ \ \ \ \mbox{and} \ \ \ \
\frac{\totald}{\totald t} y_2 = -\theta y_1.
\]
which can be written as a state function as
\[
f(y,\theta,t) = [y_2 \ \ \ -\theta y_1].
\]



\subsection{Sensitivities of Solution to Inputs}

Solutions of differential equations are often evaluated for their
sensitivity to variation in parameters, i.e.,
\[
\frac{\partial}{\partial \theta} y(t),
\]
where $\theta$ is a vector of parameters, initial state, or both.
This is useful for applications in optimization or parameter
estimation and in reporting on (co)variation with respect to parameters.

\subsection{Differentiating the Integrator}

One way to compute sensitivities is to automatically differentiate the
numerical integrator.  This can be done with \code{var} instances with
a suitably templated integrator, such as Boost's odeint
\citep{AhnertMulansky:2014}. For example, \citep{WeberEtAl:2014} used
the Dormand-Prince integrator from odeint, a fifth-order Runge-Kutta
method.  Because odeint provides only a single template parameter, all
input variables (time, initial state, system parameters) were promoted
to \code{var}.

Although possible, differentiating the integrator is inadvisable for
two reasons.  First, it generates a large expression graph which
consumes both memory and time during automatic differentiation.  This
problem is exacerbated by the overpromotion required by limited
templating.  Second, there is no way to control the error in the
sensitivity calculations.

\subsection{Coupled System}

Both of the problems faced when attempting to differentiate the
integrator can be solved by creating a coupled system that adds state
variables for the sensitivities to the original system.  The partial
derivatives are then given by the solutions for the sensitivities.

For each state variable $n$ and each parameter or initial state
$\alpha_m$ for which sensitivities are required, a newly defined state
variable
\[
z_{n,m} = \frac{\partial}{\partial \alpha_m} y_m
\]
is added to the system.  This produces coupled systems of the
following sizes, based on whether derivatives are required with respect
to the initial state, system parameters, or both.
\begin{center}
\begin{tabular}{cc|c}
\multicolumn{2}{c|}{\it Sensitivities} &
\\
{\it Initial State} & {\it Parameters} & {\it Coupled Size}
\\ \hline
+ & - & $N \times (N + 1)$
\\
- & + & $N \times (K + 1)$ 
\\
+ & + & $N \times (N + K + 1)$
\end{tabular}
\end{center} 

With the coupled system, sensitivities are calculated by the
integrator step by step and thus can be controlled for error.  The
resulting sensitivities at solution times can then be used to
construct an autodiff variable implementation of the output expression
with respect to its inputs (initial state and/or parameters).

The Stan Math Library currently uses the Dormand-Prince
integrator implementation from Boost's odeint library
\citep{AhnertMulansky:2014}, but the design is modular, with
dependencies only on solving integration problems using \code{double}
values.  More robust integrators, in particular the CVODE integrator
\citep{CohenHindmarsh:1996} from the SUNDIALS package
\citep{HindmarshEtAl:2005}, will be useful for their ability to
automatically deal with stiff systems of equations.

\subsection{Differentiating an ODE Solution with Respect to a Parameter}

The sensitivity matrix for states ($y$) by parameters ($\theta$)
is
\[
z = \frac{\partial}{\partial \theta} y.
\]
Componentwise, that is
\[
z_{n,m} = \frac{\partial}{\partial \theta_m} y_n.
\]
The time derivatives of $z$ needed to complete the coupled system can
be given via a function $h$, defined and derived componentwise as
\begin{eqnarray*}
h_{n,m}(y,z,\theta,t)
& = & \frac{\totald}{\totald t} z_{n,m}.
\\[3pt]
& = &
\frac{\totald}{\totald t} \frac{\totald}{\totald \theta_m} y_n
\\[3pt]
& = &
\frac{\totald}{\totald \theta_m} \frac{\totald}{\totald t} y_n
\\[3pt]
& = & 
\frac{\totald}{\totald \theta_m} f_n(y,\theta)
\\[3pt]
& = & 
\frac{\partial}{\partial \theta_m} f_n(y,\theta)
+ \sum_{j = 1}^N \left( \frac{\partial}{\partial \theta_m} y_j \right) \, 
         \frac{\partial}{\partial y_j} f_n(y,\alpha)
\\[3pt]
& = & 
\frac{\partial}{\partial \theta_k} f_n(y,\theta)
+ \sum_{j=1}^N z_{j,m} \frac{\partial}{\partial y_j} f_n(y,\theta).
\end{eqnarray*}
The sensitivity of a parameter is the derivative of the state of the
system with respect to that parameter, with all other parameters held
constant (but not the states).  Thus the sensitivities act as partial
derivatives with respect to the parameters but total derivatives with
respect to the states, because of the need to take into account the
change in solution as parameters change.

The coupled system will also need new initial states $\chi_{n,m}$ for
the sensitivities $z_{n,m}$, all of which work out to be zero, because
\[
\chi_{n,m} = \frac{\partial}{\partial \theta_m} \xi_n = 0.
\]

The final system that couples the original state with sensitivities
with respect to parameters has state $(y,z)$, initial conditions
$\xi,\chi$, and system function $(f,h)$.

\subsection{Differentiating an ODE Solution with Respect to the Initial State}

The next form of coupling will be of initial states, with new state
variables
\[
w = \frac{\partial}{\partial \xi} y,
\]
which works out componentwise to
\[
w_{n,k} = \frac{\partial}{\partial \xi_{k}} y_n.
\]
Sensitivities can be worked out in this case by defining a new
system with state variables offset by the initial condition,
\[
u = y - \xi.
\]
This defines a new system function $g$ with the original parameters
$\theta$ and original initial state $\xi$ now both treated as
parameters,
\[
g(u,(\theta,\xi)) = f(u + \xi, \theta).
\]
The new initial state is a zero vector by construction.  The
derivatives are now with respect to the parameters of the revised
system with state $u$, system function $g$, and parameters
$\theta,\xi$, and work out to
\begin{eqnarray*}
\frac{\totald}{\totald t} w_{n,k}
& = & \frac{\totald}{\totald t} \frac{\totald}{\totald \xi_{k}} y_n
\\[3pt]
& = & \frac{\totald}{\totald \xi_k} \frac{\totald}{\totald t} y_n
\\[3pt]
& = & \frac{\totald}{\totald \xi_k} f_n(y,\theta)
\\[3pt]
& = & \frac{\totald}{\totald \xi_k} f_n(u + \xi, \theta)
\\[3pt]
& = & \frac{\partial}{\partial \xi_k} f_n(u + \xi, \theta)
\\[3pt]
& & 
{ } + \sum_{j=1}^N 
        \left( \frac{\partial}{\partial \xi_k} u_j \right)
           \frac{\partial}{\partial u_j} f_n(u + \xi, \theta)
\\
& & 
 { } + \sum_{j=1}^N
         \left( \frac{\partial}{\partial \xi_k} \xi_j \right)
         \frac{\partial}{\partial \xi_j} f_n(u + \xi, \theta)
\\[6pt]
& = & 
\frac{\partial}{\partial \xi_k} f_n(u + \xi, \theta)
\\
& & { } + \sum_{j=1}^N \left( u_j + \frac{\partial}{\partial \xi_k} \xi_j
\right)
\frac{\partial}{\partial y_j} f_n(y,\theta).
\end{eqnarray*}
The derivative $\partial \xi_j / \partial \xi_k$ on the last line is
equal to 1 if $j = k$ and equal to 0 otherwise.

\subsection{Computing Sensitivities with Nested Automatic Differentiation}

In order to add sensitivities with respect to parameters and/or
initial states to the system, Jacobians of the differential equation
system function $f$ are required with respect to the parameters (where
the parameters may include the initial states, as shown in the last
section).  

To allow nested evaluation of system Jacobians, the Stan Math Library
allows nested derivative evaluations.  When the system derivatives are
required to solve the system of ODEs, the current stack location is
recorded, autodiff of the system takes place on the top of the stack,
and then derivative propagation stops at the recorded stack location.
This allows arbitrary reverse-mode automatic differentiation to be
nested inside other reverse-mode calculations.  The top of the stack
can even be reused for Jacobian calculations without rebuilding an
expression graph for the outer system being differentiated.

\subsection{Putting Results Back Together}

When the numerical integrator solves the coupled system, the solution
is for the original state variables $y$ along with sensitivities
$\frac{\totald}{\totald \theta} y$ and/or $\frac{\totald}{\totald \xi}
y$. 

Given a number of solution times requested, $t_1,\ldots,t_J$, the
state solutions and sensitivity solutions are used to create a
\code{vari} instance for each $y_n(t_j)$.  Each of these variables is
then connected via its sensitivity to each of the input parameters
and/or initial states.  This requires storing the sensitivities as
part of the result variables in a general (not ODE specific)
precomputed-gradients \code{vari} structure.

\subsection{System Function}

Given an initial condition and a set of requested solution times, the
Stan Math Library can integrate an ODE defined in terms of a system
function.  The system function must be able to be instantiated by the
following signature (where \code{vector} is \code{std::vector})
\begin{smallcode}
vector<var>
operator()(double t, 
           const vector<double>& y,
           const vector<var>& theta) const;
\end{smallcode}
The return value is the vector of time derivatives evaluated at time
\code{t} and system state \code{y}, given parameters \code{theta},
continuous and integer data \code{x} and \code{x\_int}, and an output
stream \code{o} for messages.  The function must be constant.

The simple harmonic oscillator could be implemented as the following
functor.
\begin{smallcode}
struct sho {
  template <typename T>
  vector<T> operator()(double t, 
                       const vector<double>& y
                       const vector<var>& theta) const
    vector<T> dy_dt(2);
    dy_dt[0] = y[1];
    dy_dt[1] = -theta[0] * y[2];
    return dy_dt;
  }
};
\end{smallcode}

\subsection{Integration Function}

The function provided by the Stan Math Library to compute solutions to
ODEs and support sensitivity calculations through autodiff has the
following signature.
\begin{smallcode}
template <typename F, typename T1, typename T2>
vector<vector<typename promote_args<T1, T2>::type> >
integrate_ode(const F& f,
              const std::vector<T1> y0,
              const double t0,
              const std::vector<double>& ts,
              const std::vector<T2>& theta
              const std::vector<double>& x,
              const std::vector<int>& x_int,
              std::ostream* msgs);
\end{smallcode}
The argument \code{f} is the system function, and it must be
implemented with enough generality to allow autodiff with respect to
the parameters and/or initial state.  The variable \code{y0} is the
initial state and \code{t0} is the initial time (which may be
different than 0); its scalar type parameter \code{T1} may be either
\code{double} or \code{var}, with \code{var} being used to autodiff
with respect to the initial state.  The vector \code{ts} is a set of
solution times requested and must have elements greater than \code{t0}
arranged in strictly ascending order.  The vector \code{theta} is for
system parameters; its scalar type parameter \code{T2} can be either
\code{double} or \code{var}, with \code{var} used to autodiff with
respect to parameters.  There are two additional arguments, \code{x}
and \code{x\_int}, used for real and double-valued data respectively.
These are provided so that data does not need to be either hard coded
in the system or promoted to \code{var} (which would add unnecessary
components to the expression graph, increasing time and memory used).
Finally, an output stream pointer can be provided for messages printed
by the integrator.

Given the harmonic oscillator class \code{sho}, and initial values
$(-1,0)$, the solutions with sensitivities for the harmonic oscillator
at times $1{:}10$ with parameter $\theta = 0.35$, initial state $y(0)
= (-1,0)$, can be obtained as follows.
\begin{smallcode}
double t0 = 0.0;   
vector<double> ts;  
for (int t = 1; t <= 10; ++t) ts.push_back(t);
var theta = 0.35;
vector<var> y0(2);  y[0] = 0;  y[1] = -1.0;
vector<var> ys 
  = integrate_ode(sho(), y0, t0, ts, theta);
for (int i = 0; i < 2; ++i) {
  if (i > 0) set_zero_all_adjoints();
  y.grad();
  for (int n = 0; n <= ys.size(); ++n) 
    printf("y(
    printf("sens: theta=
           theta.adj(), y0[0].adj(), y0[1].adj());
  }
}
\end{smallcode}

\section{Probability Functions and Traits}\label{probability-functions.section}\label{traits.section}

The primary application for which the Stan Math Library was developed
is fitting statistical models, with optimization for maximum
likelihood estimates and Markov chain Monte Carlo (MCMC) for Bayesian
posterior sampling \citep{GelmanEtAl:2013}.  For optimization,
gradient descent and quasi-Newton methods both depend on being able to
calculate derivatives of the objective function (in the primary
application, a (penalized) log likelihood or Bayesian posterior).  For
MCMC, Hamiltonian Monte Carlo (HMC) \citep{Duane:1987, Neal:2011,
  BetancourtEtAl:2014} requires gradients in the leapfrog integrator
it uses to solve the Hamiltonian equations and thus simulate an
efficient flow through the posterior.

In both of these applications, probability functions need only be
calculated up to a constant term.  In order to do this efficiently and
automatically, the Stan Math Library uses traits-based metaprogramming
and template parameters to configure its log probability functions.
For example, consider the log normal density again
\[
\log \distro{Normal}(y | \mu, \sigma)
= -\log \left( 2 \pi \right)
  - \log \sigma
  - \frac{1}{2} \left( \frac{y - \mu}{\sigma} \right)^2.
\]

In all cases, the first term, $-\log \left( 2 \pi \right)$ can be
dropped because it's constant.  If $\sigma$ is a constant, then the
second term can be dropped as well.  The final term must be preserved
unless all of the arguments, $y$, $\mu$, and $\sigma$, are constants.  

For maximum generality, each log probability function has a leading
boolean template argument \code{propto}, which will have a true value
if the calculation is being done up to a proportion and a false value
otherwise.  Calculations up to a proportion drop constant terms.  Each
of the remaining arguments is templated out separately, to allow all
eight combinations of constants and variables as arguments.  The
return type is the promotion of the argument types.  The fully
specified signature is
\begin{smallcode}
template <bool propto, typename T1, typename T2, typename T3>
typename return_type<T1, T2, T3>::type
normal_log(const T1& y, const T2& mu, const T3& sigma) {
  ...
\end{smallcode}
For scalar arguments, the \code{return\_type} traits metaprogram has
the same behavior as Boost's \code{promote\_args}; generalizations to
container arguments such as vectors and arrays will be discussed in
\refsection{vectorization}.

Without going into the details of error handling, the body of the
function uses a further metaprogram to decide which terms to include
in the result.  Not considering generalizations to containers and
analytic gradients, the body of the function behaves as
follows:
\begin{smallcode}
  include stan::math::square;
  include std::log;
  typename return_type<T1, T2, T3>::type lp = 0;
  if (include_summand<propto>::value)
    lp += NEGATIVE_LOG_2_PI;
  if (include_summand<propto,T3>::value)
    lp -= log(sigma);
  if (include_summand<propto, T1, T2, T3>::value)
    lp -= 0.5 * square((y - mu) / sigma);
  return lp;
}
\end{smallcode}
The \code{include\_summand} traits metaprogram is quite simple,
defining an enum with a true value if any of the types is not
primitive (integer or floating point).  Although it allows up to 10
arguments and supports container types, the following implementation
of \code{include\_summand} suffices for the current example.
\begin{smallcode}
template <boolean propto,
          typename T1 = double, 
          typename T2 = double, 
          typename T3 = double>
struct include_summand {
  enum { value  = !propto  
                  || !is_constant<T1>::value
                  || !is_constant<T2>::value
                  || !is_constant<T3>::value;  }
};
\end{smallcode}
The template specification requires a boolean template parameter
\code{propto}, which is true if calculations are to be done up to a
constant multiplier.  The remaining template typename parameters have
default values of \code{double}, which are the base return value type.
The structure defines a single enum \code{value}, the value of which
will be true if the \code{propto} is false or if any of the remaining
template parameters are not constant values.  For scalar types,
\code{is\_constant} behaves like Boost's \code{is\_arithmetic} trait
metaprogram (from the TypeTraits library); it is extended for
container types as described in \refsection{vectorization}.

\section{Vectorization of Reductions}\label{vectorization.section}

The Stan Math Library was developed to facilitate probabilistic
modeling and inference by making it easy to define density functions
on the log scale with derivative support.  A common use case is
modeling a sequence of independent observations $y_1,\ldots,y_N$ from
some parametric likelihood function $p(y_n|\theta)$ with shared
parameters $\theta$.  In this case, the joint likelihood function for
all the observations is
\[
p(y_1,\ldots,y_N|\theta) = \prod_{n=1}^N p(y_n|\theta),
\]
or on the log scale,
\[
\log p(y_1,\ldots,y_N|\theta) = \sum_{n=1}^N \log p(y_n|\theta).
\]
Thus the log joint likelihood is the sum of the log likelihoods
for the observations $y_n$.

\subsection{Argument Broadcasting}

The normal log density function takes three arguments, the variate or
outcome along with a location (mean) and scale (standard deviation)
parameter.  Without vectorization, each of these arguments could be a
scalar of type \code{int}, \code{double}, or \code{stan::math::var}.

With vectorization, each argument may also be a standard library
vector, \code{vector<T>}, an Eigen matrix, \code{Matrix<T,~Dynamic,~1>},
or an Eigen row matrix, \code{Matrix<T,~1,~Dynamic>}, where \code{T} is
any of the scalar types.  All container arguments must be the same
size, and any scalar arguments are broadcast to behave as if they were
containers with the scalar in every position.

\subsection{Vector Views}

Rather than implement 27 different versions of the normal log density,
the Stan Math Library introduces an expression template allowing any
of the types to be treated as a vector-like container holding its
contents.  The view provides a scalar typedef, a size, and an operator
to access elements using bracket notation.  Fundamentally, it stores a
pointer, which will in practice be to a single element or to an array.
Simplifying a bit, the code is as follows.
\begin{smallcode}
template <typename T,
          bool is_array = stan::is_vector_like<T>::value>
class VectorView {
public: 
  typedef typename scalar_type<T>::type scalar_t;

  VectorView(scalar_t& c) : x_(&c) { }

  VectorView(std::vector<scalar_t>& v) : x_(&v[0]) { }

  template <int R, int C>
  VectorView(Eigen::Matrix<scalar_t,R,C>& m) : x_(&m(0)) { }

  VectorView(scalar_t* x) : x_(x) { }

  scalar_t& operator[](int i) {
    if (is_array) return x_[i];
    else return x_[0]; 
  }
private:
  scalar_t* x_;
};
\end{smallcode}
There are two template parameters, \code{T} being the type of object
being viewed and \code{is\_array} being true if the object being
viewed is to be treated like a vector when accessing members. The
second argument can always be set by default; it acts as a typedef to
enable a simpler definition of \code{operator[]}.  The traits
metaprogram \code{is\_vector\_like<T>::value} determines if the viewed
type \code{T} acts like a container.

The typedef \code{scalar\_t} uses the metaprogram
\code{scalar\_type<T>::type} to calculate the underlying scalar type
for the viewed type \code{T}.  

The constructor takes either a scalar reference, a standard vector, or
an Eigen matrix and stores either a pointer to the scalar or the first
element of an array in private member variable \code{x\_}.

There is a single member function, \code{operator[]}, which defines
how the view returns elements for a given index \code{i}.  If the
viewed type is a container as defined by template parameter
\code{is\_array}, then it returns the element at index \code{i}; if it
is a scalar type, then the value is returned (\code{x\_[0]} is
equivalent to \code{*x\_}).

Once a view is constructed, its \code{operator[]} can be used just
like any other container.   This pattern would be easy to extend to
further containers, such as Boost sequences.  After this code is
executed, 
\begin{smallcode}
vector<var> y(3); ...
VectorView<vector<var> > y_vw(y);
\end{smallcode}
the value of \code{y\_vw[i]} tracks \code{y[i]}, and
the typedef \code{y\_vw::scalar\_t} is \code{var}.  

For a scalar, the construction is similar.  After this code is
executed,
\begin{smallcode}
double z;
VectorView<double> z_vw(z);
\end{smallcode}
the value of \code{z\_vw[i]} tracks the value of \code{z} no matter
what \code{i} is---the index is simply ignored.  

No copy is made of the contents of \code{y} and it is up to the
constructor of \code{VectorView} to ensure that the contents of
\code{y} does not change, for instance by resizing.  Furthermore, a
non-constant reference is returned by the \code{operator[]}, meaning
that clients of the view can change the contained object that is being
viewed.  This is all standard in such a view pattern, examples of
which are the \code{block}, \code{head}, and \code{tail} functions in
Eigen, which provide mutable views of matrices or vectors.

\subsection{Value Extractor}

The helper function \code{value\_of} is defined to extract the
\code{double}-based value of either an autodiff variable or a
primitive value.  The definition for primitive types is in the
\code{stan::math} namespace.
\begin{smallcode}
template <typename T>
inline double value_of(const T x) {
  return static_cast<double>(x);
}
\end{smallcode}
The function is overloaded for autodiff variables.
\begin{smallcode}
inline double value_of(const var& v) {
  return v.vi_->val_;
}
\end{smallcode}
The definition for autodiff variables is in the \code{stan::math}
namespace to allow argument-dependent lookup.

It is crucial for many of the templated definitions to be able to pull
out \code{double} values no matter what the type of argument is.  
The function \code{value\_of} allows
\begin{smallcode}
T x;
double x_d = value_of(x);
\end{smallcode}
with \code{T} instantiated to \code{stan::math::var}, \code{double}, or
\code{int}.  

\subsection{Vector Builder Generic Container}

By itself, the vector view does not provide a way to generate
intermediate quantities needed in calculations and store them in an
efficient way.  For the normal density, if $\sigma$ is a single value,
then $\log \sigma$ and $\sigma^{-1}$ can be computed once and reused.

The class \code{VectorBuilder} is used as a container for intermediate
values.  In the following code, from the normal density, it is used as
the type of intermediate containers for $\sigma^{-1}$ and $\log
\sigma$.%
\footnote{Stan's version of these functions is slightly more
  complicated in that they also support forward-mode autodiff.  The
  details of return type manipulation for forward-mode is sidestepped
  here by dropping template parameters and their helper template
  metaprograms and fixing types as \code{double}.}
\begin{smallcode}
VectorBuilder<true> inv_sigma(length(sigma));

VectorBuilder<include_summand<propto,T_scale>::value>
  log_sigma(length(sigma));
\end{smallcode}
The template parameter indicates whether the variable needs to be
stored or not.  If the parameter is false, no storage will be
allocated.  The intermediate value \code{inv\_sigma} is always
computed because it will be needed for the normal density even if no
derivatives are taken.  The intermediate value \code{log\_sigma}, on
the other hand, is only computed and memory is only allocated for it
if a term only involving a type \code{T\_scale} is included; see
\refsection{probability-functions} for the definition of
\code{include\_summand}.

The code for \code{VectorBuilder} itself is relatively straightforward
\begin{smallcode}
template <bool used, typename T1, 
          typename T2=double, typename T3=double>
struct VectorBuilder {
  VectorBuilderHelper<used, contains_vector<T1, T2, T3>::value> a_;

  VectorBuilder(size_t n) : a_(n) { }

  T1& operator[](size_t i) { return a_[i]; }
};
\end{smallcode}
The template struct \code{VectorBuilderHelper} provides the type of
the value \code{a}; it is passed a boolean template parameter
indicating whether any of the types is a vector and thus requires a
vector rather than scalar to be allocated for storage.  The constructor for
\code{VectorBuilder} just passes the requested size to the helper's
constructor.  This will either construct a container of the requested
size or return a dummy, depending on the value of the template
parameter \code{used}.  The definition of \code{operator[]} just
returns the value at index \code{i} produced by the helper.

There are three use cases for efficiency for the vector.  The simplest
is the dummy, which is employed when \code{used} is false.  It also
defines the primary template structure.
\begin{smallcode}
template <bool used, bool is_vec>
struct VectorBuilderHelper {
  VectorBuilderHelper(size_t /* n */) { }

  double& operator[](size_t /* i */) {
    throw std::logic_error("used is false.");
  }
};
\end{smallcode}
The two template parameters are booleans indicating whether the
storage is used, and if it is used, whether it is a vector or not.
For the base case, there is no allocation.  Because there are no
virtual functions declared, the size of \code{VectorBuilderHelper} is
zero bytes and it can be optimized out of the definition of
\code{VectorBuilder}.  The definition of the operator raises a
standard library logic error because it should never be called;  it is
only defined because it is required to be so that it can be used in
alternation with helper implementations that do return values.

For the case where \code{used} is true but \code{is\_vec} is false,
the following specialization stores a \code{double} value.
\begin{smallcode}
template <>
struct VectorBuilderHelper<true,false> {
  double x_;

  VectorBuilderHelper(size_t /* n */) : x_(0.0) { }

  T1& operator[](size_t /* i */) {
    return x_;
  }
};
\end{smallcode}
Note that like \code{VectorView}, it returns the value \code{x\_} no
matter what index is provided.

For the case where \code{used} is true and \code{is\_vec} is true, a
standard library vector is used to store values, allocating the memory
in the constructor.
\begin{smallcode}
template <>
struct VectorBuilderHelper<true,true> {
  std::vector<T1> x_;

  VectorBuilderHelper(size_t n) : x_(n) { }

  T1& operator[](size_t i) {
    return x_[i];
  }
};
\end{smallcode}
Instances only live on the C++ stack, so the memory is managed
implicitly when instances of \code{VectorBuilder} go out of scope.

\subsection{Operands and Partials}

The last piece of the puzzle in defining flexible functions for
automatic differentiation is a general purpose structure
\code{operands\_and\_partials}.  The key to this structure is that it
stores the operands to a function or operator as an array of
\code{vari} and stores the partial of the result with respect to the
operand in a parallel array of \code{double} values.  That is, it
allows general eager programming of partials, which allows
straightforward metaprograms in many contexts such as vectorized
density functions.  A simplified definition is as follows.%
\footnote{As with other aspects of the probability functions, the
  implementation in Stan is more general, allowing for forward-mode
  autodiff variables of varying order as well as reverse-mode autodiff
  variables and primitives.  It also allows more or fewer template
  parameters, giving all but the first default \code{double} values.}
\begin{smallcode}
template <typename T1, typename T2, typename T3>
struct OperandsAndPartials {
  size_t n_;
  vari** ops_;
  double* partials_;
  VectorView<is_vector<T1>::value, 
             is_constant_struct<T1>::value> d_x1_;
  VectorView<is_vector<T2>::value, 
             is_constant_struct<T2>::value> d_x2_;
  VectorView<is_vector<T3>::value, 
             is_constant_struct<T3>::value> d_x3_;
  ...
};
\end{smallcode}
The vector views are views into the \code{partial\_} array.  The
constructor is as follows.
\begin{smallcode}
OperandsAndPartials(const T1& x1, const T2& x2, const T3& x3)
: n_(!is_constant_struct<T1>::value * length(x1)
     + !is_constant_struct<T2>::value * length(x2)
     + !is_constant_struct<T3>::value * length(x3)),
  ops_(memalloc_.array_alloc<vari*>(n_)),
  partials_(memalloc_.array_alloc<double>(n_)),
  d_x1_(partials_),
  d_x2_(partials_
        + !is_constant_struct<T1>::value * length(x1)),
  d_x3_(partials_
        + !is_constant_struct<T1>::value * length(x1)),
        + !is_constant_struct<T2>::value * length(x2)) 
{ 
  size_t base = 0;
  if (!is_constant_struct<T1>::value)
    base += set_varis<T1>::set(&ops_[base], x1);
  if (!is_constant_struct<T2>::value)
    base += set_varis<T2>::set(&ops_[base], x2);
  if (!is_constant_struct<T3>::value)
    base += set_varis<T3>::set(&ops_[base], x3);

  std::fill(partials_, partials_ + n, 0);
}
\end{smallcode}
The class template parameters \code{T1}, \code{T2}, and \code{T3}
define the argument types for the constructor; these will be the
arguments to a probability function and may be vectors or scalars of
either primitive or autodiff type.  The size \code{n\_} of the
operands and partials arrays is computed by summing the non-constant
argument sizes; the function \code{length} is trivial and not shown.
The operands \code{ops\_} and partials \code{partials\_} are allocated
in the arena using a template function that allocates the right size
memory and casts it to an array with elements of the specified
template type.  The first partials view, \code{d\_x1\_}, starts at the
start of \code{partials\_} itself.  Each subsequent view is started
by incrementing the start of the last view;  this could perhaps be
made more efficient by extracting the pointer from the previous view.

The body of the constructor sets the operator values in \code{ops\_}
and fills the partials array with zero values.  The operands are set
using the static function \code{set} in the template class
\code{set\_varis}, which extracts the \code{vari*} from the
arguments.  

The member function \code{to\_var} returns a \code{var} with
an implementation of the following class, which stores the
operands and partials and uses them for the chain rule.
\begin{smallcode}
struct partials_vari : public vari {
  const size_t N_;
  vari** operands_;   
  double* partials_;

  partials_vari(double value, size_t N,
                vari** operands, double* partials)
  : vari(value), N_(N),
    operands_(operands), partials_(partials) { }

  void chain() {
    for (size_t n = 0; n < N_; ++n)
      operands_[n]->adj_ += adj_ * partials_[n];
  }
};
\end{smallcode}
The constructor passes the value to the superclass's constructor,
\code{vari(double)}, then stores the size, operands, and partials.
The \code{chain()} implementation increments each operand's
adjoint by the result's adjoint times the partial derivative of the
result relative to the operand.

\subsection{Example: Vectorization of the Normal Log Density}

Continuing with the example of the normal probability density function
introduced in \refsection{probability-functions}, the pieces are now
all in place to see how \code{normal\_log} is defined for reverse-mode
autodiff with vectorization.

The function signature is as follows
\begin{smallcode}
template <bool propto, 
          typename T_y, typename T_loc, typename T_scale>
typename return_type<T_y,T_loc,T_scale>::type
normal_log(const T_y& y, const T_loc& mu, const T_scale& sigma);
... 
\end{smallcode}
The first boolean template parameter will be true if calculations are
allowed to drop constant terms.  The remaining template parameters are
the types of the arguments for the variable and the location and scale
parameters.  The result is the calculated return type, which will be
\code{var} if any of the arguments is a (container of) \code{var} and
\code{double} otherwise.

The function begins by validating all of the inputs.
\begin{smallcode}
  static const char* function("stan::prob::normal_log");
  check_not_nan(function, "Random variable", y);
  check_finite(function, "Location parameter", mu);
  check_positive(function, "Scale parameter", sigma);
  check_consistent_sizes(function, "Random variable", y,
      "Location parameter", mu, "Scale parameter", sigma);
\end{smallcode}
The function name itself is provided as a static constant.  These
functions raise exceptions with warning messages indicating where the
problem arose.  The consistent size check ensures that all of the
container types (if any) are of the same size.  For example, if both
\code{y} and \code{mu} are vectors and \code{sigma} is a scalar, then
\code{y} and \code{mu} must be of the same size.

Next, the function returns zero if any of the container sizes is zero,
or if none of the argument types is an autodiff variable and
\code{propto} is true.  In either case, the function returns zero.
\begin{smallcode}
  if (!(stan::length(y) && stan::length(mu) 
        && stan::length(sigma)))
    return 0.0;

  if (!include_summand<propto,T_y,T_loc,T_scale>::value)
    return 0.0;
\end{smallcode}
Only then is the accumulator \code{logp} for the result initialized.
\begin{smallcode}
  double logp = 0;
\end{smallcode}
Next, the operands and partials accumulator is initialized along with
vector views of each of the arguments.  The size \code{N} is set to
the maximum of the argument sizes (scalars are treated as size 1).
\begin{smallcode}
  OperandsAndPartials<T_y, T_loc, T_scale> 
    operands_and_partials(y, mu, sigma);

  VectorView<const T_y> y_vec(y);
  VectorView<const T_loc> mu_vec(mu);
  VectorView<const T_scale> sigma_vec(sigma);

  size_t N = max_size(y, mu, sigma);
\end{smallcode}
Next, the vector builders for $\sigma^{-1}$ and $\log \sigma$ are
constructed and filled.
\begin{smallcode}
  VectorBuilder<true, T_partials_return, T_scale> 
    inv_sigma(length(sigma));
  for (size_t i = 0; i < length(sigma); i++)
    inv_sigma[i] = 1.0 / value_of(sigma_vec[i]);
\end{smallcode}
Although \code{inv\_sigma} is always filled, \code{log\_sigma} is not
filled if the calculation is being done up to a proportion and the
scale parameter is a constant.
\begin{smallcode}
  VectorBuilder<include_summand<propto,T_scale>::value, 
                T_partials_return, T_scale> 
    log_sigma(length(sigma));
  if (include_summand<propto,T_scale>::value)
    for (size_t i = 0; i < length(sigma); i++)
      log_sigma[i] = log(value_of(sigma_vec[i]));
\end{smallcode}
The value of \code{length(sigma)} will be 1 if \code{sigma} is a
scalar and the size of the container otherwise.  Because the
\code{include\_summand} traits metaprogram is evaluated statically,
the compiler is smart enough to simply drop this whole loop if the
summand should not be included.  As a result, exactly as many
logarithms and inversions are calculated as necessary.  These values
are then used in a loop over \code{N}, the size of the largest
argument (1 if they are all scalars).
\begin{smallcode}
  for (size_t n = 0; n < N; n++) {
    double y_dbl = value_of(y_vec[n]);
    double mu_dbl = value_of(mu_vec[n]);

    double y_minus_mu_over_sigma 
      = (y_dbl - mu_dbl) * inv_sigma[n];
    double y_minus_mu_over_sigma_squared 
      = y_minus_mu_over_sigma * y_minus_mu_over_sigma;

    if (include_summand<propto>::value)
      logp += NEG_LOG_SQRT_TWO_PI;
    if (include_summand<propto,T_scale>::value)
      logp -= log_sigma[n];
    if (include_summand<propto,T_y,T_loc,T_scale>::value)
      logp += NEGATIVE_HALF * y_minus_mu_over_sigma_squared;

    double scaled_diff = inv_sigma[n] * y_minus_mu_over_sigma;
    if (!is_constant_struct<T_y>::value)
      operands_and_partials.d_x1_[n] -= scaled_diff;
    if (!is_constant_struct<T_loc>::value)
      operands_and_partials.d_x2_[n] += scaled_diff;
    if (!is_constant_struct<T_scale>::value)
      operands_and_partials.d_x3_[n] 
        += inv_sigma[n] * (y_minus_mu_over_sigma_squared - 1);
  }
  return operands_and_partials.to_var(logp,y,mu,sigma);
}
\end{smallcode}
In each iteration, the value of \code{y[n]} and \code{mu[n]} is
extracted as a \code{double}.  Then the intermediate terms are
calculated.  These are needed for every iteration (as long as not all
arguments are constants, which was checked earlier in the function).  
Then depending on the argument types, various terms are added or
subtracted from the log density accumulator \code{logp}.  The
normalizing term is only included if the \code{propto} template
parameter is false.  The $\log \sigma$ term is only included if
$\sigma$ is not a constant.  The remaining conditional should always
succeed, but is written this way for consistency;  the compiler will
remove it as its condition is evaluated statically.

After the result is calculated, the derivatives are calculalted.  These
are added to the \code{OperandsAndPartials} data structure using the
views \code{d\_x1\_}, \code{d\_x2\_}, and \code{d\_x3\_}.  These are
all analytical partial derivatives of the normal density with respect
to its arguments.  The operands and partials structure initialized all
derivatives to zero.  Here, they are incremented (or decremented).  If
any of the arguments is a scalar, then the view is always of the same
element and this effectively increments the single derivative.  Thus
the same code works for a scalar or vector \code{sigma}, either
breaking the partial across each argument, or reusing \code{sigma} and
incrementing the partial.

The final line just converts the result, with value and arguments, to
a \code{var} for return.







\section{Evaluation}

In this section, Stan's reverse-mode automatic differentation is
evaluated for speed and memory usage.  The evaluated version of Stan
is version 2.6.3.

\subsection{Systems and Versions}

In addition to Stan, the following operator overloading, reverse-mode
automatic differentiation systems are evaluated.
\begin{itemize}
\item Adept, version 1.0:
  \smallurl{http://www.met.reading.ac.uk/clouds/adept}
\\ see \citep{Hogan:2014}
\item Adol-C, version 2.5.2:
  \smallurl{https://projects.coin-or.org/ADOL-C}
\\ see \citep{GriewankWalther:2008}
\item CppAD, version 1.5: \smallurl{http://www.coin-or.org/CppAD}
\\ see \citep{Bell:2012}
\item Sacado (Trilinos), version 11.14: \smallurl{http://trilinos.org}
\\ see \citep{Gay:2005}
\item Stan, version 2.6.3: \smallurl{http://mc-stan.org}
\\ see this paper
\end{itemize}

Like Stan, CppAD is purely header only.  Although Sacado is
distributed as part of the enormous (150MB compressed) Trilinos
library, which comes with a complex-to-configure CMake file, Sacado
itself is header only and can be run as such by including the proper
header files.  Adept requires a library archive to be built and linked
with client code.  Adol-C requires a straightforward CMake
configuration step to generate a makefile, which then compiles object
files from C and C++ code; the object files are then linked with the
client code.  Detailed instructions for building all of these
libraries from source are included with the evaluation code for this
paper and the source versions used for the evaluations are included in
the Git repository.

\subsubsection{Systems Excluded}

Other C++ systems for computing derivatives were excluded from the
evaluation for various reasons:
\begin{itemize}
\item lack of reverse-mode automatic differentation
\begin{itemize}
\item ADEL: \smallurl{https://github.com/eleks/ADEL}
\item CeresSolver: \smallurl{http://ceres-solver.org}
\item CTaylor: \smallurl{https://ctaylor.codeplex.com}
\item FAD:
  \smallurl{http://pierre.aubert.free.fr/software/software.php3}
\end{itemize}
\item lack of operator overloading to allow differentiation of
  existing C++ programs
\begin{itemize}
\item ADIC2: \smallurl{http://www.mcs.anl.gov/adic}
\item ADNumber: \smallurl{https://code.google.com/p/adnumber}
\item AutoDiff\_Library: \smallurl{https://github.com/fqiang/autodiff_library}
\item CasADi: \smallurl{http://casadi.org}
\item CppAdCodeGen:
  \smallurl{https://github.com/joaoleal/CppADCodeGen}
\item Rapsodia: \smallurl{http://www.mcs.anl.gov/Rapsodia}
\item TAPENADE:
  \smallurl{http://tapenade.inria.fr:8080/tapenade/index.jsp}
\item Theano: \smallurl{http://deeplearning.net/software/theano}
\end{itemize}
\item require graphics processing unit (GPU)
\begin{itemize}
\item AD4CL: \smallurl{https://github.com/msupernaw/AD4CL}
\end{itemize}
\item lack of open-source licensing
\begin{itemize}
\item ADC: \smallurl{http://www.vivlabs.com}
\item AMPL: \smallurl{http://ampl.com}
\item COSY INFINITY: \smallurl{http://cosy.pa.msu.edu}
\item FADBAD++: \smallurl{http://www.imm.dtu.dk/~kajm/FADBAD}
\end{itemize}
\end{itemize}

\subsection{What is being Evaluated}\label{retaping.section}

The evaluations in this paper are based on simple gradients with
retaping for each evaluation.

\subsubsection{Gradients vs.\ Jacobians}

All of the evaluations are for simple gradient calculations for
functions $f:\reals^N \rightarrow \reals$ with multiple inputs and a
single output.  That is the primary use case for which Stan's
automatic differentiation was designed.

Stan's reverse-mode automatic differentiation can be used to compute
Jacobians as shown in \refsection{jacobians}.

Stan's lazy evaluation of partial derivatives in the reverse pass over
expression graph is designed to save memory in the gradient case, but
requires recomputations of gradients when calculating Jacobians of
functions $g:\reals^N \rightarrow \reals^M$ with multiple inputs and
multiple outputs.

\subsubsection{Retaping}

Adol-C and CppAD allow the expression graph created in a forward pass
through the code (which they call a ``tape'') to be reused.  This can
speed up subsequent reverse-mode passes.  The drawback to reusing the
expression graph is that if there are conditionals or while loops in
the program being evaluated, the expression grpah cannot be reused.  The
second drawback is that to evaluate the function and gradients
requires what is effectively an interpreted forward pass to be rerun.
A major advantage comes in Jacobian calculations, where all of the
expression derivatives computed in the initial expression graph
construction can be reused.

Stan is not ideally set up to exploit retaping because of the lazy
evaluation of partial derivatives.  For the lazy evaluation cases
involving expensive functions (multiplications, transcendentals,
iterative algorihtms, etc.), this would have to be carried out each
time, just as in the Jacobian case.

CppAD goes even further in allowing functions with static expression
graphs to be taped once and reused by gluing them together into larger
functions.%
\footnote{CppAD calls this checkpointing; see
  \url{http://www.coin-or.org/CppAD/Doc/checkpoint.htm} for details.}
An related approach to compiling small pieces of larger
functions is provided by the expression templates of Adept.

\subsubsection{Memory}

Stan is very conservative with memory usage compared to other systems.
We do not actually evaluate memory because we don't know how to do it.
We can analytically evaluate the amount of memory required.  In
continuous runs, Stan's underlying memory allocation is stored by
default and reused, so that there should not actually be any
underlying system memory thrashing other than for the Eigen and
standard library vectors, which manage their own memory in the C++
heap.

\subsubsection{Compile Time}

We also have not evaluated compile time.  Stan is relatively slow to
compile, especially for the probability functions and matrix
operations, because of its extensive use of templating.  But this cost
is only paid once in the system.  Once the code is compiled, it can be
reused.  The use case for which Stan is designed typically involves
tens of thousands or even millions of gradient calculations, for which
compilation is only performed once.

\subsubsection{Systems Still under Consideration}

The following systems are open source and provide operator
overloading, but the authors of this paper have not (yet) been able to
figure out how to install them and get a simple example working.
\begin{itemize}
\item AUTODIF (ADMB): \smallurl{http://admb-project.org}
\item OpenAD: \smallurl{http://www.mcs.anl.gov/OpenAD/}
\end{itemize}

\subsection{Functors to Differentiate}\label{functors-to-diff.section}

To make evaluations easy and to ensure the same code is being
evaluated for each system, a functional is provided for each
system being evaluated that allows it to compute gradients of a
functor.  

The functors to differentiate define the following method signature.
\begin{smallcode}
template <typename T>
T operator()(const Eigen::Matrix<T, Eigen::Dynamic, 1>& x) const;
\end{smallcode}
Each functor also defines a static name function used in displaying results.
\begin{smallcode}
static std::string name() const; 
\end{smallcode}
Each functor further defines a static function that fills in the
values to differentiate.
\begin{smallcode}
template <typename T>
static void fill(Eigen::Matrix<T, Eigen::Dynnamic, 1>& x);
\end{smallcode}
The \code{fill()} function is called once for each size before timing.

For example, the following functor sums the elements of its argument vector.
\begin{smallcode}
struct sum_fun {
  template <typename T>
  T operator()(const Eigen::Matrix<T, Eigen::Dynamic, 1>& x)
    const {

    T sum_x = 0;
    for (int i = 0; i < x.size(); ++i)
      sum_x += x(i);
    return sum;
  }

  static void fill(Eigen::VectorXd& x) const {
    for (int i = 0; i < x.size(); ++i)
      x(i) = i;
  }

  static std::string name() const {
    return "sum";
  }
};
\end{smallcode}

\subsection{Functionals for Differentiation}

The functionals to perform the differentiation of functors for each
system are defined as follows.

\subsubsection{Adept Gradient Functional}

The functional using Adept to calculate gradients of functors is
defined as follows.

\begin{smallcode}
template <typename F>
void adept_gradient(const F& f,
                    const Eigen::VectorXd& x,
                    double& fx,
                    Eigen::VectorXd& grad_fx) {
  Eigen::Matrix<adept::adouble, Eigen::Dynamic, 1> x_ad(x.size());
  for (int i = 0; i < x.size(); ++i)
    x_ad(i) = x(i);
  adept::active_stack()->new_recording();
  adept::adouble fx_ad = f(x_ad);
  fx = fx_ad.value();
  fx_ad.set_gradient(1.0);
  adept::active_stack()->compute_adjoint();
  grad_fx.resize(x.size());
  for (int i = 0; i < x.size(); ++i)
    grad_fx(i) = x_ad[i].get_gradient();
}
\end{smallcode}
A new tape is explicitly started using \code{new\_recording()}, the
top variable's adjoint is set to 1 using \code{set\_gradient()}, and
derivatives are propagated by calling \code{compute\_adjoint()}.  At
this point, the gradients can be read out of the automatic
diffeentiation variables.

\subsubsection{Adol-C Gradient Functional}

The functional using Adol-C to calculate gradients of functors is
defined as follows.

\begin{smallcode}
template <typename F>
void adolc_gradient(const F& f,
                    const Eigen::VectorXd& x,
                    double& fx,
                    Eigen::VectorXd& grad_fx) {
  grad_fx.resize(x.size());
  trace_on(1);
  Eigen::Matrix<adouble, Eigen::Dynamic, 1> x_ad(x.size());
  for (int n = 0; n < x.size(); ++n)
    x_ad(n) <<= x(n);
  adouble y = f(x_ad);
  y >>= fx;
  trace_off();
  gradient(1, x.size(), &x(0), &grad_fx(0));
}
\end{smallcode}
The Adol-C library signals a new tape recording with the
\code{trace\_on()} function.  The operator \Verb|<<=| is overloaded to
create new dependent variables, which are then used to compute the
result.  The result is written into value \code{fx} using the
\Verb|>>=| operator to signal the dependent variable and the recording
is turned off using \code{trace\_off()}.  Then a function
\code{gradient()} calculates the gradients from the recording.

\subsubsection{CppAD Gradient Functional}

The functional using CppAD to calculate gradients of functors is
defined as follows.

\begin{smallcode}
template <typename F>
void cppad_gradient(const F& f,
                    const Eigen::VectorXd& x,
                    double& fx,
                    Eigen::VectorXd& grad_fx) {
  Eigen::Matrix<CppAD::AD<double>, Eigen::Dynamic, 1>
    x_ad(x.size());
  for (int n = 0; n < x.size(); ++n)
    x_ad(n) = x[n];
  Eigen::Matrix<CppAD::AD<double>, Eigen::Dynamic, 1> y(1);
  CppAD::Independent(x_ad);
  y[0] = f(x_ad);
  CppAD::ADFun<double> g = CppAD::ADFun<double>(x_ad, y);
  fx = Value(y[0]);
  Eigen::VectorXd w(1);
  w(0) = 1.0;
  grad_fx =  g.Reverse(1, w);
}
\end{smallcode}
CppAD builds-in utilities for working directly with Eigen vectors.
CppAD requires a declaration of the independent (input) variables with
\code{Independent()}.  Then a function-like object is defined by
constructing a \code{CppAD::ADFun}.  The value is extracted using
\code{Value()} and gradients are calculated using the \code{Reverse}
method of the \code{ADFun} constructed, \code{g}.

\subsubsection{Sacado Gradient Functional}

The functional using Sacado to calculate gradients of functors is
defined as follows.
\begin{smallcode}
template <typename F>
void sacado_gradient(const F& f,
                     const Eigen::VectorXd& x,
                     double& fx,
                     Eigen::VectorXd& grad_fx) {
  Eigen::Matrix<Sacado::Rad::ADvar<double>, Eigen::Dynamic, 1> 
    x_ad(x.size());
  for (int n = 0; n < x.size(); ++n)
    x_ad(n) = x[n];
  fx = f(x_ad).val();
  Sacado::Rad::ADvar<double>::Gradcomp();

  grad_fx.resize(x.size());
  for (int n = 0; n < x.size(); ++n)
    grad_fx(n) = x_ad(n).adj();
}
\end{smallcode}
The execution and memory management of Sacado are very similar to
those of Stan.  Nothing is required to start recoding other than the
use of automatic differentiation variables (here
\code{ADvar<double>}).  Functors are applied as expected and values
extracted using the method \code{val()}. Then gradients are computed with a global
function call (here \code{Gradcomp()}).  This functional was
implemented without the \code{try}-\code{catch} logic for recovering
memory shown in the Stan version, because memory is managed by Sacado
inside the call to \code{Gradcomp()}.

\subsubsection{Stan Gradient Functional}

The functional using Stan to calculate gradients of functors was
defined in \refsection{functionals}.

\subsection{Test Harness}

A single file with the test harness code is provided.  The test itself
is run with the following function, which includes a template
parameter for the type of the functor being differentiated.
\begin{smallcode}
template <typename F>
inline void run_test() {
  adept::Stack adept_global_stack_;
  F f;
  std::string file_name = F::name() + "_eval.csv";
  std::fstream fs(file_name, std::fstream::out);
  print_results_header(fs);
  int max = 16 * 1024;
  for (int N = 1; N <= max; N *= 2) {
    std::cout << "N = " << N << std::endl;
    Eigen::VectorXd x(N);
    F::fill(x);
    time_gradients(f, x, fs);
  }
  fs.close();
}
\end{smallcode}
Adept requires a global stack to be allocated by calling its nullary
constructor, which is done at the very top of the \code{run\_test()}
function.  

The key feature here is that a \code{double}-based vector \code{x} is
defined of size \code{N}, starting at 1 and doubling through size
$2^{14}$ (16,384) to show how the speed varies as a function of
problem size; larger sizes are not provided because $2^{14}$ was
enough to establish the trends with larger data.  For each size
\code{N}, the functor's static \code{fill()} function is applied to
\code{x}, then the gradients are timed.
The routine \code{time\_gradients}, which is called for each size of
input, is defined as follows.
\begin{smallcode}
template <typename F>
inline void time_gradients(const F& f, const Eigen::VectorXd& x, 
                           std::ostream& os) {
  int N = x.size();
  Eigen::VectorXd grad_fx(N);
  double fx = 0;
  clock_t start;
  std::string f_name = F::name();
  double z = 0;

  z = 0;
  start = clock();
  for (int m = 0; m < NUM_CALLS; ++m) {
    adept_gradient(f, x, fx, grad_fx);
    z += fx;
  }
  print_result(start, F::name(), "adept", N, os);

  z = 0;
  start = clock();
  for (int m = 0; m < NUM_CALLS; ++m) {
    adolc_gradient(f, x, fx, grad_fx);
    z += fx;
  }
  print_result(start, F::name(), "adolc", N, os);

  z = 0;
  start = clock();
  for (int m = 0; m < NUM_CALLS; ++m) {
    cppad_gradient(f, x, fx, grad_fx);
    z += fx;
  }
  print_result(start, F::name(), "cppad", N, os);

  z = 0;
  start = clock();
  for (int m = 0; m < NUM_CALLS; ++m) {
    sacado_gradient(f, x, fx, grad_fx);
    z += fx;
  }
  print_result(start, F::name(), "sacado", N, os);

  z = 0;
  start = clock();
  for (int m = 0; m < NUM_CALLS; ++m) {
    stan::math::gradient(f, x, fx, grad_fx);
    z += fx;
  }
  print_result(start, F::name(), "stan", N, os);

  z = 0;
  start = clock();
  for (int m = 0; m < NUM_CALLS; ++m)
    z += f(x);
  print_result(start, F::name(), "double", N, os);
}
\end{smallcode}
This function defines the necessary local variables for timing
and printing results to a file output stream, prints the header for
the output.  Then t times each system's functional call to the functor to
differentiate and prints the results.  

Timing is performed using the \code{ctime} library \code{clock()}
function, with 100,000 (or 10,000) repeated calls to each automatic differentiation
system with the only user programs executing being the Mac OS X
Terminal (version 2.5.3).

A variable is defined to pull the value out to ensure that the entire
function is not compiled away because results are not used.  The final
call applies the functor to a vector of \code{double} values without
computing gradients to provide a baseline measurement of function
execution time.%
\footnote{This doesn't seem to make a difference with any compilers;
  each function was also run in such a way to print the result to
  ensure that each system was properly configured to compute
  gradients.}

The Adol-C and CppAD systems have alternative versions that do not
recompute the ``tape'' for subsequent function calls, but this is not
the use case we are evaluating.  There is some savings for doing this
because Adol-C in particular is relatively slow during the taping
stage;  see \refsection{retaping}.

\subsection{Test Hardware and Compilation}

This section provides actual performance numbers for the various
systems using 
\begin{itemize}
\item Hardware: Macbook Pro computer (Retina, Mid 2012), with a 2.3
GHz Intel Core i7 with 16 GB of 1600 MHz DDR3 memory
\item Compiler: clang version 3.7.0 (trunk 233481)
\item Compiler Flags: \\
{\small \Verb|-O3 -DNDEBUG -DEIGEN_NO_DEBUG -DADEPT_STACK_THREAD_UNSAFE|}
\item Libraries:  Eigen 3.2.4, Boost 1.55.0
\end{itemize}
The compiler flags turn on optimization level 3 and turn off system
and Eigen-level debugging.  They also put Adept into thread-unsafe
mode for its stack, which matches the way Stan runs; like Adept, Stan
can be run in thread safe mode at roughly a 20\% penalty in
performance by using thread-local instances of the stack representing
the expression graph.

The makefile included with the code for this paper also includes the
ability to test with GCC version 3.9.  Results were similar enough for
GCC that they are not included in this paper.

\subsection{Basic Function and Operator Evaluations}

This section provides evaluations of basic operators and functions
defined as part of the C++ language or as part of the standard
\code{cmath} library.  The next section considers evaluations of
Stan-specific functions and optimized alternatives to basic functions.

\subsubsection{Sums and Products}

The simplest functions just add or multiply a sequence of numbers.
The sum functor was defined in \refsection{functors-to-diff}.  The
remaining functors will be shown without their \code{name()}
methods.  For products, the following functor is used.
\begin{smallcode}
struct product_fun {
  template <typename T>
  T operator()(const Eigen::Matrix<T, Eigen::Dynamic, 1>& x)
    const {

    T product_x = 1;
    for (int i = 0; i < x.size(); ++i)
      product_x *= x(i);
    return product_x;
  }

  static void fill(Eigen::VectorXd& x) {
    for (int i = 0; i < x.size(); ++i) 
      x(i) = pow(1e10, 1.0 / x.size()); 
  }
};
\end{smallcode}
The fill function ensures that the total product is $10^{10}$.  

\begin{figure}
\vspace*{-6pt}
\begin{center}
\includegraphics[width=2.5in]{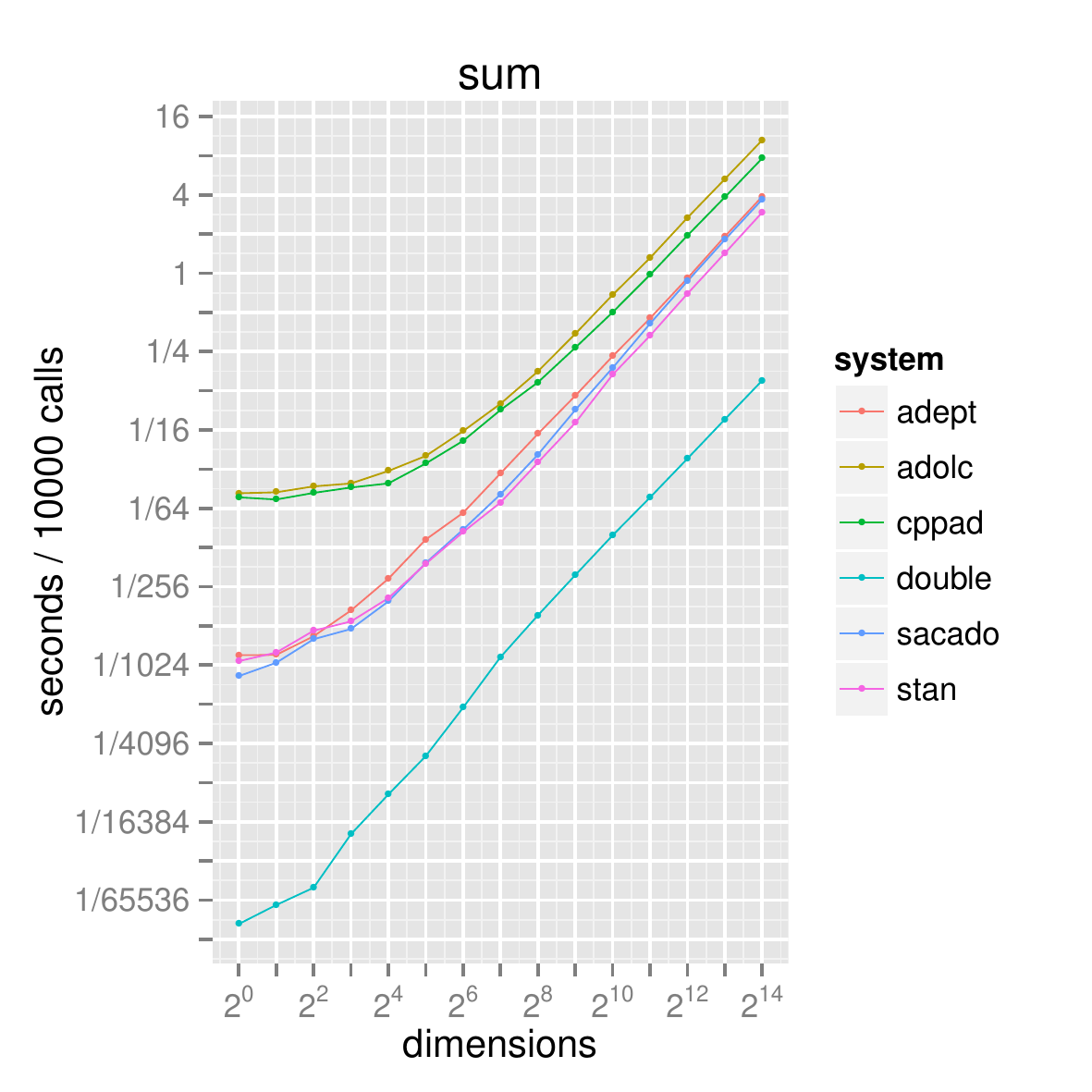}\includegraphics[width=2.5in]{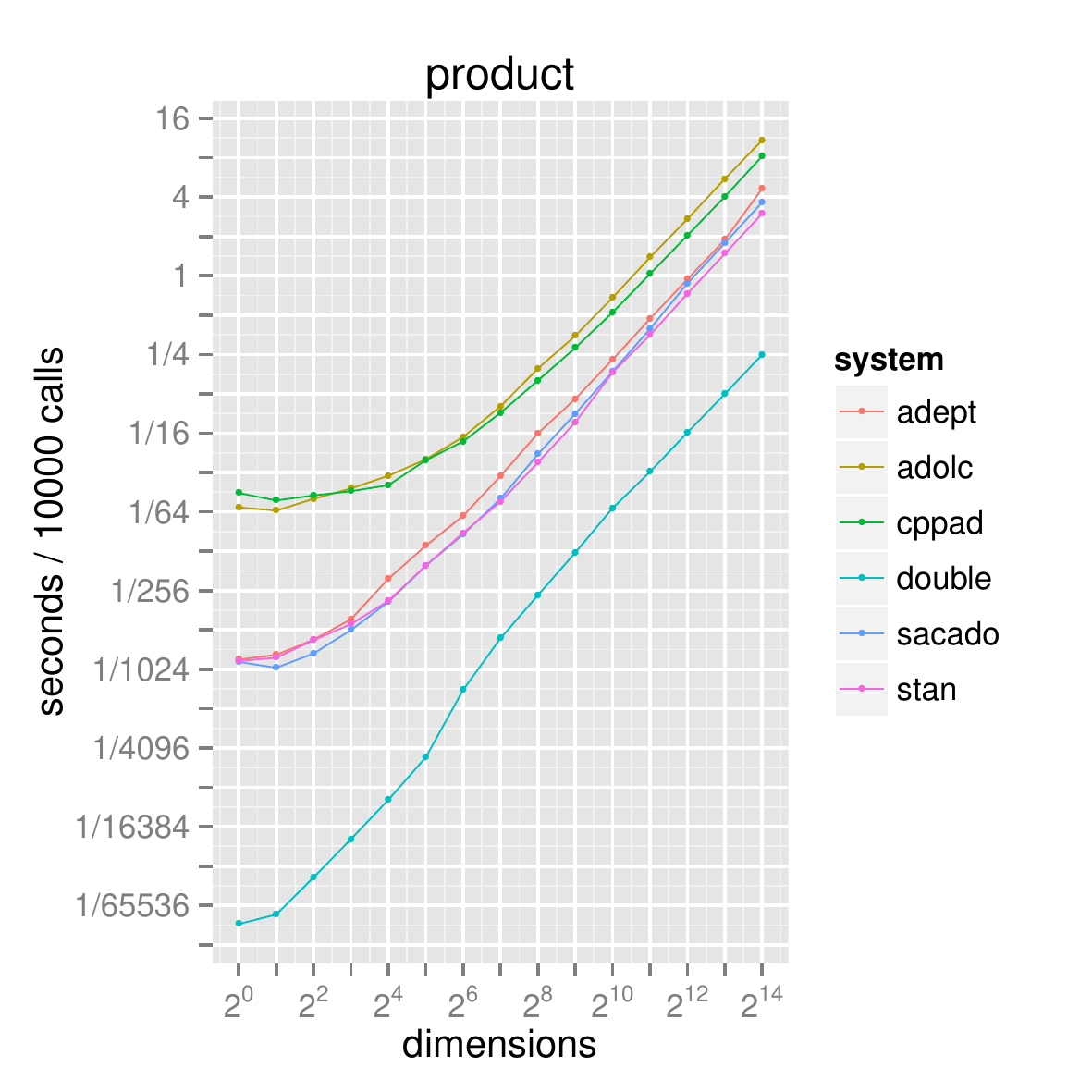}
\includegraphics[width=2.5in]{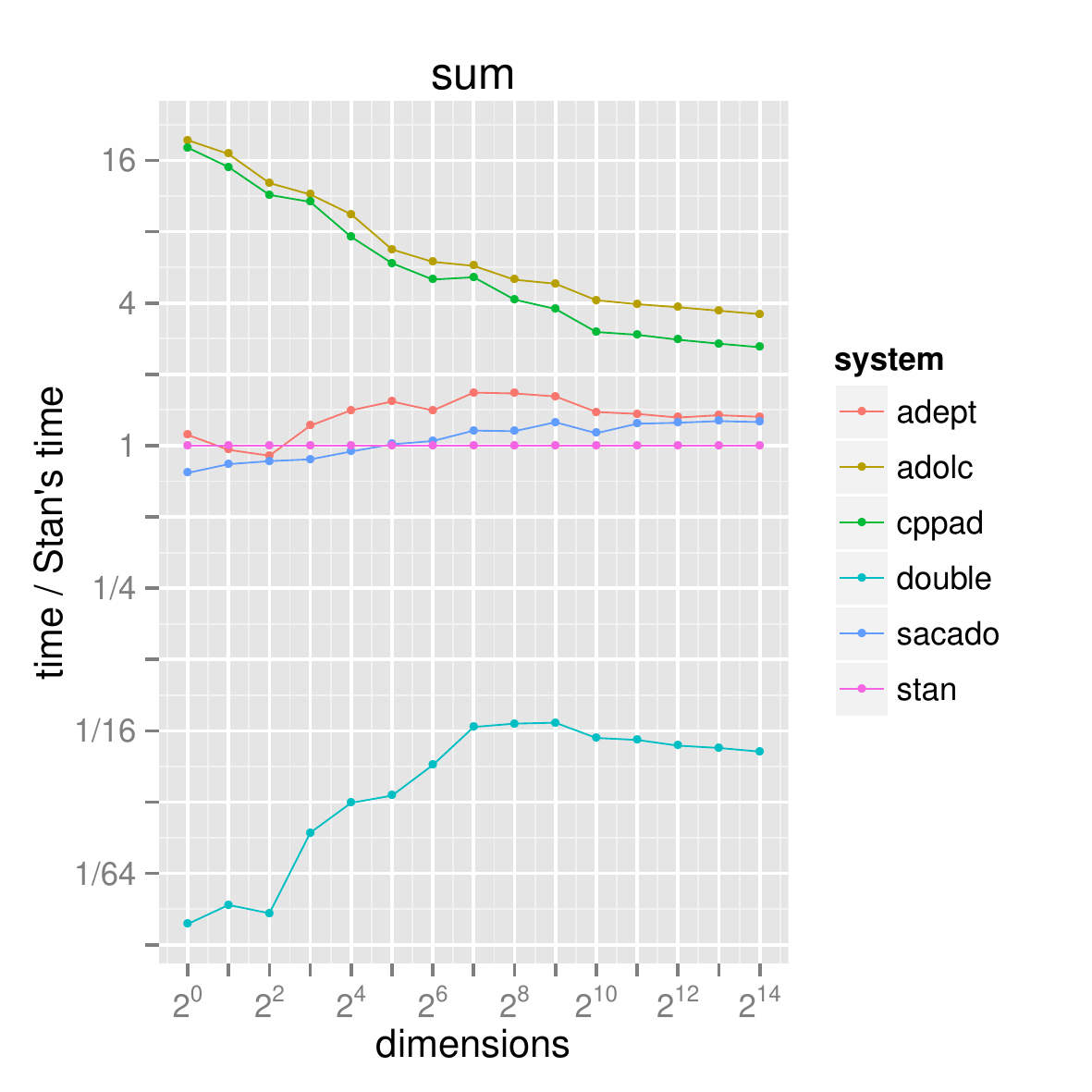}\includegraphics[width=2.5in]{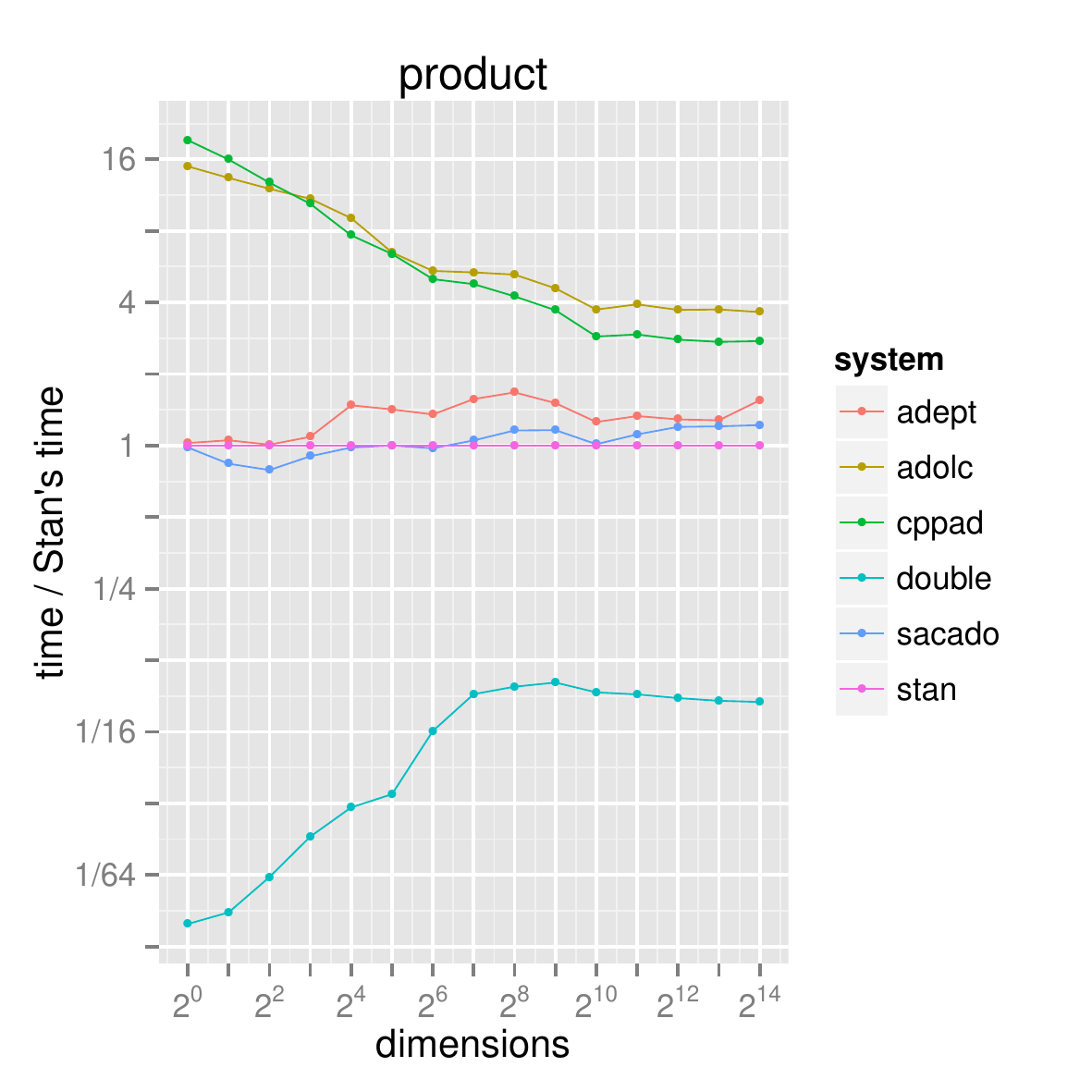}
\end{center}
\vspace*{-12pt}
\caption{\small\it Evaluation of sums (left) and products (right).
  The top plots provide a measurement of time per 100,000 gradient
  calculations.  The bottom plots shows the speed of each
  system relative to Stan's time.}\label{sum-product-eval.figure}
\end{figure}
The evaluation results for sums and products
are plotted in \reffigure{sum-product-eval}, with actual time taken
shown as well as time relative to Stan's time.%
\footnote{Given the difficulty in reading the actual time plots, only
  relative plots will be shown going forward.}
For sums and products, the time taken for \code{double}-based function
evaluation versus evaluation with gradient calculations ranges from a
factor of 100 for a handful of dimensions to roughly a factor of 10 to
15 for evaluations with 200 or more dimensions.  Sacado is the fastest
syste for problems with fewer than 8 (sums) or 16 (products)
dimensions, Stan is faster for larger problems.  It is clear from the
plots thtat both CppAD and Adol-C have large constant overheads that
make them relatively slow for smaller problems; they are still slower
than Sacado, Adept, and Stan for problems of 1000 variables or more
where their relative speed seems to stabilize.  These overall results
are fairly consistent through the evaluations.

\subsubsection{Power Function}

The following functor is evaluated to test the built-in \code{pow()}
function;  because of the structure, a long chain of derivatives is
created and both the mantissa and exponent are differentiated.  
\begin{smallcode}
struct powers_fun {
  static void fill(Eigen::VectorXd& x) {
    for (int i = 0; i < x.size(); ++i)
      x(i) = i 
  }

  template <typename T>
  T operator()(const Eigen::Matrix<T, Eigen::Dynamic, 1>& x) 
    const {

    T result = 10.0;
    for (int i = 1; i < x.size(); ++i)
      result = pow(result, x(i));
    return result;
  }
};
\end{smallcode}
\begin{figure}
\vspace*{-6pt}
\begin{center}
\includegraphics[width=2.5in]{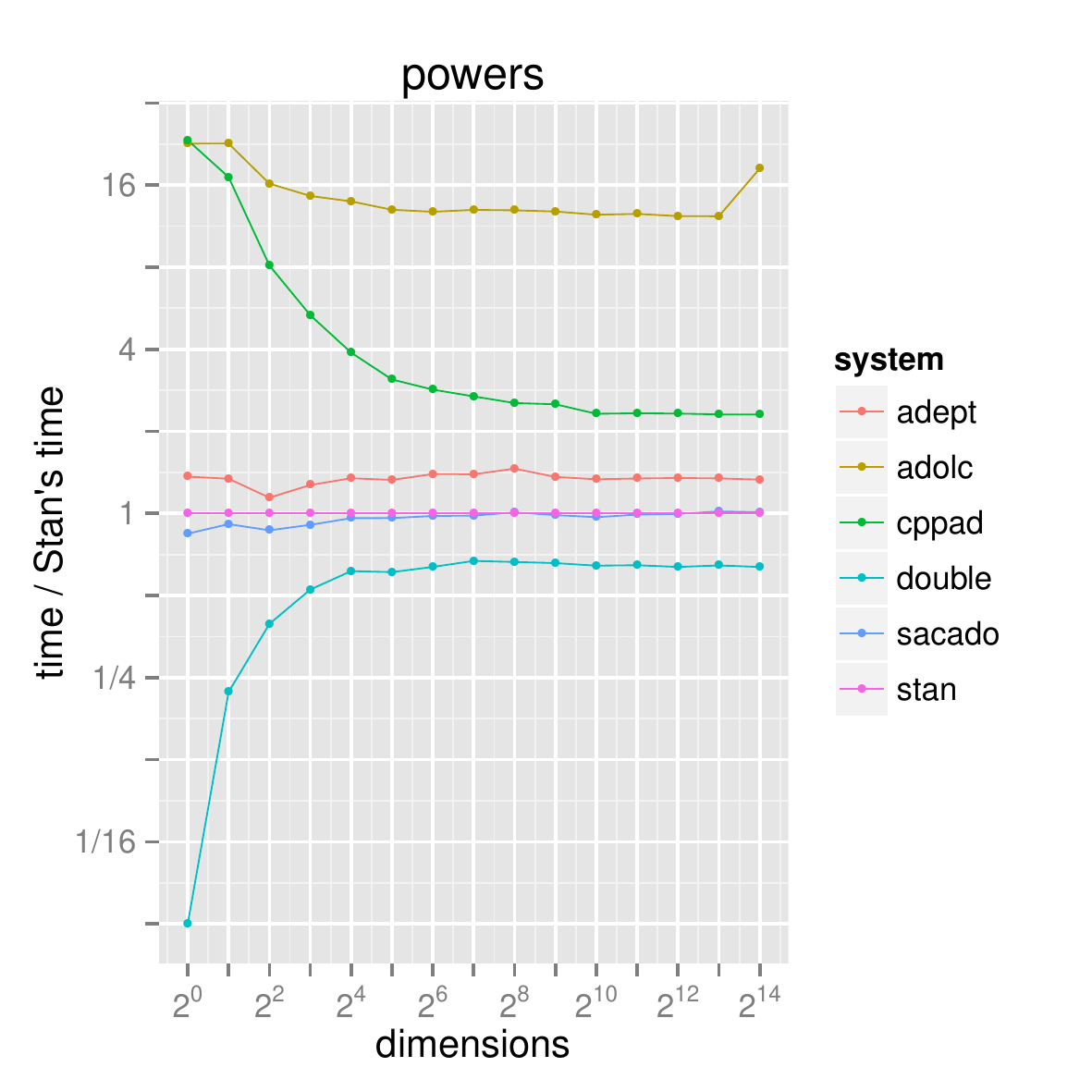}%
\includegraphics[width=2.5in]{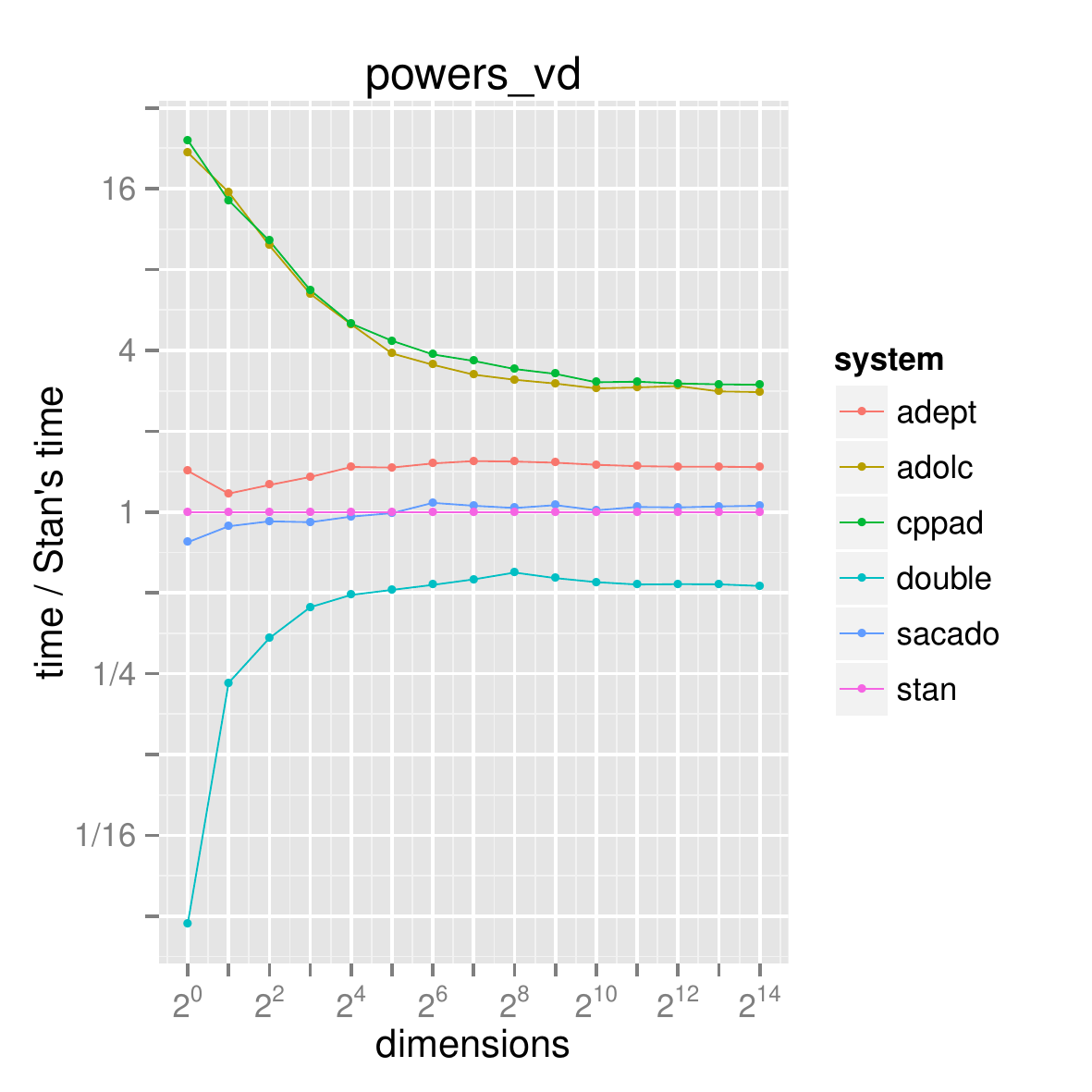}
\end{center}
\vspace*{-12pt}
\caption{\small\it Relative evaluation of power
  function.  (Left) Plot is for a function with repeated
  application of \code{pow} to two variable arguments.  (Right) 
  The sum of a sequence of \code{pow} applications to a variable with
  a fixed exponent.}\label{powers-eval.figure}
\end{figure}
The evaluation is shown in \reffigure{powers-eval}.  Here, Sacado and
Stan have almost identical performance, which is about 40\% faster
than Adept and much faster than CppAD or Adol-C.  Adol-C seems to
particularly struggle with this example, being roughly 15 times slower
than Stan.  Because of the time taken for the \code{double}-based
calculation of \code{pow()} for fractional powers, for problems of
more than 8 dimensions, the gradients are calculated in only about
50\% more time than the \code{double}-based function itself.

The second evaluation of powers is for a fixed exponent with a
summation of the results, as would be found in an iterative
algorithm.  
\begin{smallcode}
struct powers_fun {
  template <typename T>
  T operator()(const Eigen::Matrix<T, Eigen::Dynamic, 1>& x)
    const {

    T result = 10.0;
    for (int i = 1; i < x.size(); ++i)
      result = pow(result, x(i));
    return result;
  }

  static void fill(Eigen::VectorXd& x) {
    for (int i = 0; i < x.size(); ++i)
      x(i) = i 
  }
};
\end{smallcode}
The evaluation of this gradient calculation is shown on the right side
of \reffigure{powers-eval}.  The relative speeds are similar other
than for Adol-C, which is much faster in this case.  In this case, the
gradient calculations in Stan take about twice as long as the function
itself.

\subsubsection{Logarithm and Exponentiation Functions}

\begin{figure}
\vspace*{-6pt}
\begin{center}
\includegraphics[width=2.5in]{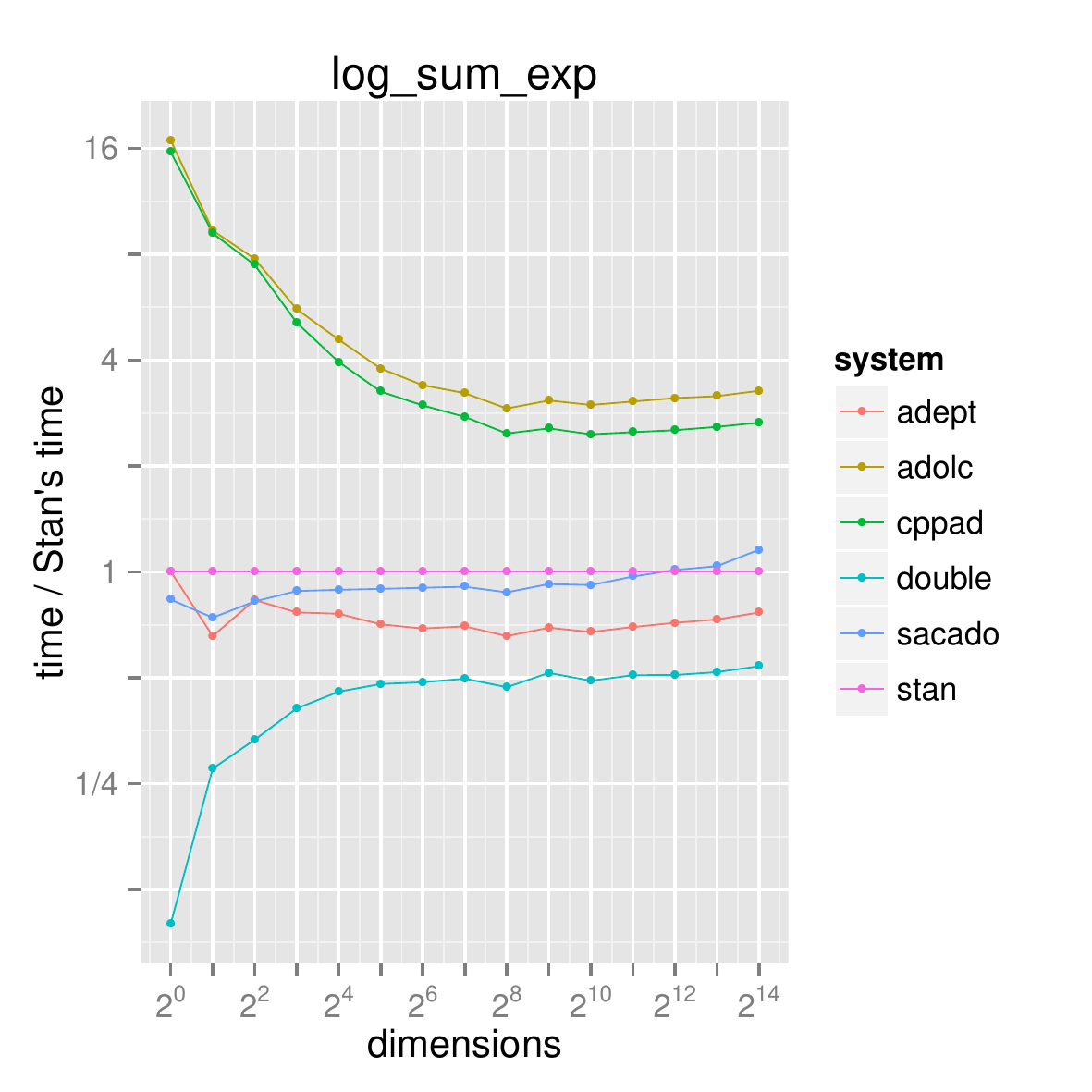}%
\includegraphics[width=2.5in]{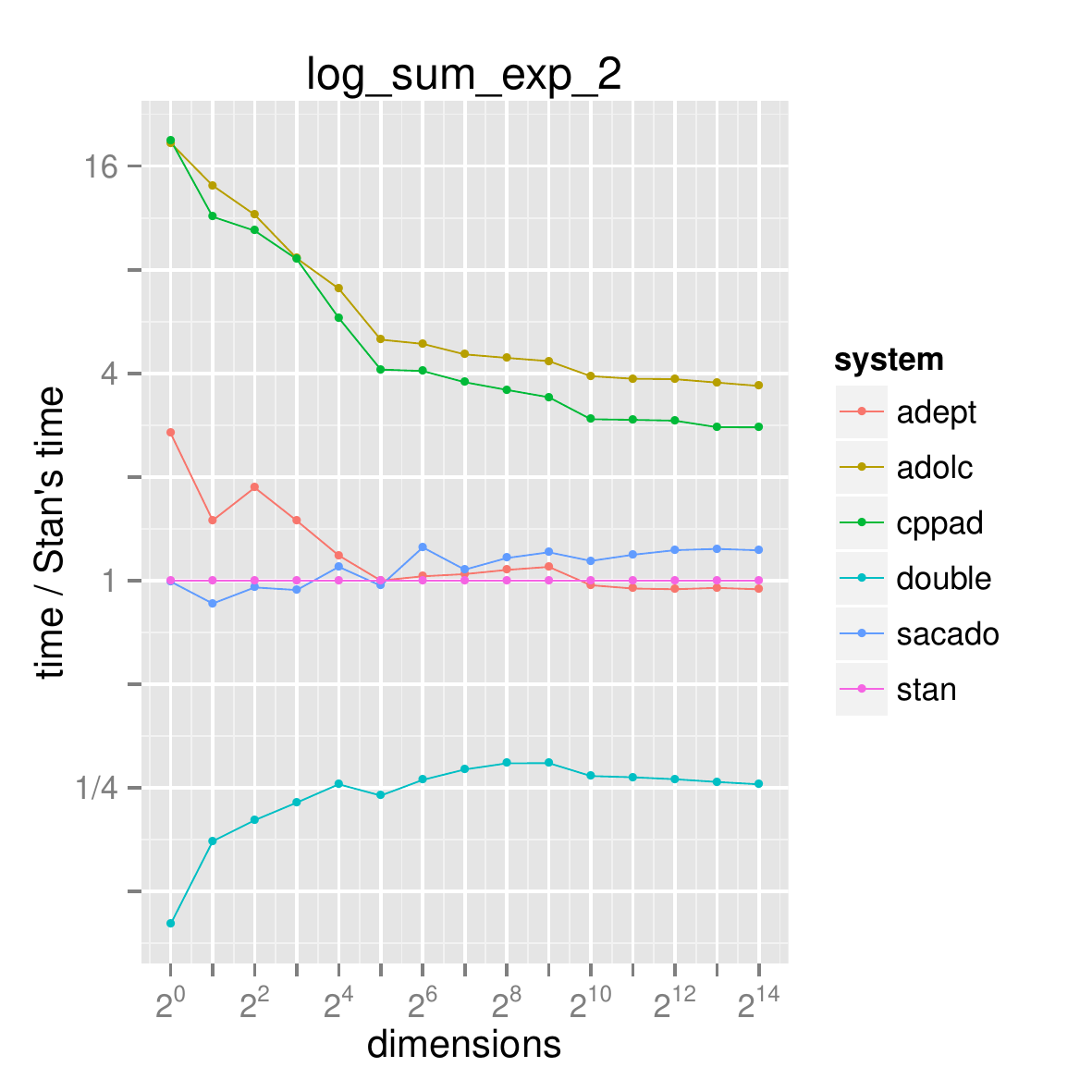}
\end{center}
\vspace*{-12pt}
\caption{\small\it Relative evaluation of the log sum of exponents
  function written recursively (left) and directy
  (right).}\label{log-sum-exp-eval.figure}
\end{figure}
The log sum of exponents function is a commonly used function in
numerical computing to avoid overflow when adding two numbers on a log
scale.  Here, a simpler form of it is defined that does not attempt to
avoid overflow, though the functor is set up so that results will not
overflow.  
\begin{smallcode}
struct log_sum_exp_fun {
  template <typename T>
  T operator()(const Eigen::Matrix<T, Eigen::Dynamic, 1>& x)
    const {

    T total = 0.0;
    for (int i = 0; i < x.size(); ++i)
      total = log(exp(total) + exp(x(i)));
    return total;
  }

  static void fill(Eigen::VectorXd& x) {
    for (int i = 0; i < x.size(); ++i) 
      x(i) = i / static_cast<double>(x.size());
  }
};
\end{smallcode}
Results are shown in \reffigure{log-sum-exp-eval}.  For this operation, the
expression templates used in Adept prove their worth and it is about
40\% faster than Stan.  Sacado is a bit faster than Stan, and again,
Adol-C and CppAd are more than twice as slow.  Because each
calculation is so slow on \code{double} values, for problems of more
than eight dimensions, the gradients are calculated in about double
the time it takes to evaluate the function itself on \code{double}
values.  

To see that it's Adept's expression templates that make the difference
and to illustrate how important the way a function is formulated is,
consider this alternative implementation of the same log sum of
exponents function.
\begin{smallcode}
struct log_sum_exp_2_fun {
  template <typename T>
  T operator()(const Eigen::Matrix<T, Eigen::Dynamic, 1>& x)
    const {

    T total = 0.0;
    for (int i = 0; i < x.size(); ++i)
      total += exp(x(i));
    return log(total);
  }
  ...
}
\end{smallcode}
The result is shown in \reffigure{log-sum-exp-eval}; with this
implementation, performance of Stan and Adept are similar.  The second
(direct) implementation is also much faster for \code{double}-based
function evaluation because of fewer applications of the expensive
\code{log} and \code{exp} functions.

\subsubsection{Matrix Products}

This section provides evaluations of taking gradients of matrix
products.  The first two evaluations are for a naive looping
implementation.  The first evaluation involves differentiating both
matrices in the product.
\begin{smallcode}
struct matrix_product_fun {
  template <typename T>
  T operator()(const Eigen::Matrix<T, Eigen::Dynamic, 1>& x) 
    const {

    using Eigen::Matrix;
    using Eigen::Dynamic;
    using Eigen::Map;
    int N = static_cast<int>(std::sqrt(x.size() / 2));
    Matrix<T, Dynamic, Dynamic> a(N,N);
    Matrix<T, Dynamic, Dynamic> b(N,N);
    int i = 0;
    for (int m = 0; m < N; ++m) {
      for (int n = 0; n < N; ++n) {
        a(m,n) = x(i++);
        b(m,n) = x(i++);
      }
    }
    Matrix<T, Dynamic, Dynamic> ab(N,N);
    for (int m = 0; m < N; ++m) {
      for (int n = 0; n < N; ++n) {
        ab(m,n) = 0;
        for (int k = 0; k < N; ++k)
          ab(m,n) += a(m,k) * b(k,n);
      }
    }
    T sum = 0;
    for (int m = 0; m < N; ++m)
      for (int n = 0; n < N; ++n)
        sum += ab(m,n);
    return sum;
  }

  static void fill(Eigen::VectorXd& x) {
    int N = static_cast<int>(sqrt(x.size() / 2));
    if (N < 1) N = 1;
    x.resize(N * N * 2);
    for (int i = 0; i < x.size(); ++i)
      x(i) = static_cast<double>(i + 1) / (x.size() + 1);
  }
};
\end{smallcode}
The \code{fill()} implementation is different than what came before,
because the evaluation requires two square matrices.  Thus the vector
to be filled is resized to the largest vector of variables that can
fill two matrices.  For a case involving $2^{12}$ variables, the
matrices being multiplied $45 \times 45$, because $45 = \lfloor
\sqrt{2^{12} / 2} \rfloor$.

The implementation of matrix product itself is just straightforward
looping, first to compute the matrix product and assign it to the
matrix \code{ab}, then to reduce the matrix to a single value through
summation.  The resulting expression graph has much higher
connectivity, with each variable being repeated $N$ times for 
an $N \times N$ matrix product.  

For the product of an $N \times N$ matrix, there are $N^2$ parameters
if one matrix has \code{double} scalar values and $2N^2$ if they are
both automatic differentiation variables.  From the definition of
matrix products, it is clear that there are a total of $N^3$ products
and $N^2$ sums required to multiply two $N \times N$ matrices.  Thus
multiplying two $45 \times 45$ matrices requires over 90,000
products to be calculated, each of which involves a further
multiplication during automatic differentiation.  

When considering the evaluations, the number of dimensions is the
number of parameters---the matrices have dimensionality equal to the
square root of that size. 

\begin{figure}
\vspace*{-6pt}
\begin{center}
\includegraphics[width=2.5in]{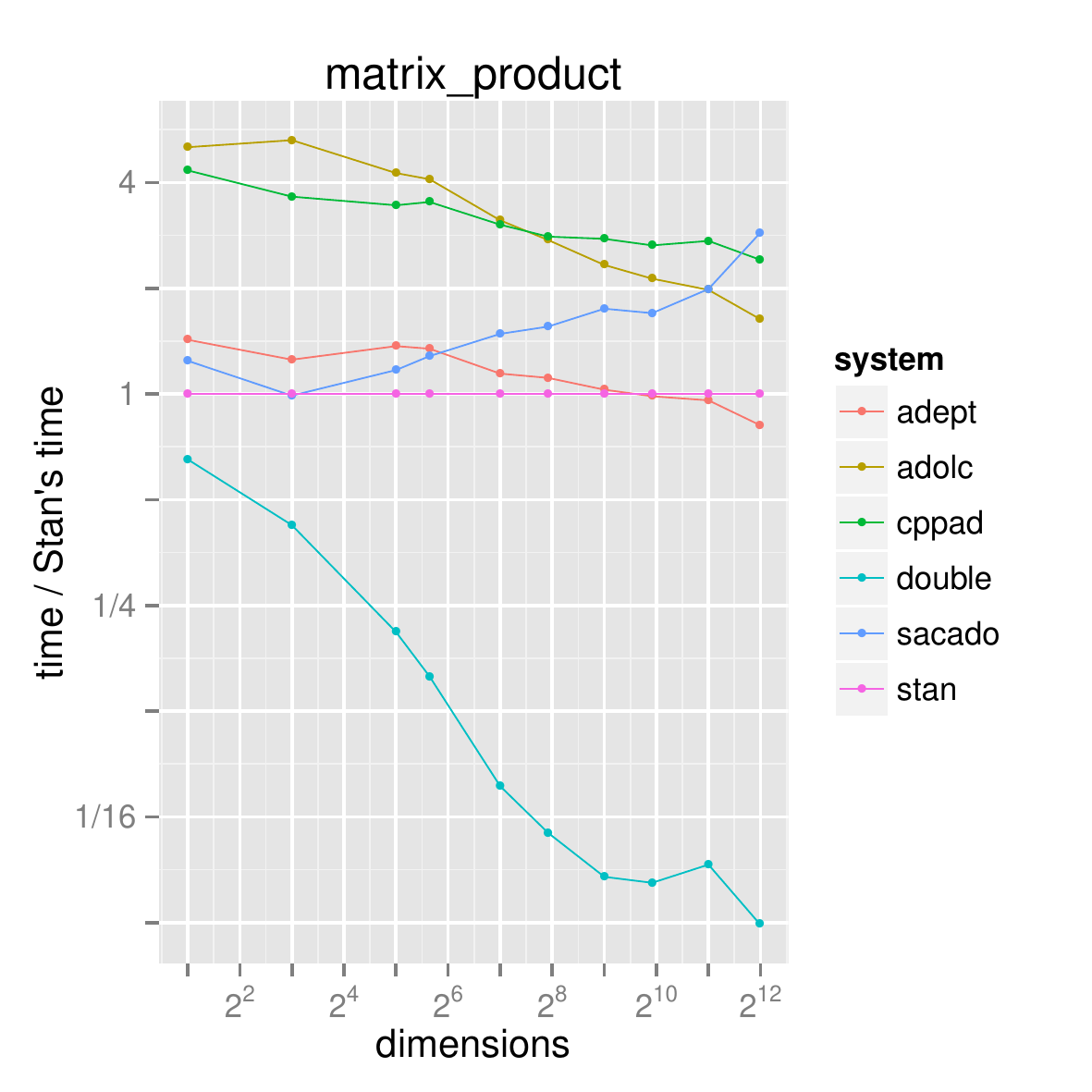}%
\includegraphics[width=2.5in]{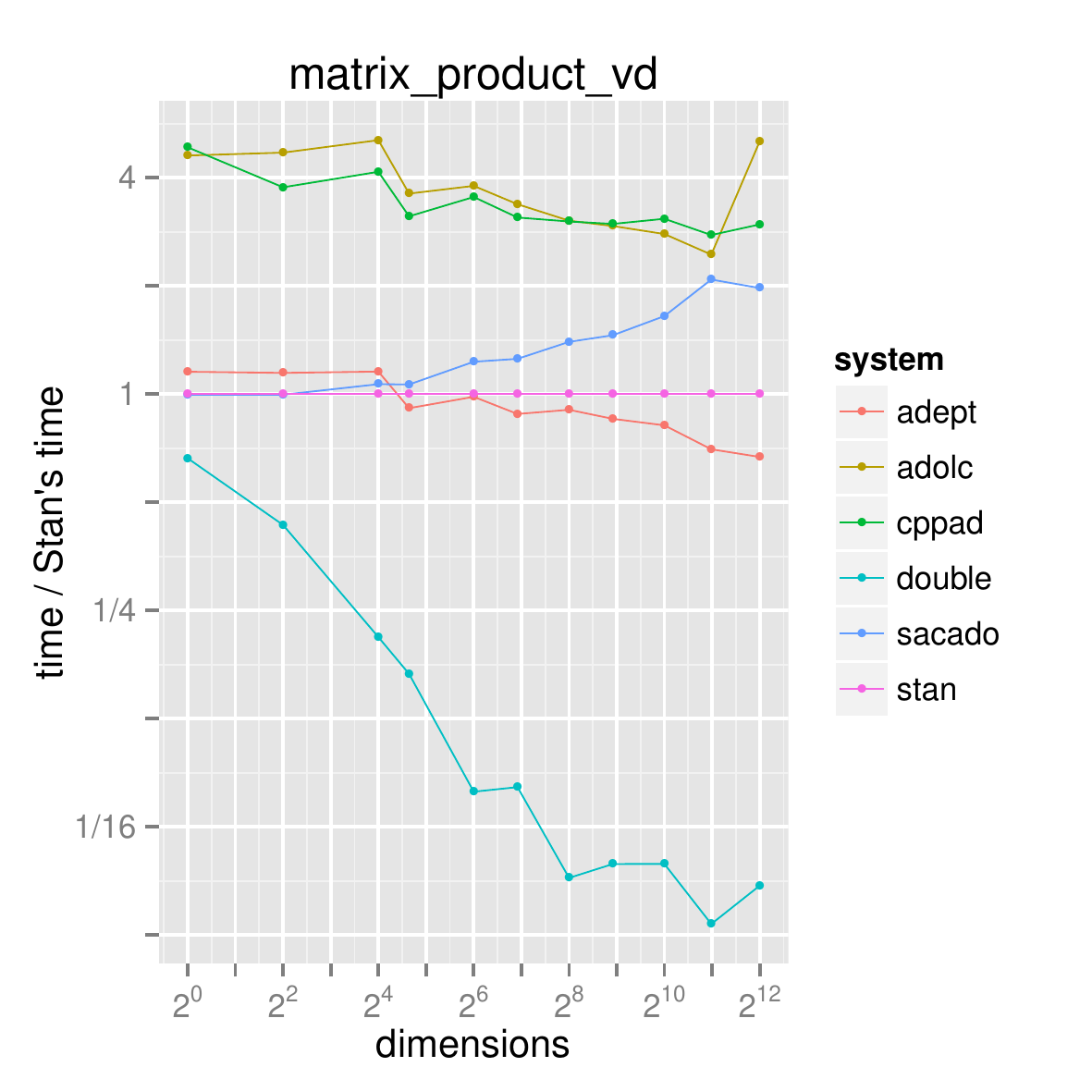}
\end{center}
\vspace*{-12pt}
\caption{\small\it Relative evaluation of naive loop-based matrix
  products with derivatives of both matrix (left) and just the first
  matrix (right).  The number of dimensions on the $x$ axis is the total
  number of entries in the matrix; the number of subexpressions
  evaluated grows proportionally to the square root of the number of
  entries.}\label{matrix-product-eval.figure}
\end{figure}
The relative timing results are shown in
\reffigure{matrix-product-eval}.  Because the matrices are resized to
accomodate two square matrices, the points do not fall exactly on even
powers of two.  Also, because iterating through powers of two sizes,
the largest pair of matrices is the same for successive lower orders,
so there are duplicated evaluations, which show some variation due to
the relatively small number of loops evaluated.  It can be seen that
Stan is faster for small matrices, with Adept being faster for larger
matrices.  Sacado fares relatively poorly compared to the less
connected evaluations.  

The following functor is for the evaluation of matrix products with
gradients taken only of the first matrix.
\begin{smallcode}
struct matrix_product_vd_fun {
  template <typename T>
  T operator()(const Eigen::Matrix<T, Eigen::Dynamic, 1>& x)
    const {

    using Eigen::Matrix;
    using Eigen::Dynamic;
    using Eigen::Map;
    int N = static_cast<int>(std::sqrt(x.size()));
    Matrix<T, Dynamic, Dynamic> a(N,N);
    Matrix<double, Dynamic, Dynamic> b(N,N);
    int i = 0;
    for (int m = 0; m < N; ++m) {
      for (int n = 0; n < N; ++n) {
        a(m,n) = x(i++);
        b(m,n) = 1.02;
      }
    }
    Matrix<T, Dynamic, Dynamic> ab(N,N);
    for (int m = 0; m < N; ++m) {
      for (int n = 0; n < N; ++n) {
        ab(m,n) = 0;
        for (int k = 0; k < N; ++k)
          ab(m,n) += a(m,k) * b(k,n);
      }
    }
    T sum = 0;
    for (int m = 0; m < N; ++m)
      for (int n = 0; n < N; ++n)
        sum += ab(m,n);
    return sum;
  }

  static void fill(Eigen::VectorXd& x) {
    int N = static_cast<int>(sqrt(x.size()));
    if (N < 1) N = 1;
    x.resize(N * N);
    for (int i = 0; i < x.size(); ++i)
      x(i) = static_cast<double>(i + 1) / (x.size() + 1);
  }
};
\end{smallcode}
Here, the resizing is to a single square matrix, so the case of $2^{12}$
variables involves two $64 \times 64$ matrices.  As with
differentiating both sides, Stan is faster for smaller matrices, with
Adept being faster for larger matrices.  

To demonstrate the utility of a less naive implementation of matrix
product, consider the results of using Eigen's built-in matrix
product, which is tuned to maximize memory locality.  The following
functor is used for the evaluation; the \code{fill()} function is the
same as the first matrix example for taking gradients of both
components.
\begin{smallcode}
struct matrix_product_eigen_fun {
  template <typename T>
  T operator()(const Eigen::Matrix<T, Eigen::Dynamic, 1>& x)
    const {

    using Eigen::Matrix;
    using Eigen::Dynamic;
    int N = static_cast<int>(std::sqrt(x.size() / 2));
    Matrix<T, Dynamic, Dynamic> a(N,N);
    Matrix<T, Dynamic, Dynamic> b(N,N);
    int i = 0;
    for (int m = 0; m < N; ++m) {
      for (int n = 0; n < N; ++n) {
        a(m,n) = x(i++);
        b(m,n) = x(i++);
      }
    }
    return (a * b).sum();
  }
};
\end{smallcode}
Both CppAD and Stan specialize \code{std::numeric\_limits}, which is
used by Eigen to calculate memory sizes and optimize memory locality
in matrix product calculations.  The difference in relative speed is
striking, as shown in \reffigure{matrix-product-eigen-eval}.
\begin{figure}
\vspace*{-6pt}
\begin{center}
\includegraphics[width=2.5in]{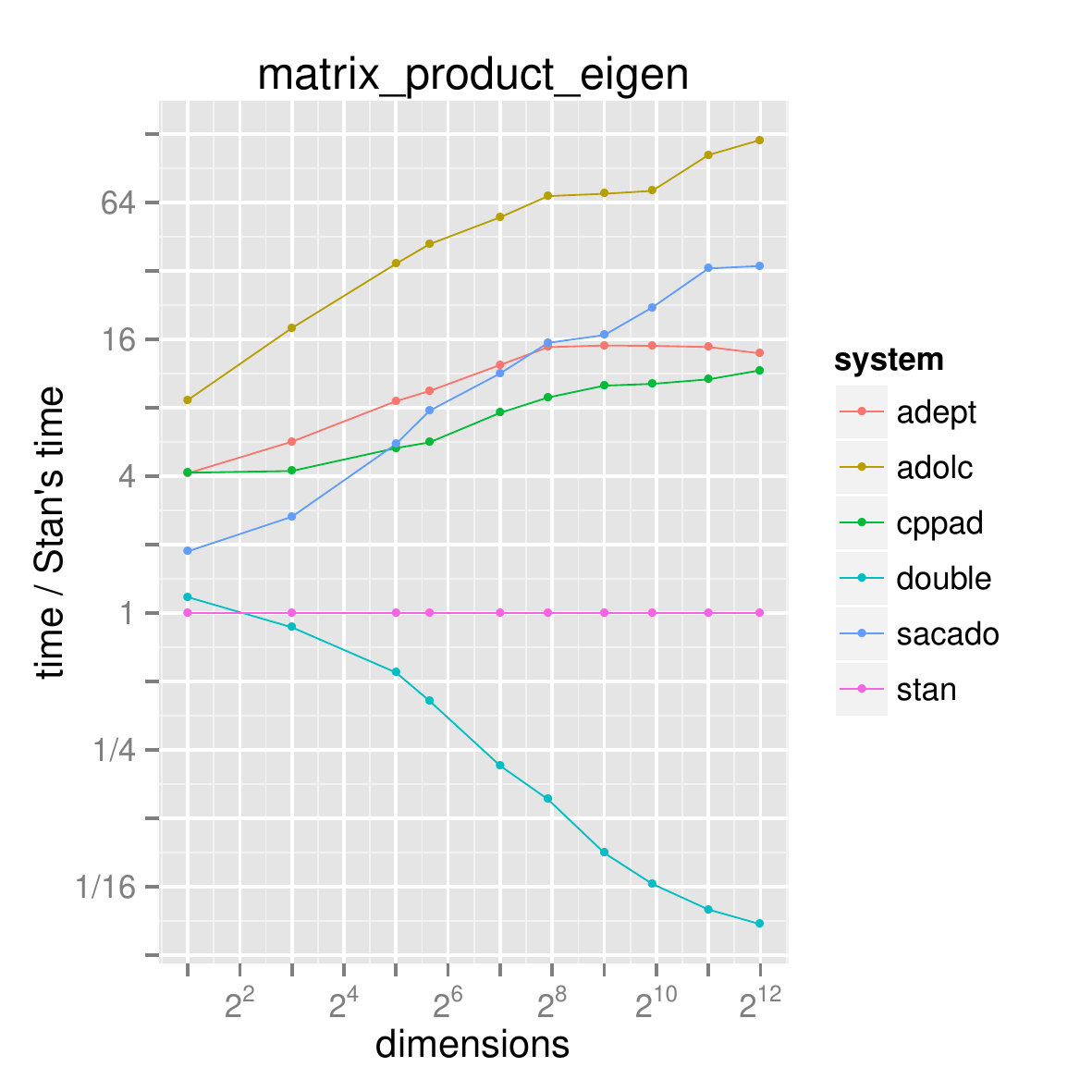}%
\includegraphics[width=2.5in]{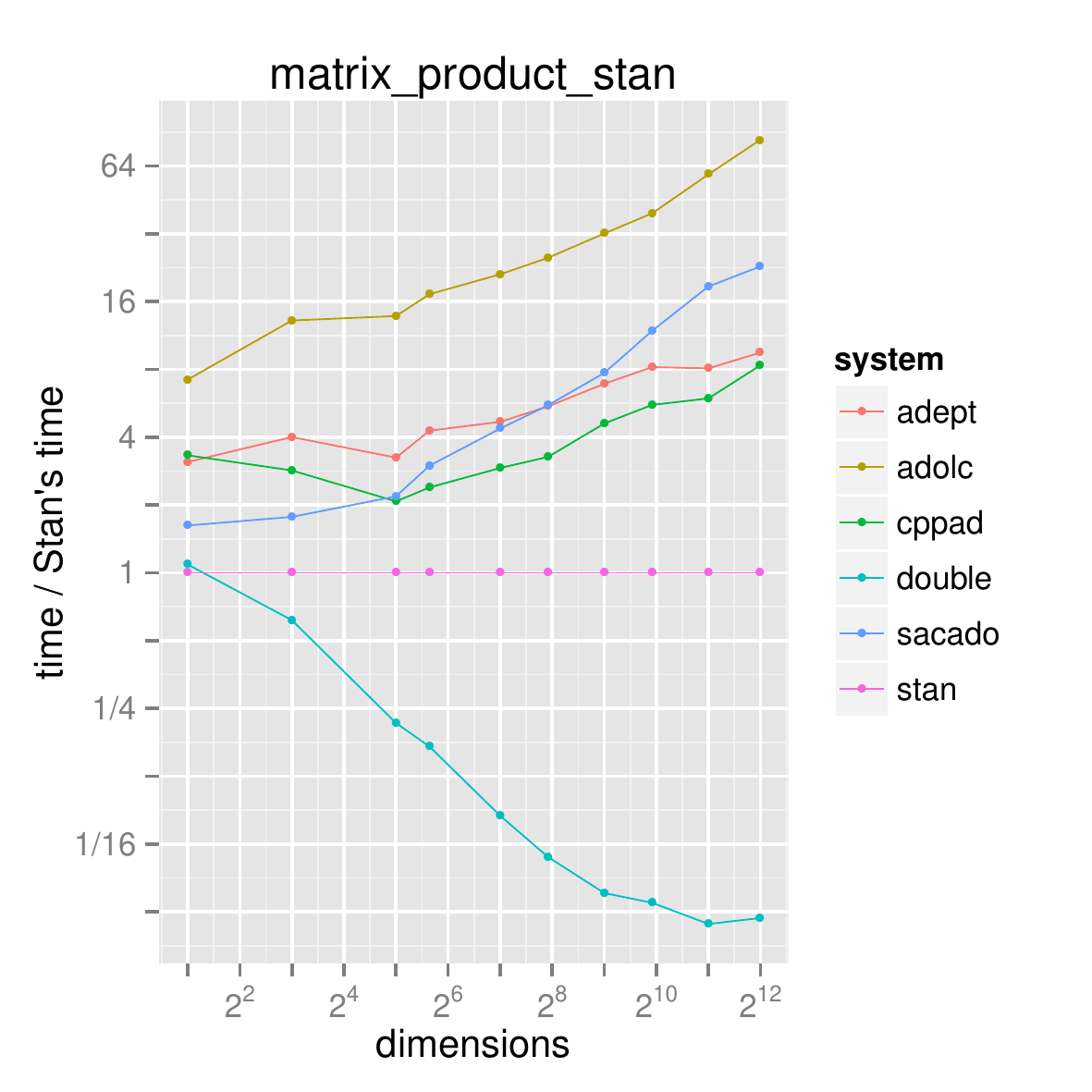}
\end{center}
\vspace*{-12pt}
\caption{\small\it Evaluation of matrix products using Eigen's
  built-in \code{operator*()} and \code{sum()} functions (left), and
  with Stan's built-in \code{multiply()} and \code{sum()} functions
  (right).  The number of dimensions on the $x$ axis is the total
  number of entries in the matrix; the number of subexpressions
  evaluated grows proportionally to the square root of the number of
  entries.}\label{matrix-product-eigen-eval.figure}
\end{figure}
Despite the rather large number of operations required for gradients,
relative speed compared to a pure \code{double}-based implementation
is better by a factor of 50\%.

The next evaluation replaces Eigen's \code{operator*} and \code{sum}
methods with customized versions in Stan.  Specifically, for Stan the
final implementation is done with
\begin{smallcode}
stan::math::sum(stan::math::multiply(a, b));
\end{smallcode}
rather than with the built-in Eigen operations, as in the previous
evaluation: 
\begin{smallcode}
(a * b).sum();
\end{smallcode}
Eigen's direct implementation is about 50\% faster, but will consume
roughly twice as much memory as Stan's custom dot-products and
summation, which rely on custom \code{vari} implementations.

\subsubsection{Normal Log Density}

The next function is closer to the applications for which Stan was
designed, being the log of the normal density function.  The functor
to be evaluated is the following.
\begin{smallcode}
struct normal_log_density_fun {
  template <typename T>
  T operator()(const Eigen::Matrix<T, Eigen::Dynamic, 1>& x)
    const {

    T mu = -0.56;
    T sigma = 1.37;
    T lp = 0;
    for (int i = 0; i < x.size(); ++i)
      lp += normal_log_density(x(i), mu, sigma);
    return lp;
  }

  static void fill(Eigen::VectorXd& x) {
    for (int i = 0; i < x.size(); ++i) 
      x(i) = static_cast<double>(i + 1 - x.size()/2) / (x.size() + 1);
  }
};
\end{smallcode}
The density function itself is defined up to an additive
constant ($-\log \sqrt{2 \pi}$) by the following function.
\begin{smallcode}
template <typename T>
inline 
T normal_log_density(const T& y, const T& mu, const T& sigma) {
  T z = (y - mu) / sigma;
  return -log(sigma) - 0.5 * z * z;
}
\end{smallcode}
\begin{figure}
\vspace*{-6pt}
\begin{center}
\includegraphics[width=2.5in]{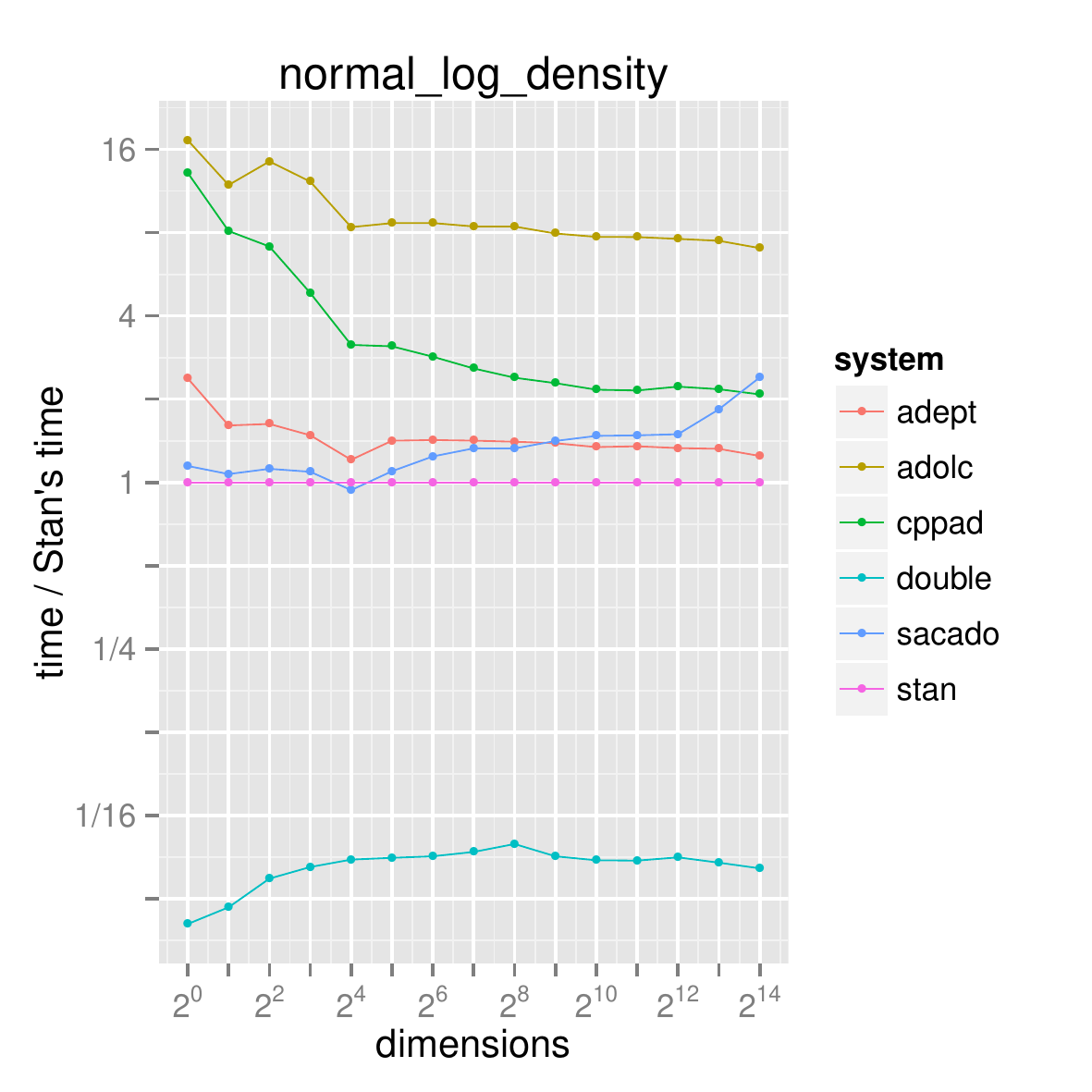}%
\includegraphics[width=2.5in]{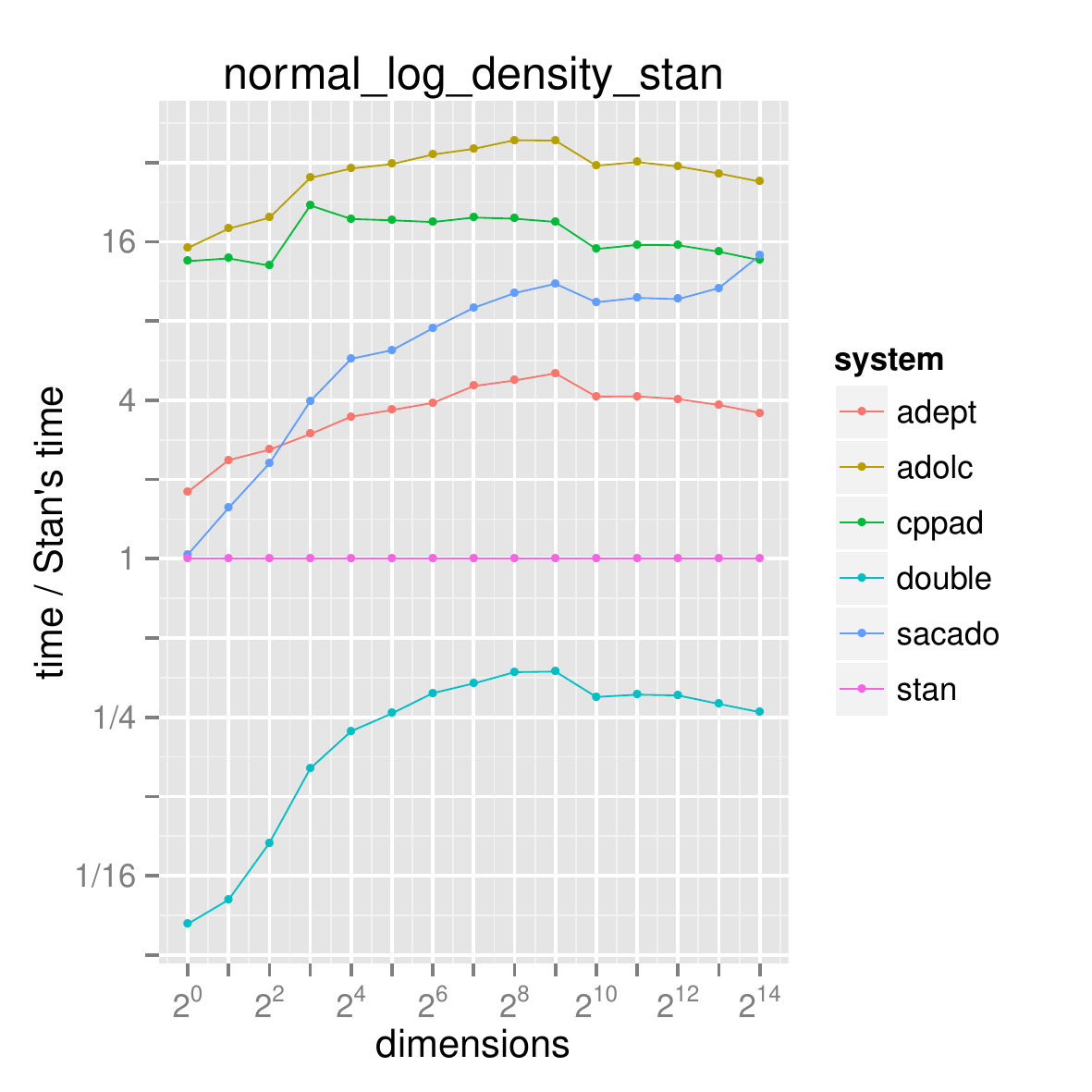}
\end{center}
\vspace*{-12pt}
\caption{\small\it Evaluation of the normal log density function
  implemented directly (left) and using Stan's built-in function
  \code{normal\_log()} for the Stan version (right). Stan's approach
  shows increasing speed relative to the naive double evaluation
  because the proportion of logarithm evaluations shrinks as
  the dimensionality grows.}\label{normal-log-density-eval.figure}
\end{figure}
The timing results are given in \reffigure{normal-log-density-eval}.
For this function, Stan is roughly 50\% faster than the next
fastest system, Adept, with the advantage declining a bit as the
problem size gets larger.  For problems with ten or more evaluations,
Stan's gradient calculation takes a bit more than twenty times as long
as the \code{double}-based evaluation.  Althoug roughly the same speed
as Stan for small problems, Sacado falls further behind as the number
of gradients evaluated increases, whereas CppAd starts very far behind
and asymptotes at roughly half the speed of Stan. 

The second plot in \reffigure{normal-log-density-eval} shows the
relative speed of Stan's built-in, vectorized version of the normal
log density function.  For this evaluation, the functor's operator is
the following;  the \code{fill()} function remains the same as before.
\begin{smallcode}
struct normal_log_density_stan_fun {
  template <typename T>
  T operator()(const Eigen::Matrix<T, Eigen::Dynamic, 1>& x)
    const {

    T mu = -0.56;
    T sigma = 1.37;
    return normal_log_density(x, mu, sigma);
  }
};
\end{smallcode}
The implementation of the log density function for Stan is the following.
\begin{smallcode}
inline stan::math::var 
normal_log_density(const Eigen::Matrix<stan::math::var, 
                                       Eigen::Dynamic, 1>& y, 
                   const stan::math::var& mu, 
                   const stan::math::var& sigma) {
  return stan::prob::normal_log<true>(y, mu, sigma);
}
\end{smallcode}
For the other systems, a standalone \code{normal\_log\_density}
provides the same definition as in the previous evaluation.
\begin{smallcode}
template <typename T>
inline T 
normal_log_density(const Eigen::Matrix<T, Eigen::Dynamic, 1>& y, 
                   const T& mu, const T& sigma) {
  T lp = 0;
  for (int i = 0; i < y.size(); ++i) {
    T z = (y(i) - mu) / sigma;
    lp += -log(sigma) - 0.5 * z * z;
  }
  return lp;
}
\end{smallcode}
The comparison with the \code{double}-based version shows that the
custom version of the normal log density is roughly a factor of six
faster than the naive implementation.  

\section{Previous Work}

Stan's basic pointer-to-implementation pattern and arena-based memory
design was based on Gay's original system RAD \cite{Gay:2005}; RAD is
also the basis for the Sacado automatic differentiation package, which
is part of the Trilinos project \cite{HerouxEtAl:2005}.  The
arena-based memory usage pattern is described in more detail in
\cite{GayAiken:2001}.  In addition to coding slightly different base
classes and memory data structures and responsibilities for recovery,
Stan uses \code{vari} specializations to allow lazy evaluation of
partials and thus save memory and reduce the number of assignments.
Stan also has a value-based return type for its client class
\code{var} rather than the reference-based return approach of RAD for
its function and operator overloads.

\section{Summary and Future Work}

This paper demonstrated the usability of Stan's automatic
differentiation library through simple examples, which showed how to
use the data structures directly or fully encapsulate gradient
calculations of templated C++ functors using Stan's gradient
functional.  The efficiency of Stan was compared to other open-source
C++ libraries, which showed varying relative performance drops
compared to Stan in the different problems evaluated.

The memory usage of Stan's gradient calculations was described
in detail, with various techniques being employed to reduce memory
usage through lazy evaluation and vectorization of operations such as
log density functions.  The Stan automatic differentiation library is
part of the sandalone Stan Math Library, which is extensively tested
for accuracy and instantiability, and has presented a stable interface
through dozens of releases of the larger Stan package.  The Stan Math
Library is distributed under the BSD license, which is compatible with
other open-source licesense such as GPL.  Stan's open-source
development community ensures continued growth of the library in the
future.

Adept's use of expression templates to unfold derivative propagations
at compile time makes it more efficient than Stan at run time in cases
where there are complex expressions on the right-hand sides of
assignments.  The advantage grows as the right-hand side expression
size grows.  There is no reason in principle why the expression
templates of Adept and the underlying efficiency of Stan's data
structures and memory management could not be combined to improve both
systems.

Although not the use case that Stan was developed for, CppAD and
Adol-C's ability to re-use a ``tape'' is potentially useful in some
applications where there are no general while loops or conditionals
that can evaluate differently on different evaluations.

It is clear from the matrix evaluations that there is also more
performance gains to be had for larger matrices, presumably through
enhanced memory locality considerations in both expression graph
construction and in derivative propagation.

The next major features planned for the stan Math Library are inverse
cumulative distribution functions and stiff differential equation
solver(s).  In order to avoid expensive, imprecise, and unstable finite
difference calculations, the latter requires second-order automatic
differentiation to compute Jacobians of a differential equation system
coupled with its sensitivities.

\section*{Acknowledgements}

We would like to thank Brad Bell (author of CppAD), Robin Hogan
(author of Adept), and Andrew Walther (author of Adol-C) for useful
comments on our evaluations of their systems and comments on a draft
of this paper.  We never heard back from the authors of Sacado.

\clearpage
\appendix

\section{Stan Functions}

The following is a comprehensive list of functions supported by Stan
as of version 2.6.  

\subsection{Type Conventions}

\subsubsection{Scalar Functions}

Where argument type \code{real} occurs, any of \code{double},
\code{int}, or \code{stan::math::var} may be used.  Result types with
\code{real} denote \code{var} if any of the arguments contain a
\code{var} and \code{double} otherwise.

\subsubsection{Vector or Array Functions}

Where value type \code{reals} is used, any of the following types may
be used.
\begin{itemize}
\item  Scalars: \
{\small
\code{double},  \
\code{int}, \
\code{var} 
}
\item  Standard vectors: \
{\small
\code{vector<double>}, \
\code{vector<int>}, \
\code{vector<var>}
}
\item Eigen vectors: \
{\small
\code{Matrix<double,\,Dynamic,\,1>}, \
\code{Matrix<var,\,Dynamic,\,1>}
}
\item Eigen row vectors: \\
{\small
\code{Matrix<double,\,Dynamic,\,1>}, \
\code{Matrix<var,\,1,\,Dynamic>}
}
\end{itemize}
If any of the arguments contains a \code{var}, the return type
\code{real} is a \code{var}, otherwise it is \code{double}.

\subsection{C++ Built-in Arithmetic Operators}

{\footnotesize
\begin{tabular}{lll}
{\it function} & {\it arguments} & {\it return} \\
\hline
\code{operator*} & \code{(real x, real y)} & \code{real} \\
\code{operator+} & \code{(real)} & \code{real} \\
\code{operator+} & \code{(real x, real y)} & \code{real} \\
\code{operator-} & \code{(real x)} & \code{real} \\
\code{operator-} & \code{(real x, real y)} & \code{real} \\
\code{operator/} & \code{(real x, real y)} & \code{real} \\
\end{tabular}
}

\subsection{C++ Built-in Relational Operators}

{\footnotesize
\begin{tabular}{lll}
{\it function} & {\it arguments} & {\it return}
\\ \hline
\code{operator!} & \code{(real x)} & \code{int} \\
\code{operator!=} & \code{(real x, real y)} & \code{int} \\
\code{operator\&\&} & \code{(real x, real y)} & \code{int} \\
\code{operator<} & \code{(real x, real y)} & \code{int} \\
\code{operator<=} & \code{(real x, real y)} & \code{int} \\
\code{operator==} & \code{(real x, real y)} & \code{int} \\
\code{operator>} & \code{(real x, real y)} & \code{int} \\
\code{operator>=} & \code{(real x, real y)} & \code{int} \\
\code{operator$||$} & \code{(real x, real y)} & \code{int} \\
\end{tabular}
}

\subsection{C++ \code{cmath} Library Functions}

{\footnotesize
\begin{tabular}{lll}
{\it function} & {\it arguments} & {\it return} 
\\ \hline
\code{abs} & \code{(real x)} & \code{real} \\
\code{acos} & \code{(real x)} & \code{real} \\
\code{acosh} & \code{(real x)} & \code{real} \\
\code{asin} & \code{(real x)} & \code{real}  \\
\code{asinh} & \code{(real x)} & \code{real} \\
\code{atan2} & \code{(real x, real y)} & \code{real} \\
\code{atan} & \code{(real x)} & \code{real} \\
\code{atanh} & \code{(real x)} & \code{real} \\
\code{cos} & \code{(real x)} & \code{real} \\
\code{cosh} & \code{(real x)} & \code{real} \\
\code{cbrt} & \code{(real x)} & \code{real} \\
\code{ceil} & \code{(real x)} & \code{real} \\
\code{erf} & \code{(real x)} & \code{real} \\
\code{erfc} & \code{(real x)} & \code{real} \\
\code{exp2} & \code{(real x)} & \code{real} \\
\code{exp} & \code{(real x)} & \code{real} \\
\code{fdim} & \code{(real x, real y)} & \code{real} \\
\code{floor} & \code{(real x)} & \code{real} \\
\code{fma} & \code{(real x, real y, real z)} & \code{real} \\
\code{fmax} & \code{(real x, real y)} & \code{real} \\
\code{fmin} & \code{(real x, real y)} & \code{real} \\
\code{fmod} & \code{(real x, real y)} & \code{real} \\
\code{hypot} & \code{(real x, real y)} & \code{real} \\
\code{log} & \code{(real x)} & \code{real} \\
\code{log10} & \code{(real x)} & \code{real} \\
\code{log1p} & \code{(real x)} & \code{real} \\
\code{log2} & \code{(real x)} & \code{real} \\
\code{pow} & \code{(real x, real y)} & \code{real} \\
\code{round} & \code{(real x)} & \code{real} \\
\code{sin} & \code{(real x)} & \code{real} \\
\code{sinh} & \code{(real x)} & \code{real} \\
\code{sqrt} & \code{(real x)} & \code{real} \\
\code{tan} & \code{(real x)} & \code{real} \\
\code{tanh} & \code{(real x)} & \code{real} \\
\code{trunc} & \code{(real x)} & \code{real} \\
\end{tabular}
}

\subsection{Special Mathematical Functions}

{\footnotesize
\begin{tabular}{lll}
{\it function} & {\it arguments} & {\it return} 
\\ \hline
\code{bessel\_first\_kind} & \code{(int v, real x)} & \code{real} \\
\code{bessel\_second\_kind} & \code{(int v, real x)} & \code{real} \\
\code{binary\_log\_loss} & \code{(int y, real y\_hat)} & \code{real} \\
\code{binomial\_coefficient\_log} & \code{(real x, real y)} & \code{real} \\
\code{digamma} & \code{(real x)} & \code{real} \\
\code{expm1} & \code{(real x)} & \code{real} \\
\code{fabs} & \code{(real x)} & \code{real} \\
\code{falling\_factorial} & \code{(real x, real n)} & \code{real} \\
\code{gamma\_p} & \code{(real a, real z)} & \code{real} \\
\code{gamma\_q} & \code{(real a, real z)} & \code{real} \\
\code{inv} & \code{(real x)} & \code{real} \\
\code{inv\_cloglog} & \code{(real y)} & \code{real} \\
\code{inv\_logit} & \code{(real y)} & \code{real} \\
\code{inv\_sqrt} & \code{(real x)} & \code{real} \\
\code{inv\_Phi} & \code{(real x)} & \code{real} \\
\code{inv\_square} & \code{(real x)} & \code{real} \\
\code{lbeta} & \code{(real alpha, real beta)} & \code{real} \\
\code{lgamma} & \code{(real x)} & \code{real} \\
\code{lmgamma} & \code{(int n, real x)} & \code{real} \\
\code{log1m} & \code{(real x)} & \code{real} \\
\code{log1m\_exp} & \code{(real x)} & \code{real} \\
\code{log1m\_inv\_logit} & \code{(real x)} & \code{real} \\
\code{log1p\_exp} & \code{(real x)} & \code{real} \\
\code{log\_diff\_exp} & \code{(real x, real y)} & \code{real} \\
\code{log\_falling\_factorial} & \code{(real x, real n)} & \code{real} \\
\code{log\_inv\_logit} & \code{(real x)} & \code{real} \\
\code{log\_mix} & \code{(real theta, real lp1, real lp2)} & \code{real} \\
\code{log\_rising\_factorial} & \code{(real x, real n)} & \code{real} \\
\code{log\_sum\_exp} & \code{(real x, real y)} & \code{real} \\
\code{logit} & \code{(real x)} & \code{real} \\
\code{modified\_bessel\_first\_kind} & \code{(int v, real z)} & \code{real} \\
\code{modified\_bessel\_second\_kind} & \code{(int v, real z)} & \code{real} \\
code{multiply\_log} & \code{(real x, real y)} & \code{real} \\
\code{owens\_t} & \code{(real h, real a)} & \code{real} \\
\code{Phi} & \code{(real x)} & \code{real} \\
\code{Phi\_approx} & \code{(real x)} & \code{real} \\
\code{rising\_factorial} & \code{(real x, real n)} & \code{real} \\
\code{square} & \code{(real x)} & \code{real} \\
\code{step} & \code{(real x)} & \code{real} \\
\code{tgamma} & \code{(real x)} & \code{real} \\
\code{trigamma} & \code{(real x)} & \code{real} \\
\end{tabular}
}

\subsection{Special Control and Test Functions}

{\footnotesize
\begin{tabular}{lll}
{\it function} & {\it arguments} & {\it return} 
\\ \hline
\code{if\_else} & \code{(int cond, real x, real y)} & \code{real} \\
\code{int\_step} & \code{(real x)} & \code{int} \\
\code{is\_inf} & \code{(real x)} & \code{int} \\
\code{is\_nan} & \code{(real x)} & \code{int} \\
\end{tabular}
}

\subsection{Matrix Arithmetic Functions}

{\footnotesize
\begin{tabular}{lll}
{\it function} & {\it arguments} & {\it return} \\
\hline
\code{add} & \code{(matrix x, matrix y)} & \code{matrix} \\
\code{add} & \code{(matrix x, real y)} & \code{matrix} \\
\code{add} & \code{(real x, matrix y)} & \code{matrix} \\
\code{add} & \code{(real x, row\_vector y)} & \code{row\_vector} \\
\code{add} & \code{(real x, vector y)} & \code{vector} \\
\code{add} & \code{(row\_vector x, real y)} & \code{row\_vector} \\
\code{add} & \code{(row\_vector x, row\_vector y)} & \code{row\_vector} \\
\code{add} & \code{(vector x, real y)} & \code{vector} \\
\code{add} & \code{(vector x, vector y)} & \code{vector} 
\\ \hline
\code{columns\_dot\_product} & \code{(matrix x, matrix y)} & \code{row\_vector} \\
\code{columns\_dot\_product} & \code{(row\_vector x, row\_vector y)} & \code{row\_vector} \\
\code{columns\_dot\_product} & \code{(vector x, vector y)} & \code{row\_vector} \\
\hline
\code{columns\_dot\_self} & \code{(matrix x)} & \code{row\_vector} \\
\code{columns\_dot\_self} & \code{(row\_vector x)} & \code{row\_vector} \\
\code{columns\_dot\_self} & \code{(vector x)} & \code{row\_vector} \\
\hline
\code{crossprod} & \code{(matrix x)} & \code{matrix} \\
\hline
\code{diag\_post\_multiply} & \code{(matrix m, row\_vector rv)} & \code{matrix} \\
\code{diag\_post\_multiply} & \code{(matrix m, vector v)} & \code{matrix} \\
\code{diag\_pre\_multiply} & \code{(row\_vector rv, matrix m)} & \code{matrix} \\
\code{diag\_pre\_multiply} & \code{(vector v, matrix m)} & \code{matrix} \\
\hline
\code{divide} & \code{(matrix x, real y)} & \code{matrix} \\
\code{divide} & \code{(row\_vector x, real y)} & \code{row\_vector} \\
\code{divide} & \code{(vector x, real y)} & \code{vector} \\
\hline
\code{dot\_product} & \code{(row\_vector x, row\_vector y)} & \code{real} \\
\code{dot\_product} & \code{(row\_vector x, vector y)} & \code{real} \\
\code{dot\_product} & \code{(vector x, row\_vector y)} & \code{real} \\
\code{dot\_product} & \code{(vector x, vector y)} & \code{real} \\
\hline
\code{dot\_self} & \code{(row\_vector x)} & \code{real} \\
\code{dot\_self} & \code{(vector x)} & \code{real} \\
\hline
\code{elt\_divide} & \code{(matrix x, matrix y)} & \code{matrix} \\
\code{elt\_divide} & \code{(matrix x, real y)} & \code{matrix} \\
\code{elt\_divide} & \code{(real x, matrix y)} & \code{matrix} \\
\code{elt\_divide} & \code{(real x, row\_vector y)} & \code{row\_vector} \\
\code{elt\_divide} & \code{(real x, vector y)} & \code{vector} \\
\code{elt\_divide} & \code{(row\_vector x, real y)} & \code{row\_vector} \\
\code{elt\_divide} & \code{(row\_vector x, row\_vector y)} & \code{row\_vector} \\
\code{elt\_divide} & \code{(vector x, real y)} & \code{vector} \\
\code{elt\_divide} & \code{(vector x, vector y)} & \code{vector} \\
\end{tabular}
}

{\footnotesize
\begin{tabular}{lll}
{\it function} & {\it arguments} & {\it return} \\
\hline
\code{elt\_multiply} & \code{(matrix x, matrix y)} & \code{matrix} \\
\code{elt\_multiply} & \code{(row\_vector x, row\_vector y)} & \code{row\_vector} \\
\code{elt\_multiply} & \code{(vector x, vector y)} & \code{vector} \\
\hline
\code{multiply\_lower\_tri\_self\_transpose} & \code{(matrix x)} & \code{matrix} \\
\hline
\code{multiply} & \code{(matrix x, matrix y)} & \code{matrix} \\
\code{multiply} & \code{(matrix x, real y)} & \code{matrix} \\
\code{multiply} & \code{(matrix x, vector y)} & \code{vector} \\
\code{multiply} & \code{(real x, matrix y)} & \code{matrix} \\
\code{multiply} & \code{(real x, row\_vector y)} & \code{row\_vector} \\
\code{multiply} & \code{(real x, vector y)} & \code{vector} \\
\code{multiply} & \code{(row\_vector x, matrix y)} & \code{row\_vector} \\
\code{multiply} & \code{(row\_vector x, real y)} & \code{row\_vector} \\
\code{multiply} & \code{(row\_vector x, vector y)} & \code{real} \\
\code{multiply} & \code{(vector x, real y)} & \code{vector} \\
\code{multiply} & \code{(vector x, row\_vector y)} & \code{matrix} \\
\hline
\code{mdivide\_right} & \code{(matrix B, matrix A)} & \code{matrix} \\
\code{mdivide\_right} & \code{(row\_vector b, matrix A)} & \code{row\_vector} \\
\hline
\code{mdivide\_left} & \code{(matrix A, matrix B)} & \code{matrix} \\
\code{mdivide\_left} & \code{(matrix A, vector b)} & \code{vector} \\
\hline
\code{quad\_form} & \code{(matrix A, matrix B)} & \code{matrix} \\
\code{quad\_form} & \code{(matrix A, vector B)} & \code{real} \\
\hline
\code{quad\_form\_diag} & \code{(matrix m, row\_vector rv)} & \code{matrix} \\
\code{quad\_form\_diag} & \code{(matrix m, vector v)} & \code{matrix} \\
\hline
\code{quad\_form\_sym} & \code{(matrix A, matrix B)} & \code{matrix} \\
\code{quad\_form\_sym} & \code{(matrix A, vector B)} & \code{real} \\
\hline
\code{rows\_dot\_product} & \code{(matrix x, matrix y)} & \code{vector} \\
\code{rows\_dot\_product} & \code{(row\_vector x, row\_vector y)} & \code{vector} \\
\code{rows\_dot\_product} & \code{(vector x, vector y)} & \code{vector} \\
\hline
\code{rows\_dot\_self} & \code{(matrix x)} & \code{vector} \\
\code{rows\_dot\_self} & \code{(row\_vector x)} & \code{vector} \\
\code{rows\_dot\_self} & \code{(vector x)} & \code{vector} \\
\hline
\code{subtract} & \code{(matrix x)} & \code{matrix} \\
\code{subtract} & \code{(matrix x, matrix y)} & \code{matrix} \\
\code{subtract} & \code{(matrix x, real y)} & \code{matrix} \\
\code{subtract} & \code{(real x)} & \code{real} \\
\code{subtract} & \code{(real x, matrix y)} & \code{matrix} \\
\code{subtract} & \code{(real x, row\_vector y)} & \code{row\_vector} \\
\code{subtract} & \code{(real x, vector y)} & \code{vector} \\
\code{subtract} & \code{(row\_vector x)} & \code{row\_vector} \\
\code{subtract} & \code{(row\_vector x, real y)} & \code{row\_vector} \\
\code{subtract} & \code{(row\_vector x, row\_vector y)} & \code{row\_vector} \\
\code{subtract} & \code{(vector x)} & \code{vector} \\
\code{subtract} & \code{(vector x, real y)} & \code{vector} \\
\code{subtract} & \code{(vector x, vector y)} & \code{vector} \\
\hline
\code{tcrossprod} & \code{(matrix x)} & \code{matrix} \\
\end{tabular}
}

\subsection{Special Matrix Functions}

{\footnotesize
\begin{tabular}{lll}
{\it function} & {\it arguments} & {\it return}  
\\ \hline
\code{cumulative\_sum} & \code{(real[] x)} & \code{real[]} \\
\code{cumulative\_sum} & \code{(row\_vector rv)} & \code{row\_vector} \\
\code{cumulative\_sum} & \code{(vector v)} & \code{vector} \\
\hline
\code{dims} & \code{(T x)} & \code{int[]} \\
\hline
\code{distance} & \code{(row\_vector x, row\_vector y)} & \code{real} \\
\code{distance} & \code{(row\_vector x, vector y)} & \code{real} \\
\code{distance} & \code{(vector x, row\_vector y)} & \code{real} \\
\code{distance} & \code{(vector x, vector y)} & \code{real} \\
\hline
\code{exp} & \code{(matrix x)} & \code{matrix} \\
\code{exp} & \code{(row\_vector x)} & \code{row\_vector} \\
\code{exp} & \code{(vector x)} & \code{vector} \\
\hline
\code{log} & \code{(matrix x)} & \code{matrix} \\
\code{log} & \code{(row\_vector x)} & \code{row\_vector} \\
\code{log} & \code{(vector x)} & \code{vector} \\
\hline
\code{log\_softmax} & \code{(vector x)} & \code{vector} \\
\hline
\code{log\_sum\_exp} & \code{(matrix x)} & \code{real} \\
\code{log\_sum\_exp} & \code{(real x[])} & \code{real} \\
\code{log\_sum\_exp} & \code{(row\_vector x)} & \code{real} \\
\code{log\_sum\_exp} & \code{(vector x)} & \code{real} \\
\hline
\code{max} & \code{(matrix x)} & \code{real} \\
\code{max} & \code{(real x[])} & \code{real} \\
\code{max} & \code{(row\_vector x)} & \code{real} \\
\code{max} & \code{(vector x)} & \code{real} \\
\hline
\code{mean} & \code{(matrix x)} & \code{real} \\
\code{mean} & \code{(real x[])} & \code{real} \\
\code{mean} & \code{(row\_vector x)} & \code{real} \\
\code{mean} & \code{(vector x)} & \code{real} \\
\hline
\code{min} & \code{(int x[])} & \code{int} \\
\code{min} & \code{(matrix x)} & \code{real} \\
\code{min} & \code{(real x[])} & \code{real} \\
\code{min} & \code{(row\_vector x)} & \code{real} \\
\code{min} & \code{(vector x)} & \code{real} \\
\end{tabular}
}

{\footnotesize
\begin{tabular}{lll}
{\it function} & {\it arguments} & {\it return}  \\

\hline
\code{num\_elements} & \code{(T[] x)} & \code{int} \\
\code{num\_elements} & \code{(matrix x)} & \code{int} \\
\code{num\_elements} & \code{(row\_vector x)} & \code{int} \\
\code{num\_elements} & \code{(vector x)} & \code{int} \\
\hline
\code{prod} & \code{(int x[])} & \code{real} \\
\code{prod} & \code{(matrix x)} & \code{real} \\
\code{prod} & \code{(real x[])} & \code{real} \\
\code{prod} & \code{(row\_vector x)} & \code{real} \\
\code{prod} & \code{(vector x)} & \code{real} \\
\hline
\code{sd} & \code{(matrix x)} & \code{real} \\
\code{sd} & \code{(real x[])} & \code{real} \\
\code{sd} & \code{(row\_vector x)} & \code{real} \\
\code{sd} & \code{(vector x)} & \code{real} \\
\hline
\code{softmax} & \code{(vector x)} & \code{vector} \\
\hline
\code{squared\_distance} & \code{(row\_vector x, row\_vector y[])} & \code{real} \\
\code{squared\_distance} & \code{(row\_vector x, vector y[])} & \code{real} \\
\code{squared\_distance} & \code{(vector x, row\_vector y[])} & \code{real} \\
\code{squared\_distance} & \code{(vector x, vector y)} & \code{real} \\
\hline
\code{sum} & \code{(int x[])} & \code{int} \\
\code{sum} & \code{(matrix x)} & \code{real} \\
\code{sum} & \code{(real x[])} & \code{real} \\
\code{sum} & \code{(row\_vector x)} & \code{real} \\
\code{sum} & \code{(vector x)} & \code{real} \\
\hline
\code{variance} & \code{(matrix x)} & \code{real} \\
\code{variance} & \code{(real x[])} & \code{real} \\
\code{variance} & \code{(row\_vector x)} & \code{real} \\
\code{variance} & \code{(vector x)} & \code{real} \\
\end{tabular}
}

\subsection{Matrix and Array Manipulation Functions}

{\footnotesize
\begin{tabular}{lll}
{\it function} & {\it arguments} & {\it return}
\\ \hline
\code{append\_col} & \code{(matrix x, matrix y)} & \code{matrix} \\
\code{append\_col} & \code{(matrix x, vector y)} & \code{matrix}  \\
\code{append\_col} & \code{(row\_vector x, row\_vector y)} & \code{row\_vector} \\
\code{append\_col} & \code{(vector x, matrix y)} & \code{matrix}  \\
\code{append\_col} & \code{(vector x, vector y)} & \code{matrix}  \\
\hline
\code{append\_row} & \code{(matrix x, matrix y)} & \code{matrix} \\
\code{append\_row} & \code{(matrix x, row\_vector y)} & \code{matrix}   \\
\code{append\_row} & \code{(row\_vector x, matrix y)} & \code{matrix}   \\
\code{append\_row} & \code{(row\_vector x, row\_vector y)} & \code{matrix}   \\
\code{append\_row} & \code{(vector x, vector y)} & \code{vector}   \\
\hline
\code{block} & \code{(matrix x, int i, int j, int n\_rows, int n\_cols)} & \code{matrix} \\
\hline
\code{col} & \code{(matrix x, int n)} & \code{vector} \\
\code{cols} & \code{(matrix x)} & \code{int} \\
\code{cols} & \code{(row\_vector x)} & \code{int} \\
\code{cols} & \code{(vector x)} & \code{int} \\
\hline
\code{diag\_matrix} & \code{(vector x)} & \code{matrix} \\
\hline
\code{diagonal} & \code{(matrix x)} & \code{vector} \\
\hline
\code{head} & \code{(T[] sv, int n)} & \code{T[]} \\
\code{head} & \code{(row\_vector rv, int n)} & \code{row\_vector} \\
\code{head} & \code{(vector v, int n)} & \code{vector} \\
\hline
\code{transpose} & \code{(matrix x)} & \code{matrix} \\
\code{transpose} & \code{(row\_vector x)} & \code{vector} \\
\code{transpose} & \code{(vector x)} & \code{row\_vector} \\
\hline
\code{rank} & \code{(int[] v, int s)} & \code{int} \\
\code{rank} & \code{(real[] v, int s)} & \code{int} \\
\code{rank} & \code{(row\_vector v, int s)} & \code{int} \\
\code{rank} & \code{(vector v, int s)} & \code{int} \\
\hline
\code{rep\_array} & \code{(T x, int k, int m, int n)} & \code{T[]} \\
\code{rep\_array} & \code{(T x, int m, int n)} & \code{T[]} \\
\code{rep\_array} & \code{(T x, int n)} & \code{T[]} \\
\hline
\code{rep\_matrix} & \code{(real x, int m, int n)} & \code{matrix} \\
\code{rep\_matrix} & \code{(row\_vector rv, int m)} & \code{matrix} \\
\code{rep\_matrix} & \code{(vector v, int n)} & \code{matrix} \\
\hline
\code{rep\_row\_vector} & \code{(real x, int n)} & \code{row\_vector} \\
\hline
\code{rep\_vector} & \code{(real x, int m)} & \code{vector} \\
\end{tabular}
}

{\footnotesize
\begin{tabular}{lll}
{\it function} & {\it arguments} & {\it return}
\\ \hline
\code{row} & \code{(matrix x, int m)} & \code{row\_vector} \\
\hline
\code{rows} & \code{(matrix x)} & \code{int} \\
\code{rows} & \code{(row\_vector x)} & \code{int} \\
\code{rows} & \code{(vector x)} & \code{int} \\
\hline
\code{segment} & \code{(T[] sv, int i, int n)} & \code{T[]} \\
\code{segment} & \code{(row\_vector v, int i, int n)} & \code{row\_vector} \\
\code{segment} & \code{(vector v, int i, int n)} & \code{vector} \\
\hline
\code{sort\_asc} & \code{(int[] v)} & \code{int[]} \\
\code{sort\_asc} & \code{(real[] v)} & \code{real[]} \\
\code{sort\_asc} & \code{(row\_vector v)} & \code{row\_vector} \\
\code{sort\_asc} & \code{(vector v)} & \code{vector} \\
\hline
\code{sort\_desc} & \code{(int[] v)} & \code{int[]} \\
\code{sort\_desc} & \code{(real[] v)} & \code{real[]} \\
\code{sort\_desc} & \code{(row\_vector v)} & \code{row\_vector} \\
\code{sort\_desc} & \code{(vector v)} & \code{vector} \\
\hline
\code{sort\_indices\_asc} & \code{(int[] v)} & \code{int[]} \\
\code{sort\_indices\_asc} & \code{(real[] v)} & \code{int[]} \\
\code{sort\_indices\_asc} & \code{(row\_vector v)} & \code{int[]} \\
\code{sort\_indices\_asc} & \code{(vector v)} & \code{int[]} \\
\hline
\code{sort\_indices\_desc} & \code{(int[] v)} & \code{int[]} \\
\code{sort\_indices\_desc} & \code{(real[] v)} & \code{int[]} \\
\code{sort\_indices\_desc} & \code{(row\_vector v)} & \code{int[]} \\
\code{sort\_indices\_desc} & \code{(vector v)} & \code{int[]} \\
\hline
\code{sub\_col} & \code{(matrix x, int i, int j, int n\_rows)} & \code{vector} \\
\hline
\code{sub\_row} & \code{(matrix x, int i, int j, int n\_cols)} & \code{row\_vector} \\
\hline
\code{tail} & \code{(T[] sv, int n)} & \code{T[]} \\
\code{tail} & \code{(row\_vector rv, int n)} & \code{row\_vector} \\
\code{tail} & \code{(vector v, int n)} & \code{vector} \\
\hline
\code{to\_array\_1d} & \code{(int[...] a)} & \code{int[]} \\
\code{to\_array\_1d} & \code{(matrix m)} & \code{real[]} \\
\code{to\_array\_1d} & \code{(real[...] a)} & \code{real[]} \\
\code{to\_array\_1d} & \code{(row\_vector v)} & \code{real[]} \\
\code{to\_array\_1d} & \code{(vector v)} & \code{real[]} \\
\hline
\code{to\_array\_2d} & \code{(matrix m)} & \code{real[,]} \\
\hline
\code{to\_matrix} & \code{(int[,] a)} & \code{matrix} \\
\code{to\_matrix} & \code{(matrix m)} & \code{matrix} \\
\code{to\_matrix} & \code{(real[,] a)} & \code{matrix} \\
\code{to\_matrix} & \code{(row\_vector v)} & \code{matrix} \\
\code{to\_matrix} & \code{(vector v)} & \code{matrix} \\
\hline
\code{to\_row\_vector} & \code{(int[] a)} & \code{row\_vector} \\
\code{to\_row\_vector} & \code{(matrix m)} & \code{row\_vector} \\
\code{to\_row\_vector} & \code{(real[] a)} & \code{row\_vector} \\
\code{to\_row\_vector} & \code{(row\_vector v)} & \code{row\_vector} \\
\code{to\_row\_vector} & \code{(vector v)} & \code{row\_vector} \\
\hline
\code{to\_vector} & \code{(int[] a)} & \code{vector} \\
\code{to\_vector} & \code{(matrix m)} & \code{vector} \\
\code{to\_vector} & \code{(real[] a)} & \code{vector} \\
\code{to\_vector} & \code{(row\_vector v)} & \code{vector} \\
\code{to\_vector} & \code{(vector v)} & \code{vector} \\
\end{tabular}
}

\subsection{Linear Algebra Functions}

{\footnotesize
\begin{tabular}{lll}
{\it function} & {\it arguments} & {\it return} 
\\ \hline
\code{cholesky\_decompose} & \code{(matrix A)} & \code{matrix} \\
\code{determinant} & \code{(matrix A)} & \code{real} \\
\code{eigenvalues\_sym} & \code{(matrix A)} & \code{vector} \\
\code{eigenvectors\_sym} & \code{(matrix A)} & \code{matrix} \\
\code{inverse} & \code{(matrix A)} & \code{matrix} \\
\code{inverse\_spd} & \code{(matrix A)} & \code{matrix} \\
\code{log\_determinant} & \code{(matrix A)} & \code{real} \\
\code{mdivide\_left\_tri\_low} & \code{(matrix A, matrix B)} & \code{matrix} \\
\code{mdivide\_left\_tri\_low} & \code{(matrix A, vector B)} & \code{vector} \\
\code{mdivide\_right\_tri\_low} & \code{(matrix B, matrix A)} & \code{matrix} \\
\code{mdivide\_right\_tri\_low} & \code{(row\_vector B, matrix A)} & \code{row\_vector} \\
\code{qr\_Q} & \code{(matrix A)} & \code{matrix} \\
\code{qr\_R} & \code{(matrix A)} & \code{matrix} \\
\code{singular\_values} & \code{(matrix A)} & \code{vector} \\
\code{trace} & \code{(matrix A)} & \code{real} \\
\code{trace\_gen\_quad\_form} & \code{(matrix D,matrix A, matrix B)} & \code{real} \\
\code{trace\_quad\_form} & \code{(matrix A, matrix B)} & \code{real} \\
\end{tabular}
}

\subsection{Probability Functions}

{\footnotesize
\begin{tabular}{lll}
{\it function} & {\it arguments} & {\it return}
\\ \hline
\code{bernoulli\_ccdf\_log} & \code{(ints y, reals theta)} & \code{real}  \\
\code{bernoulli\_cdf} & \code{(ints y, reals theta)} & \code{real}  \\
\code{bernoulli\_cdf\_log} & \code{(ints y, reals theta)} & \code{real}  \\
\code{bernoulli\_log} & \code{(ints y, reals theta)} & \code{real} \\
\code{bernoulli\_logit\_log} & \code{(ints y, reals alpha)} & \code{real}  \\
\hline
\code{beta\_binomial\_ccdf\_log} & \code{(ints n, ints N, reals alpha, reals beta)} & \code{real} \\
\code{beta\_binomial\_cdf} & \code{(ints n, ints N, reals alpha, reals beta)} & \code{real} \\
\code{beta\_binomial\_cdf\_log} & \code{(ints n, ints N, reals alpha, reals beta)} & \code{real} \\
\code{beta\_binomial\_log} & \code{(ints n, ints N, reals alpha, reals beta)} & \code{real} \\
\hline
\code{beta\_ccdf\_log} & \code{(reals theta, reals alpha, reals beta)} & \code{real} \\
\code{beta\_cdf} & \code{(reals theta, reals alpha, reals beta)} & \code{real} \\
\code{beta\_cdf\_log} & \code{(reals theta, reals alpha, reals beta)} & \code{real} \\
\code{beta\_log} & \code{(reals theta, reals alpha, reals beta)} & \code{real} \\
\hline 
\code{binomial\_ccdf\_log} & \code{(ints n, ints N, reals theta)} & \code{real} \\
\code{binomial\_cdf} & \code{(ints n, ints N, reals theta)} & \code{real} \\
\code{binomial\_cdf\_log} & \code{(ints n, ints N, reals theta)} & \code{real} \\
\code{binomial\_log} & \code{(ints n, ints N, reals theta)} & \code{real} \\
\hline
\code{binomial\_logit\_log} & \code{(ints n, ints N, reals alpha)} & \code{real} \\
\hline
\code{categorical\_log} & \code{(ints y, vector theta)} & \code{real} \\
\hline
\code{categorical\_logit\_log} & \code{(ints y, vector beta)} & \code{real} \\
\hline
\code{cauchy\_ccdf\_log} & \code{(reals y, reals mu, reals sigma)} & \code{real} \\
\code{cauchy\_cdf} & \code{(reals y, reals mu, reals sigma)} & \code{real} \\
\code{cauchy\_cdf\_log} & \code{(reals y, reals mu, reals sigma)} & \code{real} \\
\code{cauchy\_log} & \code{(reals y, reals mu, reals sigma)} & \code{real} \\
\hline
\code{chi\_square\_ccdf\_log} & \code{(reals y, reals nu)} & \code{real} \\
\code{chi\_square\_cdf} & \code{(reals y, reals nu)} & \code{real} \\
\code{chi\_square\_cdf\_log} & \code{(reals y, reals nu)} & \code{real} \\
\code{chi\_square\_log} & \code{(reals y, reals nu)} & \code{real} \\
\hline
\code{dirichlet\_log} & \code{(vector theta, vector alpha)} & \code{real} \\
\hline
\code{double\_exponential\_ccdf\_log} & \code{(reals y, reals mu, reals sigma)} & \code{real} \\
\code{double\_exponential\_cdf} & \code{(reals y, reals mu, reals sigma)} & \code{real} \\
\code{double\_exponential\_cdf\_log} & \code{(reals y, reals mu, reals sigma)} & \code{real} \\
\code{double\_exponential\_log} & \code{(reals y, reals mu, reals sigma)} & \code{real} \\
\hline
\code{exp\_mod\_normal\_ccdf\_log} & \code{(reals y, reals mu,} \\ 
& \code{\ reals sigma reals lambda)} & \code{real} \\
\code{exp\_mod\_normal\_cdf} & \code{(reals y, reals mu,} \\
& \code{\ reals sigma reals lambda)} & \code{real} \\
\code{exp\_mod\_normal\_cdf\_log} & \code{(reals y, reals mu,} \\
& \code{\ reals sigma reals lambda)} & \code{real} \\
\code{exp\_mod\_normal\_log} & \code{(reals y, reals mu,} \\
& \code{\ reals sigma reals lambda)} & \code{real} \\
\hline
\code{exponential\_ccdf\_log} & \code{(reals y, reals beta)} & \code{real} \\
\code{exponential\_cdf} & \code{(reals y, reals beta)} & \code{real} \\
\code{exponential\_cdf\_log} & \code{(reals y, reals beta)} & \code{real} \\
\code{exponential\_log} & \code{(reals y, reals beta)} & \code{real} \\
\end{tabular}
}

{\footnotesize
\begin{tabular}{lll}
{\it function} & {\it arguments} & {\it return}
\\ \hline
\code{frechet\_ccdf\_log} & \code{(reals y, reals alpha, reals sigma)} & \code{real} \\
\code{frechet\_cdf} & \code{(reals y, reals alpha, reals sigma)} & \code{real} \\
\code{frechet\_cdf\_log} & \code{(reals y, reals alpha, reals sigma)} & \code{real} \\
\code{frechet\_log} & \code{(reals y, reals alpha, reals sigma)} & \code{real} \\
\hline
\code{gamma\_ccdf\_log} & \code{(reals y, reals alpha, reals beta)} & \code{real} \\
\code{gamma\_cdf} & \code{(reals y, reals alpha, reals beta)} & \code{real} \\
\code{gamma\_cdf\_log} & \code{(reals y, reals alpha, reals beta)} & \code{real} \\
\code{gamma\_log} & \code{(reals y, reals alpha, reals beta)} & \code{real} \\
\hline
\code{gaussian\_dlm\_obs\_log} & \code{(matrix y, matrix F, matrix G,}
\\
& \code{\ matrix V, matrix W, vector m0, matrix C0)} & \code{real} \\
\code{gaussian\_dlm\_obs\_log} & \code{(matrix y, matrix F, matrix G,} \\
& \code{\ vector V, matrix W, vector m0, matrix C0)} & \code{real} \\
\hline
\code{gumbel\_ccdf\_log} & \code{(reals y, reals mu, reals beta)} & \code{real} \\
\code{gumbel\_cdf} & \code{(reals y, reals mu, reals beta)} & \code{real} \\
\code{gumbel\_cdf\_log} & \code{(reals y, reals mu, reals beta)} & \code{real} \\
\code{gumbel\_log} & \code{(reals y, reals mu, reals beta)} & \code{real} \\
\hline
\code{hypergeometric\_log} & \code{(int n, int N, int a, int b)} & \code{real} \\
\hline
\code{inv\_chi\_square\_ccdf\_log} & \code{(reals y, reals nu)} & \code{real} \\
\code{inv\_chi\_square\_cdf} & \code{(reals y, reals nu)} & \code{real} \\
\code{inv\_chi\_square\_cdf\_log} & \code{(reals y, reals nu)} & \code{real} \\
\code{inv\_chi\_square\_log} & \code{(reals y, reals nu)} & \code{real} \\
\hline
\code{inv\_gamma\_ccdf\_log} & \code{(reals y, reals alpha, reals beta)} & \code{real} \\
\code{inv\_gamma\_cdf} & \code{(reals y, reals alpha, reals beta)} & \code{real} \\
\code{inv\_gamma\_cdf\_log} & \code{(reals y, reals alpha, reals beta)} & \code{real} \\
\code{inv\_gamma\_log} & \code{(reals y, reals alpha, reals beta)} & \code{real} \\
\hline
\code{inv\_wishart\_log} & \code{(matrix W, real nu, matrix Sigma)} & \code{real} \\
\hline
\code{lkj\_corr\_cholesky\_log} & \code{(matrix L, real eta)} & \code{real} \\
\code{lkj\_corr\_log} & \code{(matrix y, real eta)} & \code{real} \\
\hline
\code{logistic\_ccdf\_log} & \code{(reals y, reals mu, reals sigma)} & \code{real} \\
\code{logistic\_cdf} & \code{(reals y, reals mu, reals sigma)} & \code{real} \\
\code{logistic\_cdf\_log} & \code{(reals y, reals mu, reals sigma)} & \code{real} \\
\code{logistic\_log} & \code{(reals y, reals mu, reals sigma)} & \code{real} \\
\hline
\code{lognormal\_ccdf\_log} & \code{(reals y, reals mu, reals sigma)} & \code{real} \\
\code{lognormal\_cdf} & \code{(reals y, reals mu, reals sigma)} & \code{real} \\
\code{lognormal\_cdf\_log} & \code{(reals y, reals mu, reals sigma)} & \code{real} \\
\code{lognormal\_log} & \code{(reals y, reals mu, reals sigma)} & \code{real} \\
\end{tabular}
}

{\footnotesize
\begin{tabular}{lll}
{\it function} & {\it arguments} & {\it return}
\\ \hline
\code{multi\_gp\_cholesky\_log} & \code{(matrix y, matrix L, vector w)} & \code{real} \\
\code{multi\_gp\_log} & \code{(matrix y, matrix Sigma, vector w)} & \code{real} \\
\hline
\code{multi\_normal\_cholesky\_log} & \code{(row\_vectors y, row\_vectors mu, matrix L)} & \code{real} \\
\code{multi\_normal\_cholesky\_log} & \code{(row\_vectors y, vectors mu, matrix L)} & \code{real} \\
\code{multi\_normal\_cholesky\_log} & \code{(vectors y, row\_vectors mu, matrix L)} & \code{real} \\
\code{multi\_normal\_cholesky\_log} & \code{(vectors y, vectors mu, matrix L)} & \code{real} \\
\hline
\code{multi\_normal\_log} & \code{(row\_vectors y, row\_vectors mu, matrix Sigma)} & \code{real} \\
\code{multi\_normal\_log} & \code{(row\_vectors y, vectors mu, matrix Sigma)} & \code{real} \\
\code{multi\_normal\_log} & \code{(vectors y, row\_vectors mu, matrix Sigma)} & \code{real} \\
\code{multi\_normal\_log} & \code{(vectors y, vectors mu, matrix Sigma)} & \code{real} \\
\hline
\code{multi\_normal\_prec\_log} & \code{(row\_vectors y, row\_vectors mu, matrix Omega)} & \code{real} \\
\code{multi\_normal\_prec\_log} & \code{(row\_vectors y, vectors mu, matrix Omega)} & \code{real} \\
\code{multi\_normal\_prec\_log} & \code{(vectors y, row\_vectors mu, matrix Omega)} & \code{real} \\
\code{multi\_normal\_prec\_log} & \code{(vectors y, vectors mu, matrix Omega)} & \code{real} \\
\hline
\code{multi\_student\_t\_log} & \code{(row\_vectors y, real nu, row\_vectors mu,} \\
& \code{\ matrix Sigma)} & \code{real} \\
\code{multi\_student\_t\_log} & \code{(row\_vectors y, real nu, vectors mu,} \\
& \code{\ matrix Sigma)} & \code{real} \\
\code{multi\_student\_t\_log} & \code{(vectors y, real nu, row\_vectors mu,} \\
& \code{\ matrix Sigma)} & \code{real} \\
\code{multi\_student\_t\_log} & \code{(vectors y, real nu, vectors mu,} \\
& \code{\ matrix Sigma)} & \code{real} \\
\hline
\code{multinomial\_log} & \code{(int[] y, vector theta)} & \code{real} \\
\hline
\code{neg\_binomial\_2\_ccdf\_log} & \code{(ints n, reals mu, reals phi)} & \code{real} \\
\code{neg\_binomial\_2\_cdf} & \code{(ints n, reals mu, reals phi)} & \code{real} \\
\code{neg\_binomial\_2\_cdf\_log} & \code{(ints n, reals mu, reals phi)} & \code{real} \\
\code{neg\_binomial\_2\_log} & \code{(ints y, reals mu, reals phi)} & \code{real} \\
\hline
\code{neg\_binomial\_2\_log\_log} & \code{(ints y, reals eta, reals phi)} & \code{real} \\
\hline
\code{neg\_binomial\_ccdf\_log} & \code{(ints n, reals alpha, reals beta)} & \code{real} \\
\code{neg\_binomial\_cdf} & \code{(ints n, reals alpha, reals beta)} & \code{real} \\
\code{neg\_binomial\_cdf\_log} & \code{(ints n, reals alpha, reals beta)} & \code{real} \\
\code{neg\_binomial\_log} & \code{(ints n, reals alpha, reals beta)} & \code{real} \\
\hline
\code{normal\_ccdf\_log} & \code{(reals y, reals mu, reals sigma)} & \code{real} \\
\code{normal\_cdf} & \code{(reals y, reals mu, reals sigma)} & \code{real} \\
\code{normal\_cdf\_log} & \code{(reals y, reals mu, reals sigma)} & \code{real} \\
\code{normal\_log} & \code{(reals y, reals mu, reals sigma)} & \code{real} \\
\hline
\code{ordered\_logistic\_log} & \code{(int k, real eta, vector c)} & \code{real} \\
\hline
\code{pareto\_ccdf\_log} & \code{(reals y, reals y\_min, reals alpha)} & \code{real} \\
\code{pareto\_cdf} & \code{(reals y, reals y\_min, reals alpha)} & \code{real} \\
\code{pareto\_cdf\_log} & \code{(reals y, reals y\_min, reals alpha)} & \code{real} \\
\code{pareto\_log} & \code{(reals y, reals y\_min, reals alpha)} & \code{real} \\
\hline
\code{pareto\_type\_2\_ccdf\_log} & \code{(reals y, reals mu, reals lambda, reals alpha)} & \code{real} \\
\code{pareto\_type\_2\_cdf} & \code{(reals y, reals mu, reals lambda, reals alpha)} & \code{real} \\
\code{pareto\_type\_2\_cdf\_log} & \code{(reals y, reals mu, reals lambda, reals alpha)} & \code{real} \\
\code{pareto\_type\_2\_log} & \code{(reals y, reals mu, reals lambda, reals alpha)} & \code{real} \\
\end{tabular}
}

{\footnotesize
\begin{tabular}{lll}
{\it function} & {\it arguments} & {\it return}
\\ \hline
\code{poisson\_ccdf\_log} & \code{(ints n, reals lambda)} & \code{real} \\
\code{poisson\_cdf} & \code{(ints n, reals lambda)} & \code{real} \\
\code{poisson\_cdf\_log} & \code{(ints n, reals lambda)} & \code{real} \\
\code{poisson\_log} & \code{(ints n, reals lambda)} & \code{real} \\
\hline
\code{poisson\_log\_log} & \code{(ints n, reals alpha)} & \code{real} \\
\hline
\code{rayleigh\_ccdf\_log} & \code{(real y, real sigma)} & \code{real} \\
\code{rayleigh\_cdf} & \code{(real y, real sigma)} & \code{real} \\
\code{rayleigh\_cdf\_log} & \code{(real y, real sigma)} & \code{real} \\
\code{rayleigh\_log} & \code{(reals y, reals sigma)} & \code{real} \\
\hline
\code{scaled\_inv\_chi\_square\_ccdf\_log} & \code{(reals y, reals nu, reals sigma)} & \code{real} \\
\code{scaled\_inv\_chi\_square\_cdf} & \code{(reals y, reals nu, reals sigma)} & \code{real} \\
\code{scaled\_inv\_chi\_square\_cdf\_log} & \code{(reals y, reals nu, reals sigma)} & \code{real} \\
\code{scaled\_inv\_chi\_square\_log} & \code{(reals y, reals nu, reals sigma)} & \code{real} \\
\hline
\code{skew\_normal\_ccdf\_log} & \code{(reals y, reals mu, reals sigma,} \\
& \code{\ reals alpha)} & \code{real} \\
\code{skew\_normal\_cdf} & \code{(reals y, reals mu, reals sigma,} \\
\code{\ reals alpha)} & \code{real} \\
\code{skew\_normal\_cdf\_log} & \code{(reals y, reals mu, reals sigma,} \\
& \code{\ reals alpha)} & \code{real} \\
\code{skew\_normal\_log} & \code{(reals y, reals mu, reals sigma,} \\
& \code{\ reals alpha)} & \code{real} \\
\hline
\code{student\_t\_ccdf\_log} & \code{(reals y, reals nu, reals mu,} \\
& \code{\ reals sigma)} & \code{real} \\
\code{student\_t\_cdf} & \code{(reals y, reals nu, reals mu,} \\
& \code{\ reals sigma)} & \code{real} \\
\code{student\_t\_cdf\_log} & \code{(reals y, reals nu, reals mu,} \\
& \code{\ reals sigma)} & \code{real} \\
\code{student\_t\_log} & \code{(reals y, reals nu, reals mu,} \\
& \code{\ reals sigma)} & \code{real} \\
\hline
\code{uniform\_ccdf\_log} & \code{(reals y, reals alpha, reals beta)} & \code{real} \\
\code{uniform\_cdf} & \code{(reals y, reals alpha, reals beta)} & \code{real} \\
\code{uniform\_cdf\_log} & \code{(reals y, reals alpha, reals beta)} & \code{real} \\
\code{uniform\_log} & \code{(reals y, reals alpha, reals beta)} & \code{real} \\
\hline
\code{von\_mises\_log} & \code{(reals y, reals mu, reals kappa)} & \code{real} \\
\hline
\code{weibull\_ccdf\_log} & \code{(reals y, reals alpha, reals sigma)} & \code{real} \\
\code{weibull\_cdf} & \code{(reals y, reals alpha, reals sigma)} & \code{real} \\
\code{weibull\_cdf\_log} & \code{(reals y, reals alpha, reals sigma)} & \code{real} \\
\code{weibull\_log} & \code{(reals y, reals alpha, reals sigma)} & \code{real} \\
\hline
\code{wiener\_log} & \code{(reals y, reals alpha, reals tau,} 
\\
& \code{\ reals beta, reals delta)} & \code{real}
\\
\hline
\code{wishart\_log} & \code{(matrix W, real nu, matrix Sigma)} & \code{real} \\
\end{tabular}
}

\clearpage

\bibliographystyle{apalike}
\bibliography{../../bibtex/all}

\end{document}